\documentclass[aps,showpacs,preprintnumbers, amssymb,onecolumn,11pt,preprintnumbers,amsmath,amssymb,floatfix,prd,showpacs,superscriptaddress,nofootinbib,showkeys]{revtex4}

\usepackage{float}
\usepackage{graphics,epsfig}
\usepackage{graphicx}
\usepackage{dcolumn}
\usepackage{bm}
\begin{document}
\baselineskip=0.8 cm
\title{{\bf  Observational constraints on models of the Universe with time variable Gravitational and Cosmological constants along MOG }}
\author{M. Khurshudyan}
\affiliation{{Department of Theoretical Physics, Yerevan State
University, 1 Alex Manookian, 0025, Yerevan, Armenia}}

\author{N. S. Mazhari } 

\author{D. Momeni } 
\affiliation{Eurasian International Center for Theoretical Physics,
Eurasian National University, Astana 010008, Kazakhstan}

\author{R. Myrzakulov}   \affiliation{Eurasian International Center for Theoretical Physics,
Eurasian National University, Astana 010008, Kazakhstan}

\author{M. Raza}   \affiliation{Department of Mathematics, COMSATS Institute of Information Technology, Sahiwal, Pakistan
 } \affiliation{Centre for Optical and Electromagnetic Research, Department of Electrical Engineering, Zhejiang University, Hangzhou,
China}

\vspace*{0.2cm}
\begin{abstract}
The subject of this paper is to investigate the weak regime
covariant scalar-tensor-vector gravity (STVG) theory, known as the
MOdified gravity (MOG)  theory of gravity. First, we show that the
MOG in the absence of scalar fields is converted into
$\Lambda(t),G(t)$ models. Time evolution of the cosmological
parameters for a family of viable models have been investigated.
Numerical results with the cosmological data have been adjusted.
We've introduced a model for dark energy (DE) density and
cosmological constant which involves first order derivatives of
Hubble parameter. To extend this model, correction terms including
the gravitational constant are added. In our scenario, the
cosmological constant is a function of time. To complete the model,
interaction terms between dark energy and dark matter (DM) manually
entered in phenomenological form.  Instead of using the dust model
for DM, we have proposed DM equivalent to a barotropic fluid.  Time
evolution of DM is a function of other cosmological parameters.
Using sophisticated algorithms, the behavior of various quantities
including the densities, Hubble parameter, etc. have been
investigated graphically. The statefinder parameters have been used
for the classification of DE models. Consistency of the numerical
results with experimental data of $SneIa+BAO+CMB$ are studied by
numerical analysis with high accuracy.

\end{abstract}

\maketitle
\newpage
\vspace*{0.2cm}

\section{ A new synthesis of time variable $G,\Lambda$ models as MOG models}
All Cosmological data from different sources testify to the fact
that our world is made of a substance of negative pressure $73\%$ (
dark energy (DE) ), missing mass  $23\%$ ( dark matter (DM))  and
only $4\%$ conductive material
 (baryon matter)
\cite{Riess et al}. DM and DE can have interaction and the
interaction of these is not known in the physics. It is not an
electromagnetic field and metallic material interaction.
Mathematical function is determined phenomenologically because types
of interactions is unknown with an overall classification
interaction function can be written as
$Q=Q(H,\dot{H},\rho_m,\rho_{DE},\rho_{DM},...)$.

\par
Several models have been proposed to explain the universe's
accelerated expansion \cite{jamil1}-\cite{jamil7}. The models can be divided into two general
groups: the first group of models that are needed to correct the
Einstein theory of gravity with a new geometric terms is known as
geometric models. The first of these models is $f(R)$ which is
obtained by replacing the $R$ Ricci curvature with arbitrary $f(R)$
function\cite{H. A. Buchdahl}. The second group of models that are
expansion is attributed to exotic fluids with negative pressure. It
is believed that exotic fluid is a mimic dark energy equation of
state in the present era
 (for a modern review see \cite{Bamba:2013iga},\cite{Sami:2013ssa}
).  Both of these models have different applications and important results of these models are derived as alternative cosmological models
\cite{Motohashi:2010zz,Motohashi:2010tb,Motohashi:2010qj,
Appleby:2009uf,Starobinsky:2007hu}.
\par

Several properties of DE have been studied in numerous
papers\cite{Kiefer:2010pb,Shafieloo:2009ti,Sahni:2008xx}. DE can be
decay  \cite{Alam:2004jy} or reconstruct from different theoretical
models \cite{Sahni:2006pa}. There is no simple and unique model that
can have to describe this exotic energy. Models in which the
Scalar-tensor  fields used are able to solve such complex issues by
simple mathematics to the extend possible
\cite{Gannouji:2006jm,Gannouji:2007im}. So there are very attractive
models to study. A scalar-Tensor model is proposed among all the
different cosmological models. The model is able to explain the DM
and dynamic clusters of galaxies with an additional vector field and
relying only baryonic matter \cite{Moffat}. This model is known as
STVG or  MOG.  MOG can be seen as a covariant theory with
vector-tensor-scalar fields for gravity with the following action:
\begin{eqnarray}
&&S=-\frac{1}{16\pi}\int \frac{1}{G}(R+2\Lambda)\sqrt{-g}d^4x+S_{\phi}+S_{M}\\&& \nonumber-\int \frac{1}{G}\Big[\frac{1}{2}g^{\alpha\beta}\Big(\nabla_{\alpha}\log G\nabla_{\beta}\log G+\nabla_{\alpha}\log \mu\nabla_{\beta}\log \mu\Big)+U_{G}(G)+W_{\mu}(\mu)\Big]\sqrt{-g}d^4x.
\end{eqnarray}
\par
The first term of the action is Einstein-Hilbert Lagrangian. The
second term is the conventional scalar field and the last term
contains a $G$ kinetic energy field that plays the role of the
gravitational constant (However, the fields can be considered
similar to a time dependent gravitational constant by slowly time
varying fields) \cite{Moffat:2011rp}. This action classes are
written in covariant forms and are used to investigate the
astrophysical phenomena such as rotation curves of galaxies, mass
distribution of cosmic clusters or gravitational lenses. The model
might be a suitable alternative to $\Lambda$CDM model considered
\cite{Toth:2010ah}. In order to understand the role of scalar and
vector fields we write the equations of motion for  FLRW metric :
$$
ds^2=dt^2-a(t)^2[(1-kr^2)^{-1}dr^2+r^2d\Omega^2],\ \ d\Omega^2=d\theta^2+\sin^2{\theta}d\phi^2$$
Form of the equations can be rewritten as a
generalized  Friedmann equations as follow
\cite{Moffat:2007ju}:
\begin{eqnarray}
&&H^2+\frac{k}{a^2}=\frac{8\pi G\rho}{3}
-\frac{4\pi}{3}\left(\frac{\dot{G}^2}{G^2}+\frac{\dot{\mu}^2}{\mu^2}-\dot{\omega}^2-G\omega\mu^2\phi_0^2\right)\nonumber\\
&&\qquad{}+\frac{8\pi}{3}\left(
\omega GV_\phi+\frac{V_G}{G^2}+\frac{V_\mu}{\mu^2}+V_\omega
\right)
+\frac{\Lambda}{3}+H\frac{\dot{G}}{G},
\label{eq:FR1}\\
&&\frac{\ddot{a}}{a}=-\frac{4\pi G}{3}(\rho+3p)
+\frac{8\pi}{3}\left(\frac{\dot{G}^2}{G^2}+\frac{\dot{\mu}^2}{\mu^2}-\dot{\omega}^2-G\omega\mu^2\phi_0^2\right)\nonumber\\
&&\qquad{}+\frac{8\pi}{3}\left(
\omega GV_\phi+\frac{V_G}{G^2}+\frac{V_\mu}{\mu^2}+V_\omega
\right)
+\frac{\Lambda}{3}+H\frac{\dot{G}}{2G}+\frac{\ddot{G}}{2G}-\frac{\dot{G}^2}{G^2},\nonumber\\
&&\ddot{G}+3H\dot{G}-\frac{3}{2}\frac{\dot{G}^2}{G}+\frac{G}{2}\left(\frac{\dot{\mu}^2}{\mu^2}-\dot{\omega}^2\right)+\frac{3}{G}V_G-V_G'\nonumber\\
&&\qquad{}+G\left[\frac{V_\mu}{\mu^2}+V_\omega\right]
+\frac{G}{8\pi}\Lambda-\frac{3G}{8\pi}\left(\frac{\ddot{a}}{a}+H^2\right)=0,\\
&&\ddot{\mu}+3H\dot{\mu}-\frac{\dot{\mu}^2}{\mu}-\frac{\dot{G}}{G}\dot{\mu}+G\omega\mu^3\phi_0^2+\frac{2}{\mu}V_\mu-V'_\mu=0,\\
&&\ddot{\omega}+3H\dot{\omega}-\frac{\dot{G}}{G}\dot{\omega}-\frac{1}{2}G\mu^2\phi_0^2+GV_\phi+V'_\omega=0.\label{eq:omega}
\end{eqnarray}

Scalar and vector fields interaction terms of the aforementioned
classes are self interaction and they are shown by an arbitrary
mathematical functions: $V_\phi(\phi)$, $V_G(G)$,
$V_\omega(\omega)$, and $V_\mu(\mu)$. The resulting equations of
motion are highly nonlinear and there is no possibility to find
analytical solutions. The only possible way to evaluate answer is
numerical method. At the same time, we must also determine the shape
of the interaction $V_{i}$. Mathematical differences may be a good
solution for finding certain family of potentials. If we consider
the $G$ scalar field with a time variable gravitational field (G(t))
and ignore the contributions of the other fields in favor of the
G(t), and also due to the cosmological data $\frac{\dot{G}}{G}\ll
1$, time evolution of G(t) will be the major contribution.

In fact, data from the large cosmological confirm our conjecture
about just keeping the $G(t)$, and  kinetic part of $G(t)$ can be
neglected because:
\begin{eqnarray}
&&g^{\alpha\beta}\nabla_{\alpha}\log G\nabla_{\beta}\log G  \simeq (\frac{\dot{G}}{G})^2    \ll 1.
\end{eqnarray}
Regardless, second-order derivatives of additional fields which
introduced additional degrees of freedom and in the absence of
additional fields on MOG, with the approximation that the time
evolution of the fields is very slowly varying, MOG and
Einstein-Hilbert action can be considered as the same. The
difference is that now $G(t)$ is a scalar time variable field.
Equations of motion are written in the following general form, if we
consider small variation of $G(t)$ and $G(t),\Lambda$ are functions
of time \cite{Bonanno:2006xa}:
\begin{eqnarray}
S\simeq-\frac{1}{16\pi}\int \frac{1}{G}(R+2\Lambda)\sqrt{-g}d^4x+S_{M}.
\end{eqnarray}
(see for instance \cite{Abdussattar})
\begin{equation}
R_{\mu\nu}-\frac{1}{2}Rg_{\mu\nu}\approx -8 \pi G(t) \left[ T_{\mu\nu} -
\frac{\Lambda(t)}{8 \pi G(t)}g_{\mu\nu} \right]\label{FEQ},
\end{equation}
Energy-momentum function of matter fields (ordinary or exotic) is proposed as follows:
\begin{eqnarray}
T_{\mu\nu}=\mathcal{L}_{M}g_{\mu\nu}-2\frac{\delta  \mathcal{L}_{M}}{\delta g^{\mu\nu}}.
\end{eqnarray}
 Cosmological models, which were introduced by the mentioned equations of motion have been investigated several times by different authors \cite{Abdussattar,Jamil:2009sq,Lu:2009iv,Sadeghi:2013xca}.
 But we approach this problem with a more general view.  As we have shown, MOG is the limit of weak fields able to induce and introduces a gravitational field $G(t)$. So, our paper can be considered as a cosmological analysis of MOG in the weak field regime. We are particularly interested to see how cosmological data
$SneIa+BAO+CIB$  will constrain our model parameters.\par Our plan
in this paper is: In section II: introducing the cosmological
constant and dark model consist of $\{H,\dot{H},..\}$. In section
III: dynamic extraction of the model and additional equation
governing $G(t)$ and inference different densities. In section IV,
numerical analysis of the equations. In section V, statefinder
parameters $(r,s)$ analysis. In section VI, observational
constraints. The final section is devoted to the results of
references.

\section{Toy models}
A DE model of our interest is described via energy density
$\rho_{D}$ \cite{Chen}:
\begin{equation}\label{eq:rhoD}
\rho_{D}=\alpha\frac{\ddot{H}}{H}+\beta \dot{H}+\gamma H^{2},
\end{equation}
where $\beta$, $\gamma$ are positive constants, while for $\alpha$ in light of the time variable scenario, we suppose that
\begin{equation}\label{eq:alpha}
\alpha(t)=\alpha_{0}+\alpha_{1} G(t)+\alpha_{2} t \frac{\dot{G}(t)}{G(t)},
\end{equation}
where $\alpha_{0}$, $\alpha_{1}$ and $\alpha_{2}$ are positive
constants and $G(t)$ is a varying gravitational constant. Its a
generalization of Ricci dark energy scenario \cite{riccide} to
higher derivatives terms of Hubble parameter.  An interaction term
$Q$ between DE and a barotropic fluid $P_{b}=\omega_{b}\rho_{b}$is
taken to be
\begin{equation}\label{eq:Q}
Q=3Hb(\rho_{b}+\rho_{D})
\end{equation}
We propose three phenomenological models for DE as the following:
\begin{enumerate}
\item The first model is the simplest one, in which we assume that time variable cosmological constant has the same order of energy as the density of DE.
$$\Lambda(t)=\rho_{D},$$ In this model, $\rho_{D}$ is determined using continuity equation with a dissipative interaction term Q.
\item Secondly, generalization of cosmological constant is proposed as a modified Ricci DE model to time variable scenario has an oscillatory form in terms of H.
$$\Lambda(t)=\rho_{b}\sin^{3}{(tH)}+\rho_{D}\cos{(tH)},$$
Note that if we think on trigonometric term as oscillatory term, the
amplitudes of the oscillations are assumed to be proportional to the
barotropic and DE components. Meanwhile these coefficients satisfy
continuity equations.
\item
The last toy model is inspired from the small variation of G(t) and
a logarithmic term of H. Here, coefficients are written in the forms
of barotropic and DE densities .
$$\Lambda(t)=\rho_{b}\ln{(tH)}+\rho_{D}\sin{\left (t\frac{\dot{G}(t)}{G(t)} \right )}.$$  In this model, a time dependent and G variable assumption is imposed.
\end{enumerate}
Following the suggested models we will study time evolution and
cosmological predictions of our cosmological model. Furthermore, we
will compare the numerical results with a package of observational
data.

\section{Dynamic of  models}
 By using the
following FRW metric for a flat Universe,
\begin{equation}\label{s2}
ds^2=-dt^2+a(t)^2\left(dr^{2}+r^{2}d\Omega^{2}\right),
\end{equation}
field equations (\ref{FEQ}) can be reduced to the following Friedmann equations,
\begin{equation}\label{eq: Fridmman vlambda}
H^{2}=\frac{\dot{a}^{2}}{a^{2}}=\frac{8\pi G(t)\rho}{3}+\frac{\Lambda(t)}{3},
\end{equation}
and,
\begin{equation}\label{eq:fridman2}
\frac{\ddot{a}}{a}=-\frac{4\pi
G(t)}{3}(\rho+3P)+\frac{\Lambda(t)}{3},
\end{equation}
where $d\Omega^{2}=d\theta^{2}+\sin^{2}\theta d\phi^{2}$, and $a(t)$
represents the scale factor. \\
Energy conservation law $T^{;j}_{ij}=0$ reads as,
\begin{equation}\label{eq:conservation}
\dot{\rho}+3H(\rho+P)=0.
\end{equation}
Combination of (\ref{eq: Fridmman vlambda}), (\ref{eq:fridman2}) and (\ref{eq:conservation}) gives the relationship between $\dot{G}(t)$ and $\dot{\Lambda}(t)$
\begin{equation}\label{eq:glambda}
\dot{G}=-\frac{\dot{\Lambda}}{8\pi\rho}.
\end{equation}
To introduce an interaction between DE and DM (\ref{eq:conservation}) we should mathematically split it into two following equations
\begin{equation}\label{eq:inteqm}
\dot{\rho}_{DM}+3H(\rho_{DM}+P_{DM})=Q,
\end{equation}
and
\begin{equation}\label{eq:inteqG}
\dot{\rho}_{DE}+3H(\rho_{DE}+P_{DE})=-Q.
\end{equation}
For the barotropic fluid with $P_{b}=\omega_{b}\rho_{b}$ (\ref{eq:inteqm}) will take following form
\begin{equation}
\dot{\rho}_{b}+3H(1+\omega_{b}-b)\rho_{b}=3Hb\rho_{D}.
\end{equation}
Pressure of the DE can be recovered from (\ref{eq:inteqG})
\begin{equation}
P_{D}=-\rho_{D}-\frac{\dot{\rho}_{D}}{3H}-b\frac{3H^{2}-\Lambda(t)}{8 \pi G(t)}.
\end{equation}
Therefore with a fixed form of $\Lambda(t)$ we will be able to
observe behavior of $P_{D}$. Cosmological parameters of our interest
are EoS parameters of DE $\omega_{D}=P_{D}/\rho_{D}$, EoS parameter
of composed fluid
$$\omega_{tot}=\frac{P_{b}+P_{D} }{\rho_{b}+\rho_{D}},$$
deceleration parameter $q$, which can be written as
\begin{equation}\label{eq:accchange}
q=\frac{1}{2}(1+3\frac{P}{\rho} ),
\end{equation}
where $P=P_{b}+P_{D}$ and $\rho=\rho_{b}+\rho_{D}$.
We have a full system of equations of motion and interaction terms. Now we are ready to investigate cosmological predictions of our model.
\section{Numerical analysis of the  Cosmological parameters}
In next sections we fully analyze time evolution of three models of
DE. Using numerical integration, we will show that how cosmological
parameters $H,G(t),q,w_{\text{tot}}$, and time decay rate
$\frac{d\log G}{dt}$ and densities $\rho_D,..$ change. We fit
parameters like $H_0,$ etc from observational data.

\subsection{Model 1: $\Lambda(t)=\rho_{D}$}
\begin{figure}[h!]
 \begin{center}$
 \begin{array}{cccc}
\includegraphics[width=50 mm]{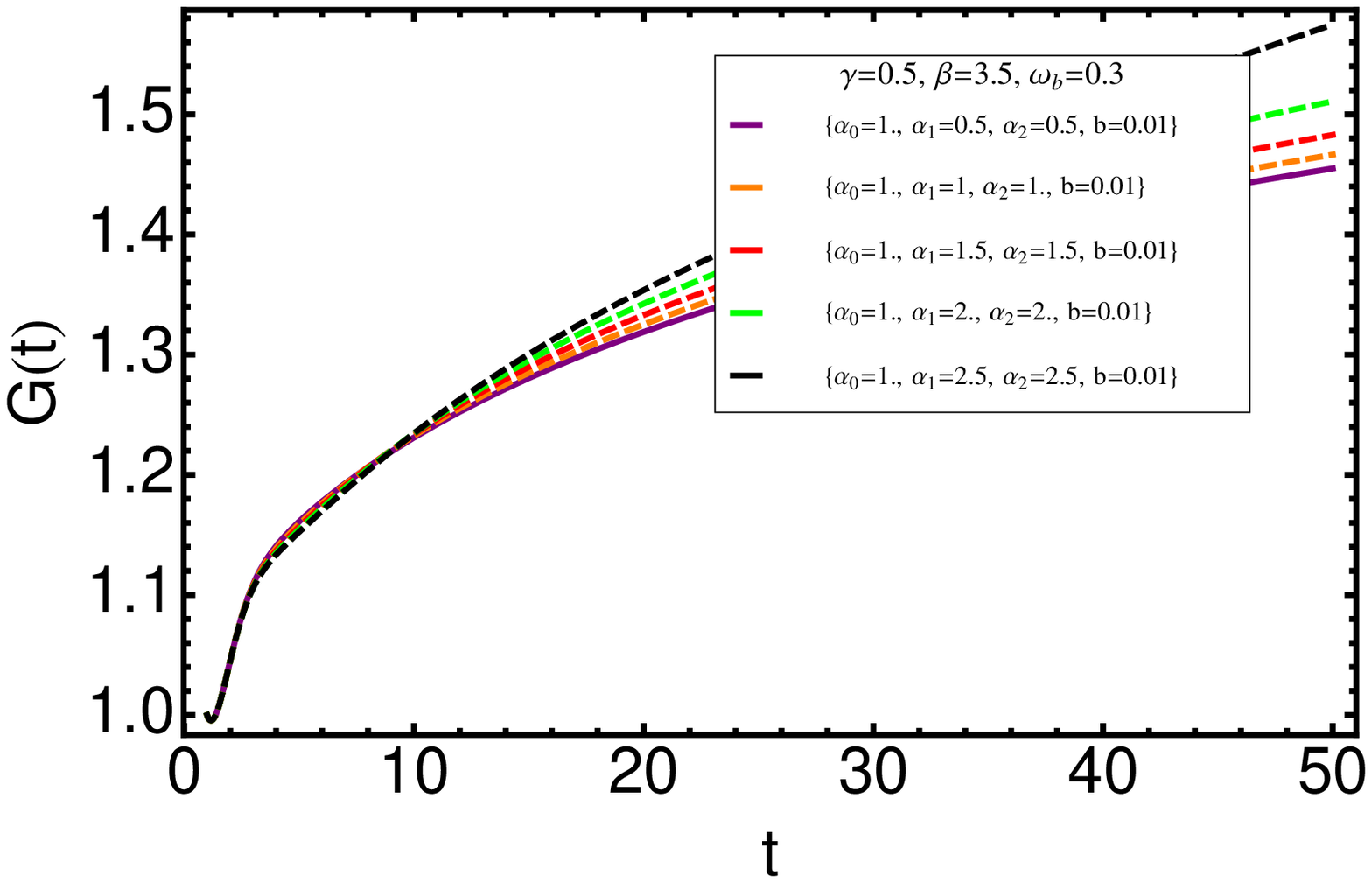} &
\includegraphics[width=50 mm]{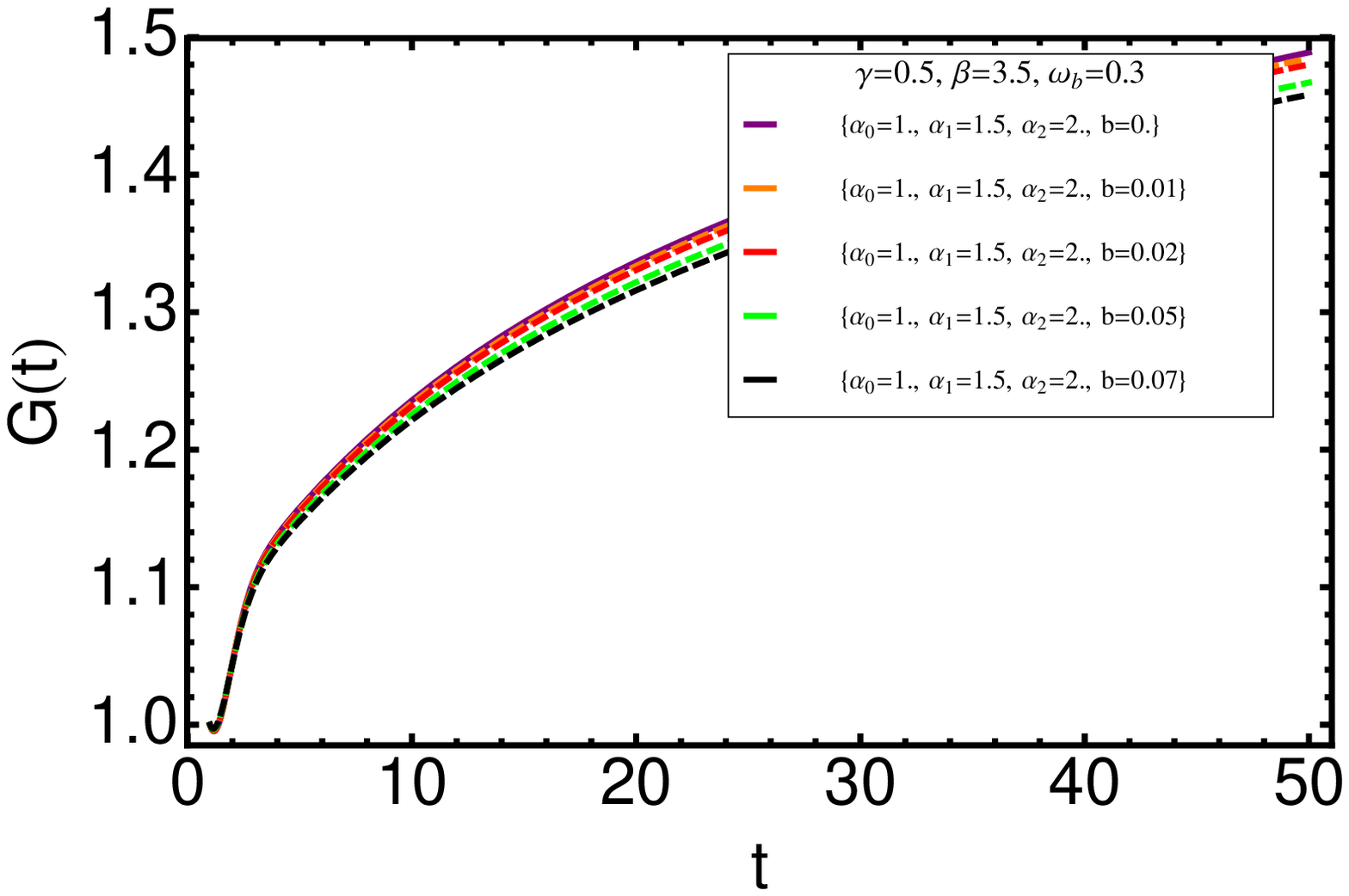}\\
\includegraphics[width=50 mm]{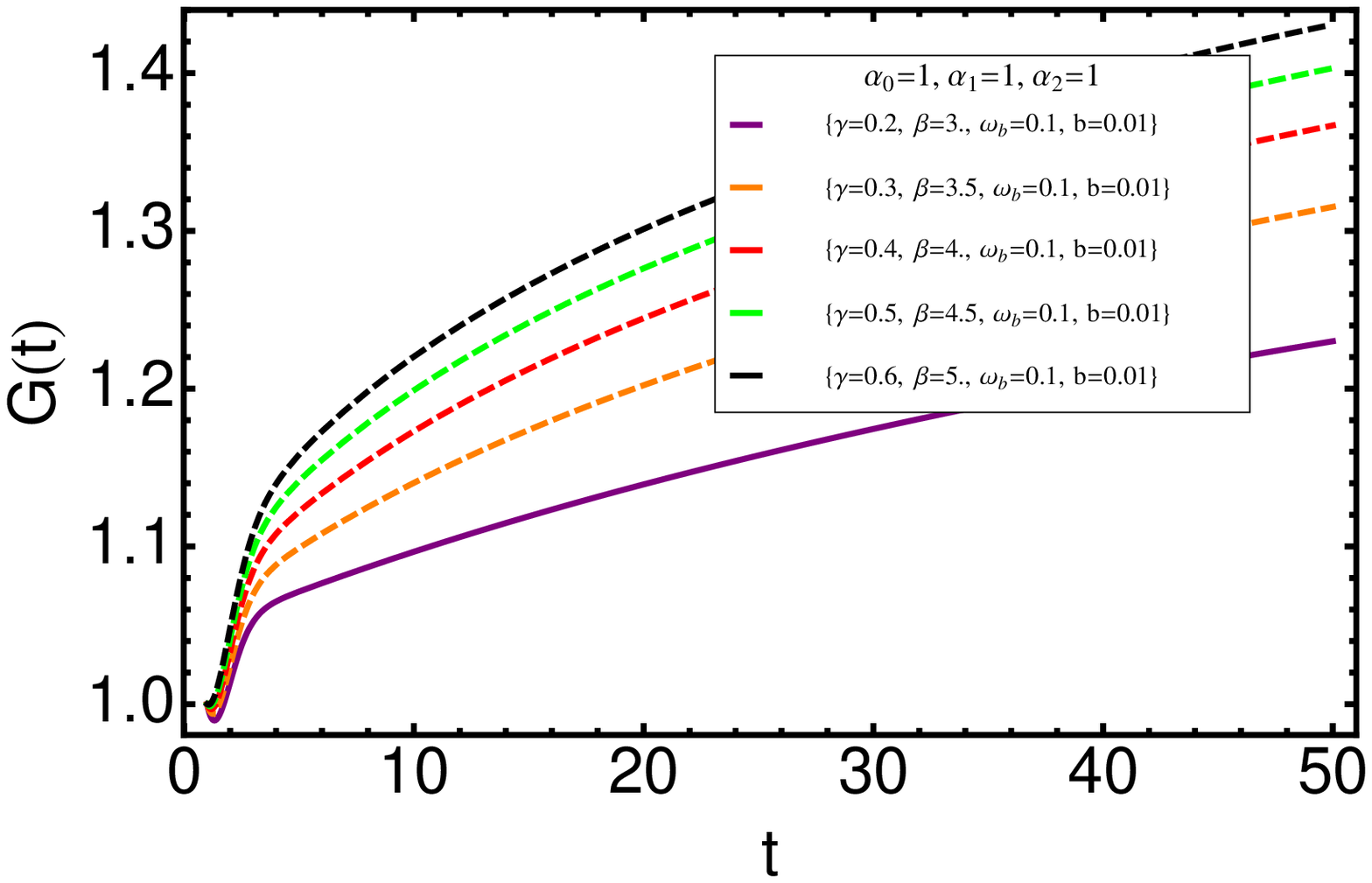} &
\includegraphics[width=50 mm]{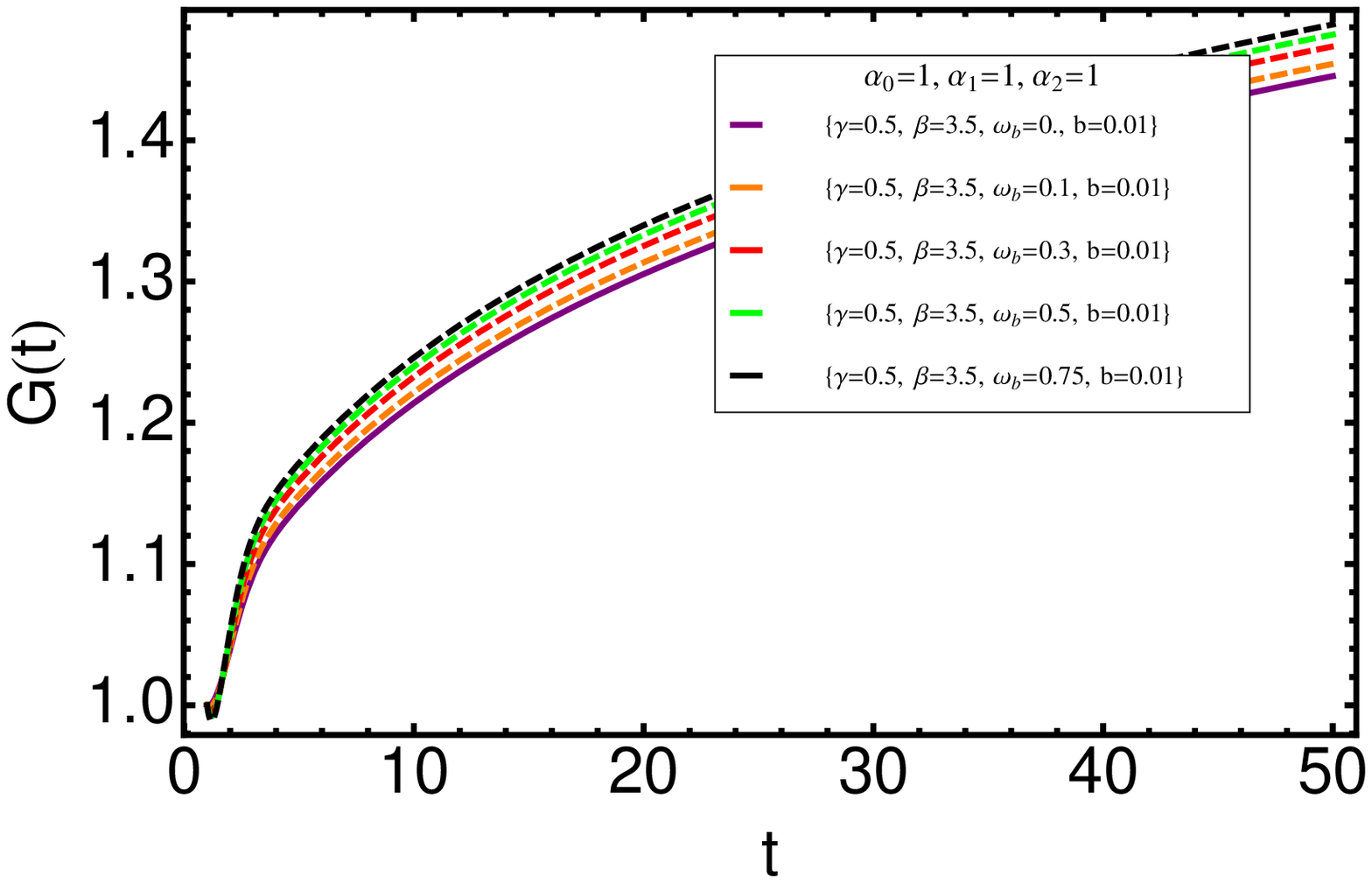}
 \end{array}$
 \end{center}
\caption{Behavior of Gravitational constant $G(t)$ against $t$ for Model 1.}
 \label{fig:1}
\end{figure}

\begin{figure}[h!]
 \begin{center}$
 \begin{array}{cccc}
\includegraphics[width=50 mm]{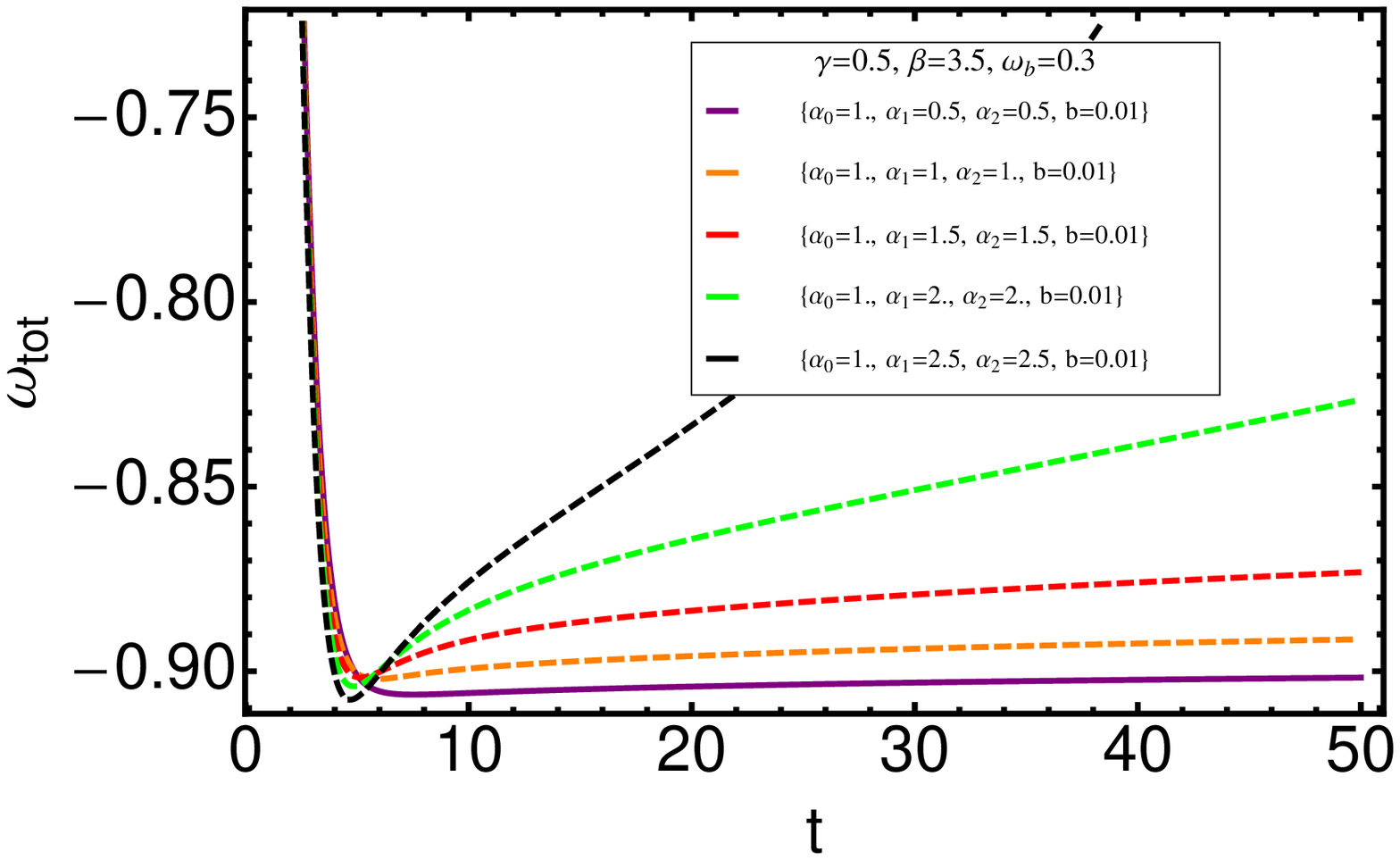} &
\includegraphics[width=50 mm]{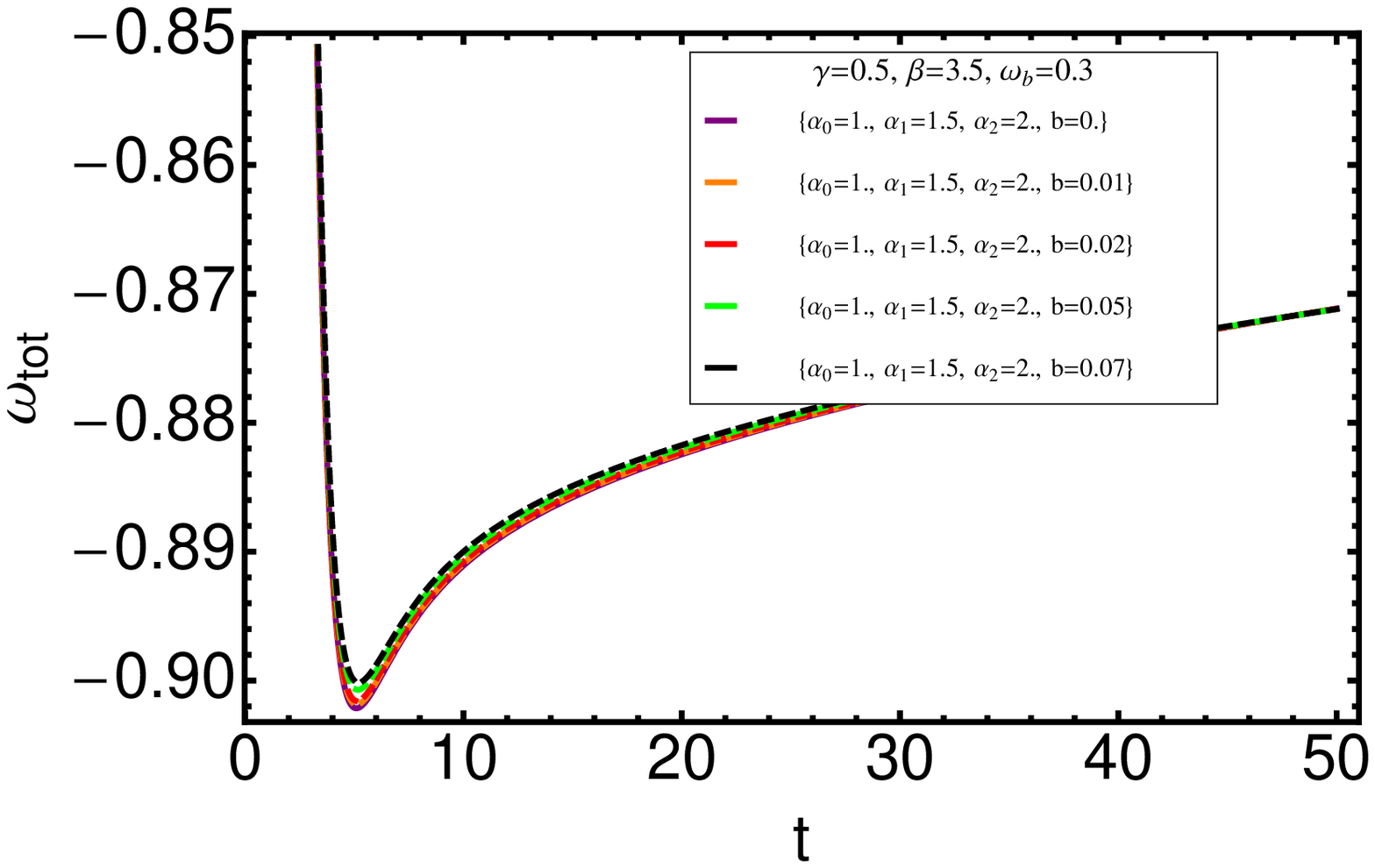}\\
\includegraphics[width=50 mm]{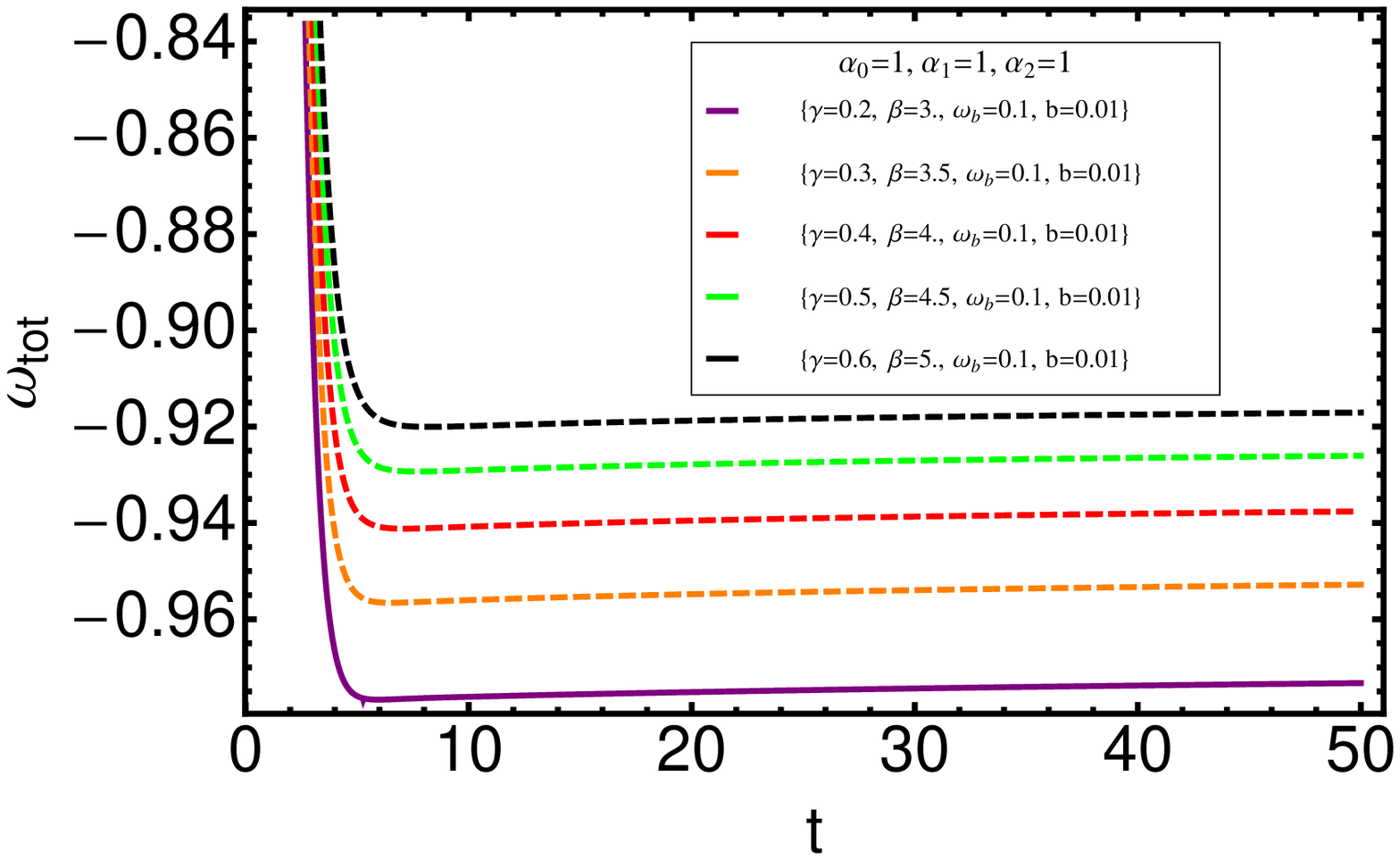} &
\includegraphics[width=50 mm]{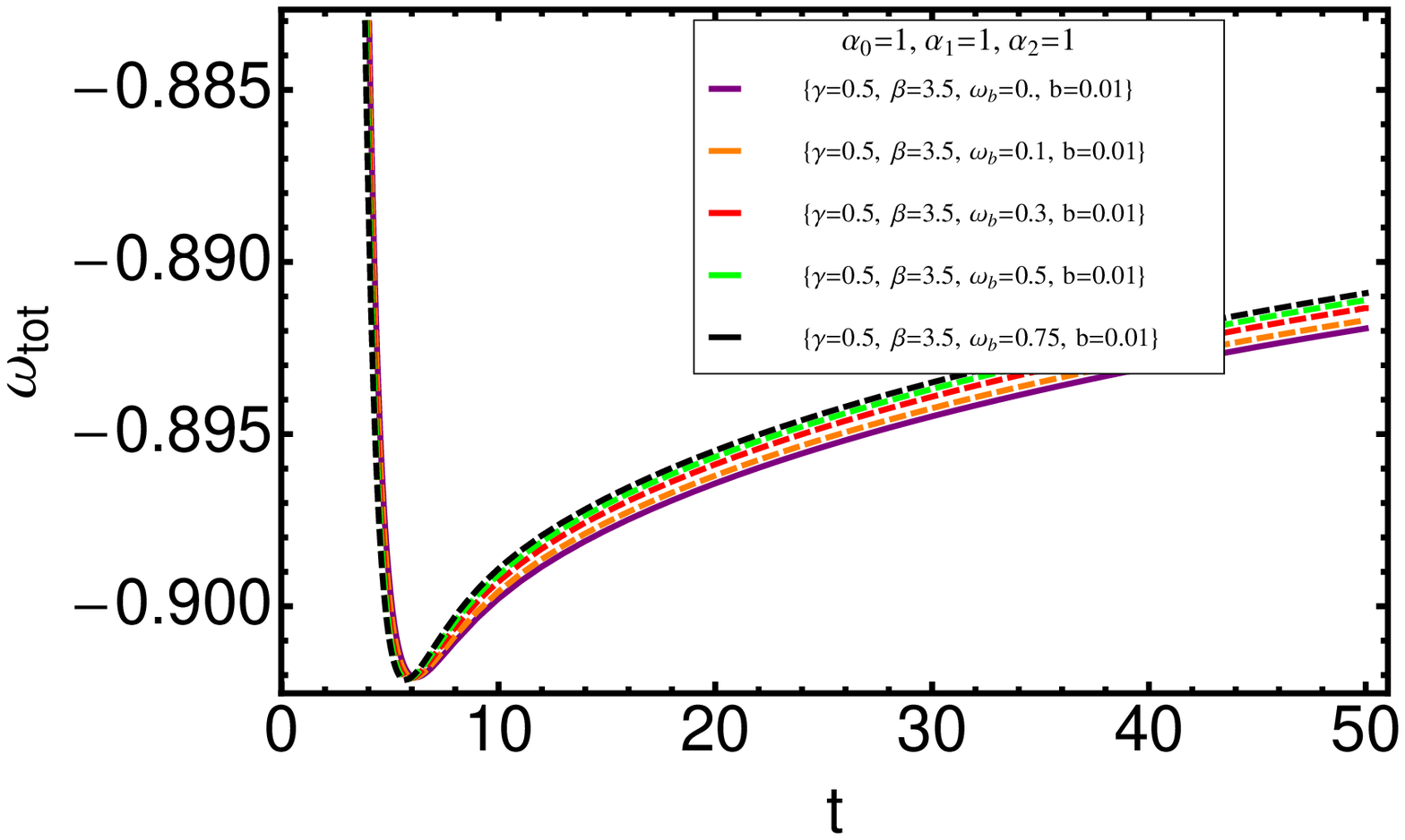}
 \end{array}$
 \end{center}
\caption{Behavior of EoS parameter $\omega_{tot}$ against $t$ for Model 1.}
 \label{fig:2}
\end{figure}

\begin{figure}[h!]
 \begin{center}$
 \begin{array}{cccc}
\includegraphics[width=50 mm]{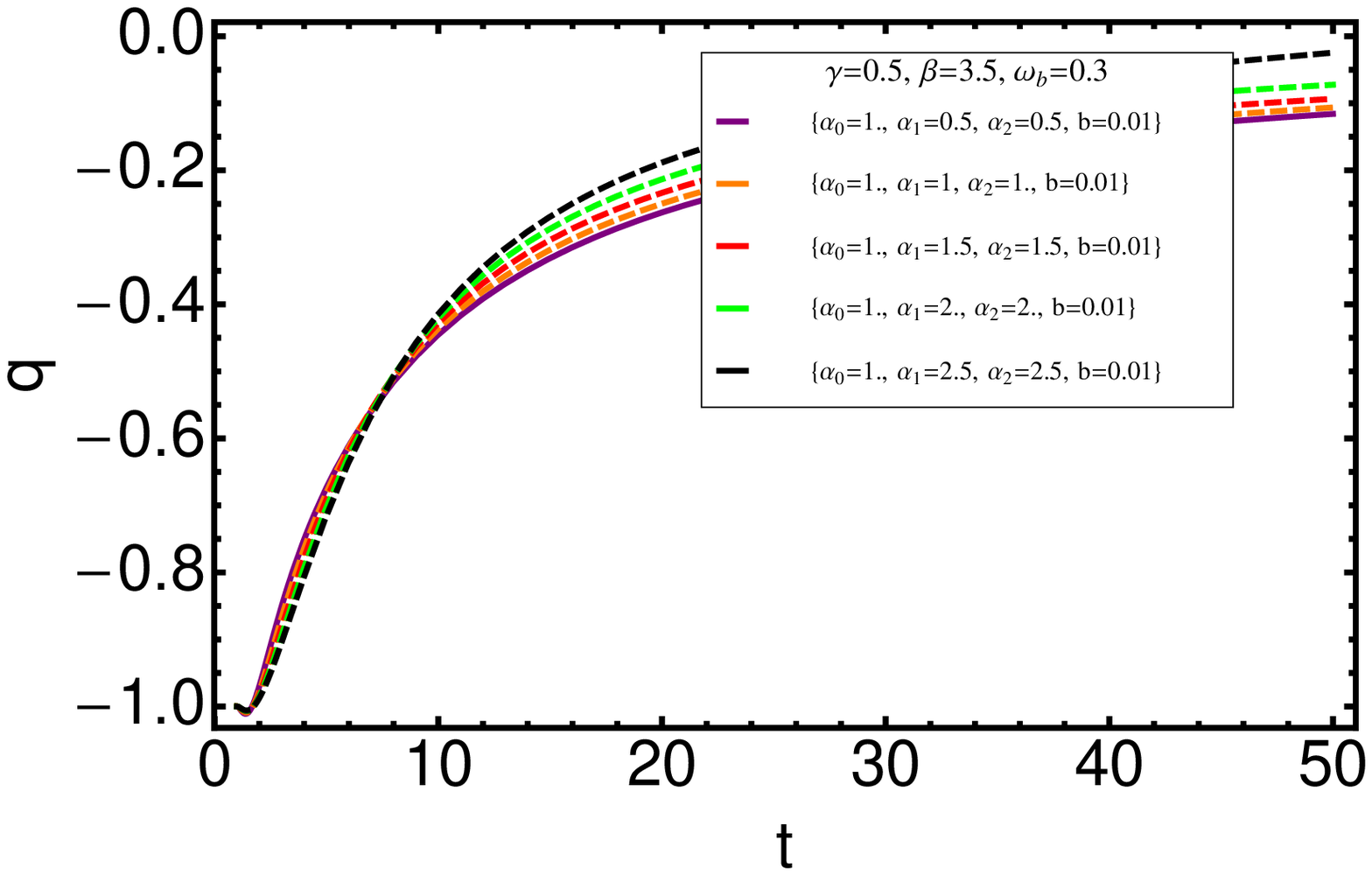} &
\includegraphics[width=50 mm]{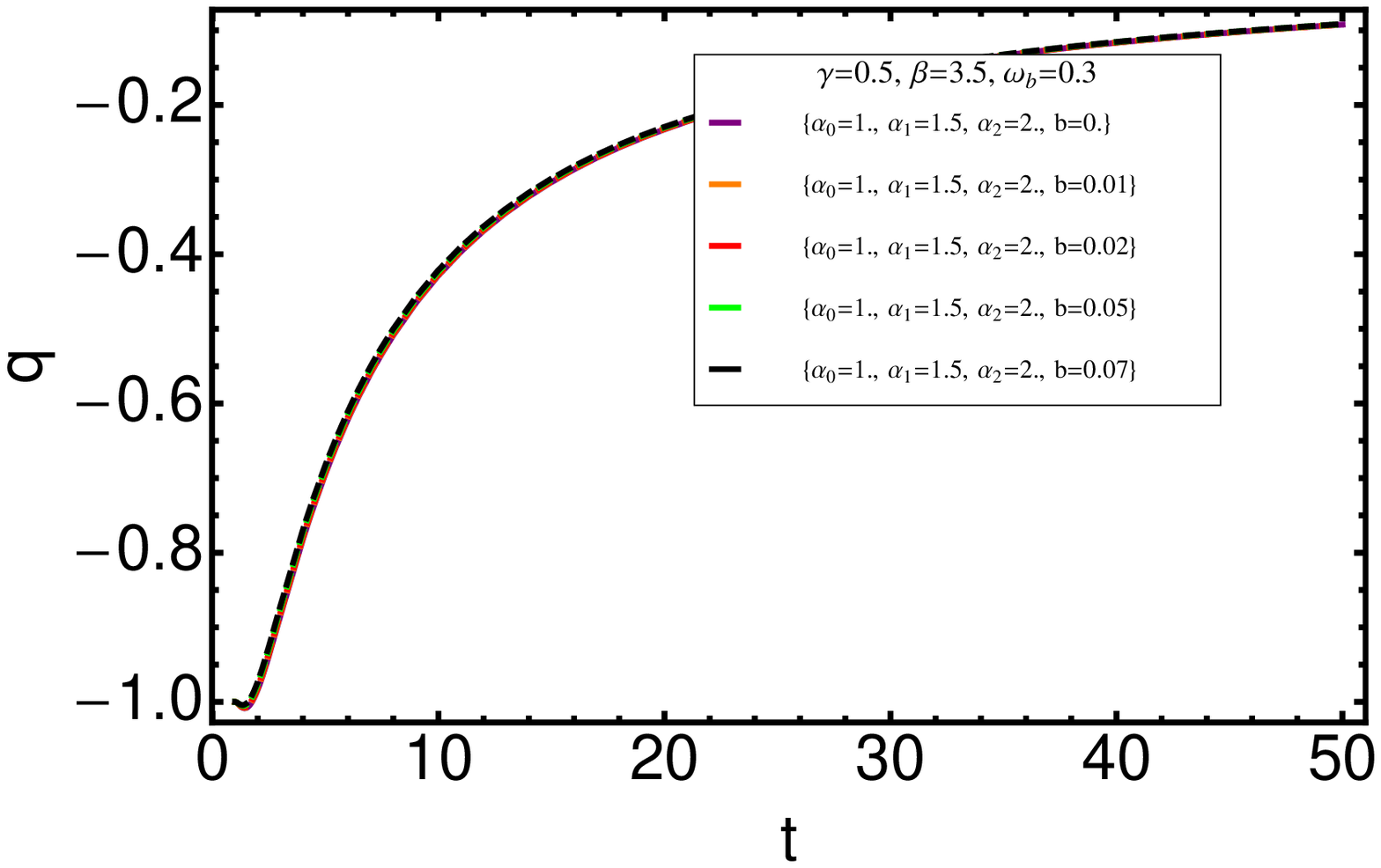}\\
\includegraphics[width=50 mm]{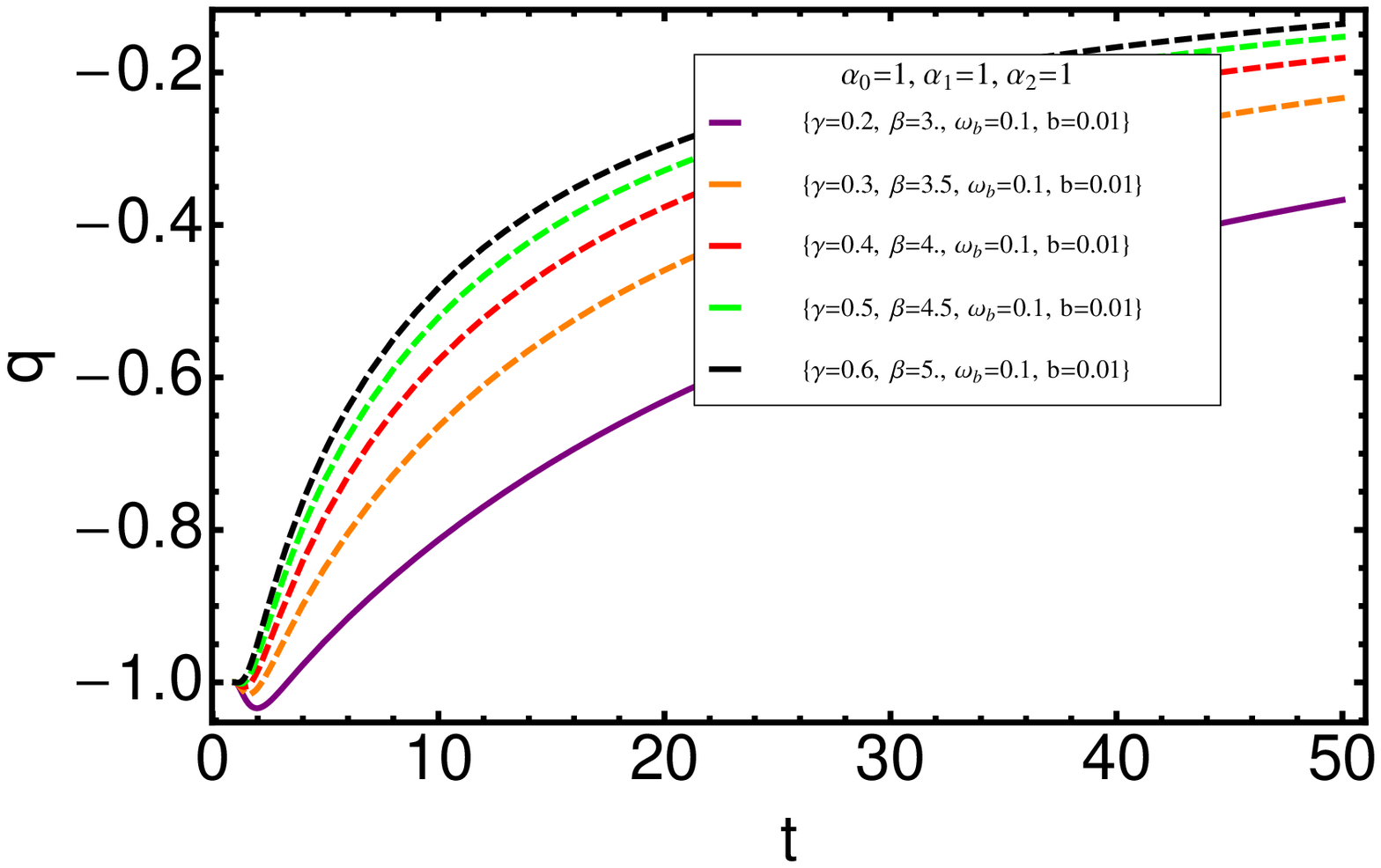} &
\includegraphics[width=50 mm]{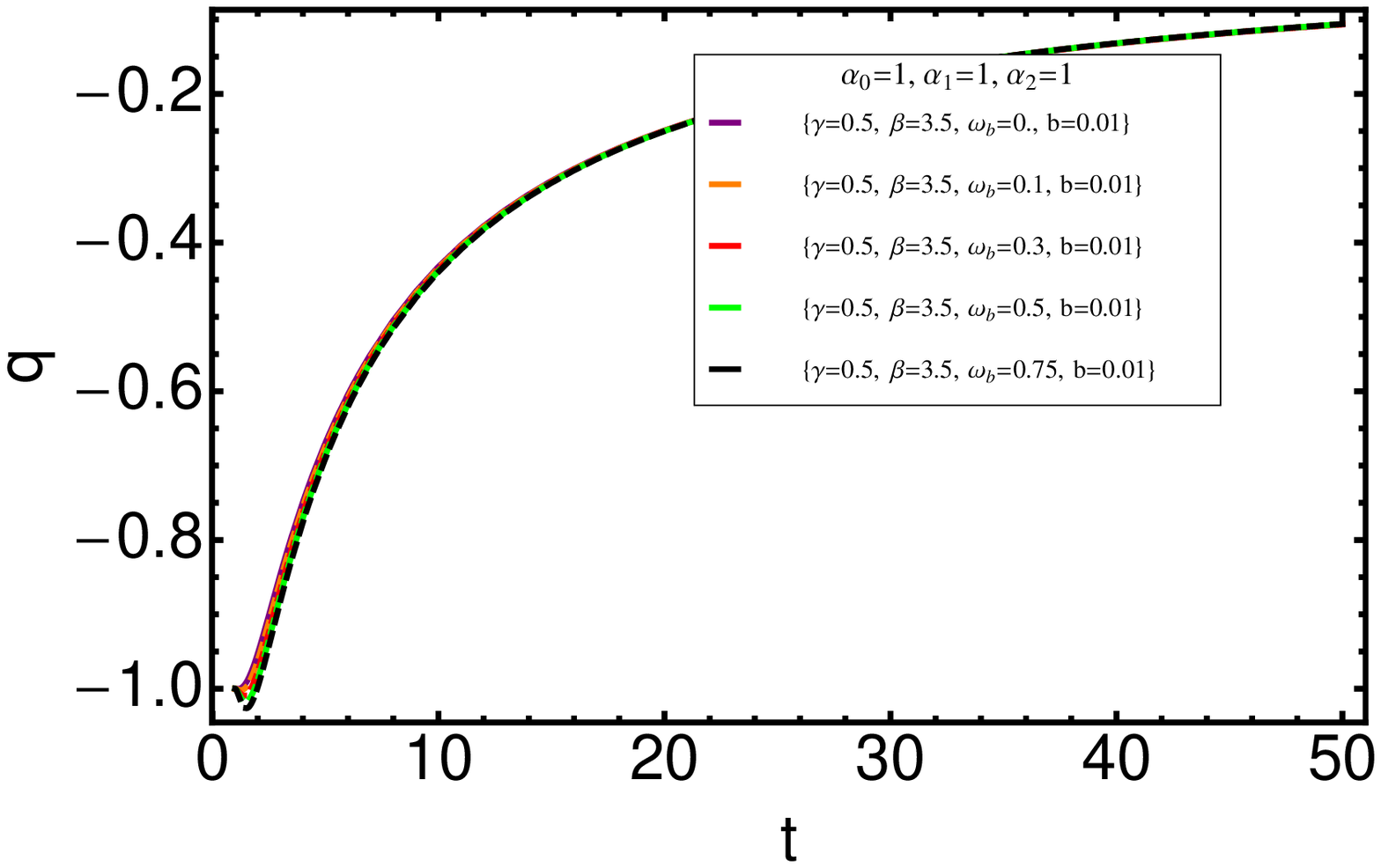}
 \end{array}$
 \end{center}
\caption{Behavior of deceleration parameter $q$ against $t$ for Model 1.}
 \label{fig:3}
\end{figure}

In this section we will consider $\Lambda(t)$ to be of the form
\begin{equation}\label{eq:lambda1}
\Lambda(t)=\rho_{D}.
\end{equation}
Therefore for the pressure of DE we will have
\begin{equation}\label{eq:P1}
P_{D}=\left( \frac{b}{8 \pi G(t)} -1 \right )\rho_{D}-\frac{\dot{\rho}_{D}}{3H}-\frac{3b}{8 \pi G(t)}H^{2}.
\end{equation}
The dynamics of $G(t)$ we will have
\begin{equation}\label{eq:G1}
\frac{\dot{G}(t)}{G(t)}+\frac{\dot{\rho}_{D}}{3H^{2}-\rho_{D}}=0.
\end{equation}
Performing a numerical analysis for the general case we recover the
graphical behavior of different cosmological parameters. Graphical
behavior of Gravitational constant $G(t)$ against time $t$ presented
in Fig.(\ref{fig:1}). We see that $G(t)$ is an increasing function.
Different plots represent behavior of $G(t)$ as a function of the
parameters of the model. For this model with the specific behavior
of $G(t)$ for Hubble parameter $H$ gives decreasing behavior over
time. It is confirmed by LCDM scenario. From the analysis of the
graphical behavior of $\omega_{tot}$ we made the following
conclusion that with $\alpha_{0}=1$, $\gamma=0.5$, $\beta=3.5$,
$\omega_{b}=0.3$, $b=0.01$ (interaction parameter) and with
increasing $\alpha_{1}$ and $\alpha_{2}$ we increase the value of
$\omega_{tot}$ for later stages of evolution, while for the early
stages, in history, it is a decreasing function. For instance, with
$\alpha_{1}=0.5$ and $\alpha_{2}=0.5$ (blue line) $\omega_{tot}$ is
a constant and $\omega_{tot} \approx -0.9$ (Top left plot in
Fig.(\ref{fig:2})). Top right plot of Fig.(\ref{fig:2}) presents
graphical behavior of $\omega_{tot}$ against time as a function of
the parameter $b$ characterizing interaction between DE and DM. We
see that for the later stages of the evolution the interaction
$Q=3hb(\rho_{b}+\rho_{D})$ does not play any role. An existence of
the interaction can be observed only for relatively early stages of
evolution and when $b$ is too much higher than the real values of it
estimated from observations. The left-bottom plot shows the
decreasing behavior of $\omega_{tot}$ at early stages of evolution
which, while for later stages, becomes a constant. This behavior is
observed for $\alpha_{0}=\alpha_{1}=\alpha_{2}=1$, $\omega_{b}=0.1$,
$b=0.01$ and for increasing $\gamma$ and $\beta$. With the increase
in $\gamma$ and $\beta$, we increase the value of $\omega_{tot}$.
The right-bottom plot represents behavior as a function of
$\omega_{b}$. In Fig.{\ref{fig:3}}, the graphical behavior of the
deceleration parameter $q$ is observed which is a negative quantity
throughout the evolution of the Universe i.e. we have an ever
accelerated Universe. Right panel (top and bottom) shows that the
behavior of $q$ does not strongly depend upon the interaction
parameter $b$ and EoS parameter $\omega_{b}$. We also see that $q$
starts its evolution from $-1$ and for a very short period of the
history it becomes smaller than $-1$, but after this $q>-1$ for
ever, giving a hope that observational facts can be modeled (for
later stages!). Right panel (top and bottom) represents the behavior
of $q$ for $\alpha_{1}=\alpha_{2}$ and \{$\gamma$, $\beta$\} (top
and bottom) respectively. With the increase in the values of the
parameters, the value of $q$ increases. Some information about
$\omega_{D}$, $\Lambda(t)$ and $\dot{G}(t)/G(t)$ can be found in
Appendix.

\begin{figure}[h!]
 \begin{center}$
 \begin{array}{cccc}
\includegraphics[width=50 mm]{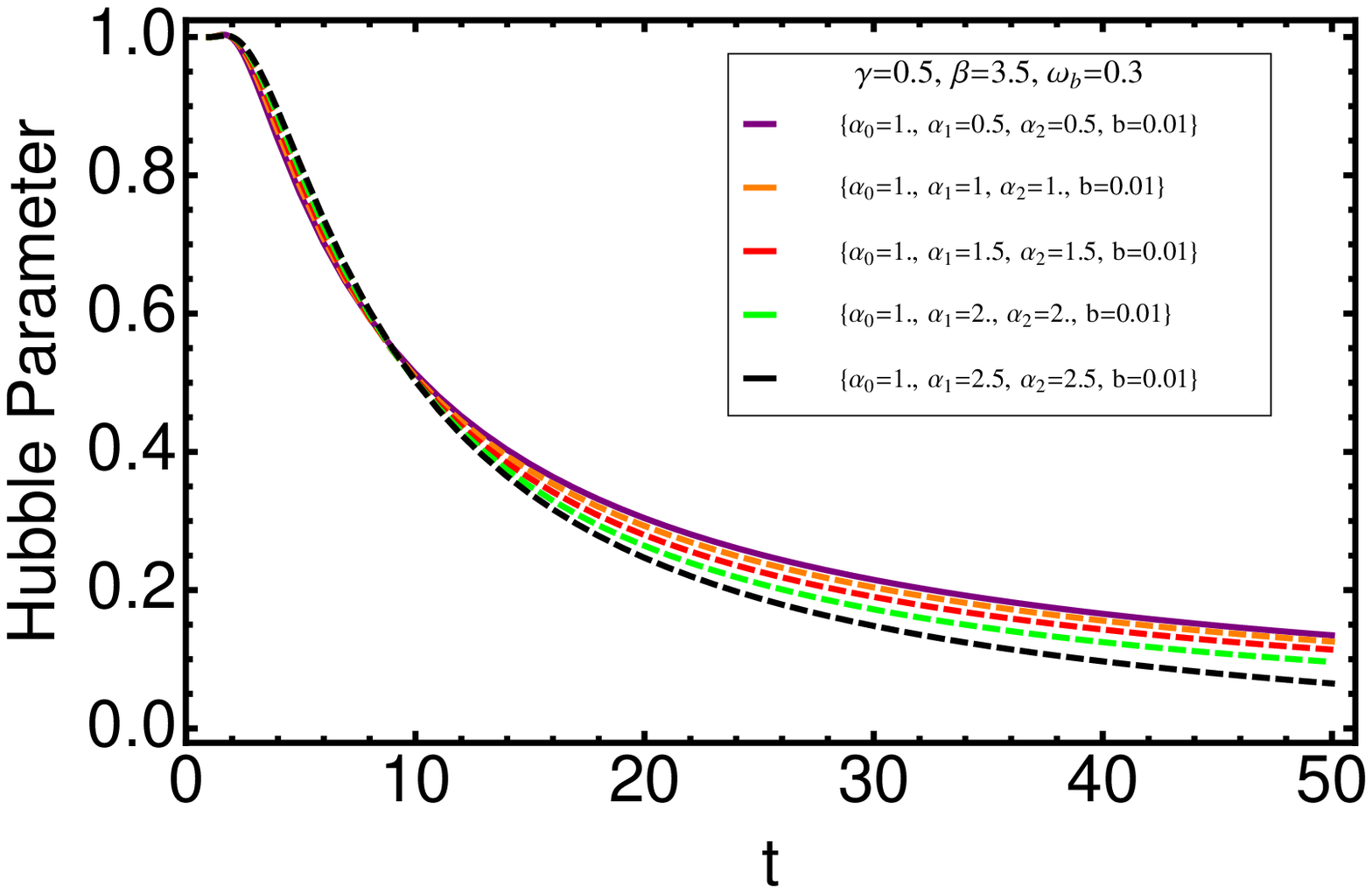} &
\includegraphics[width=50 mm]{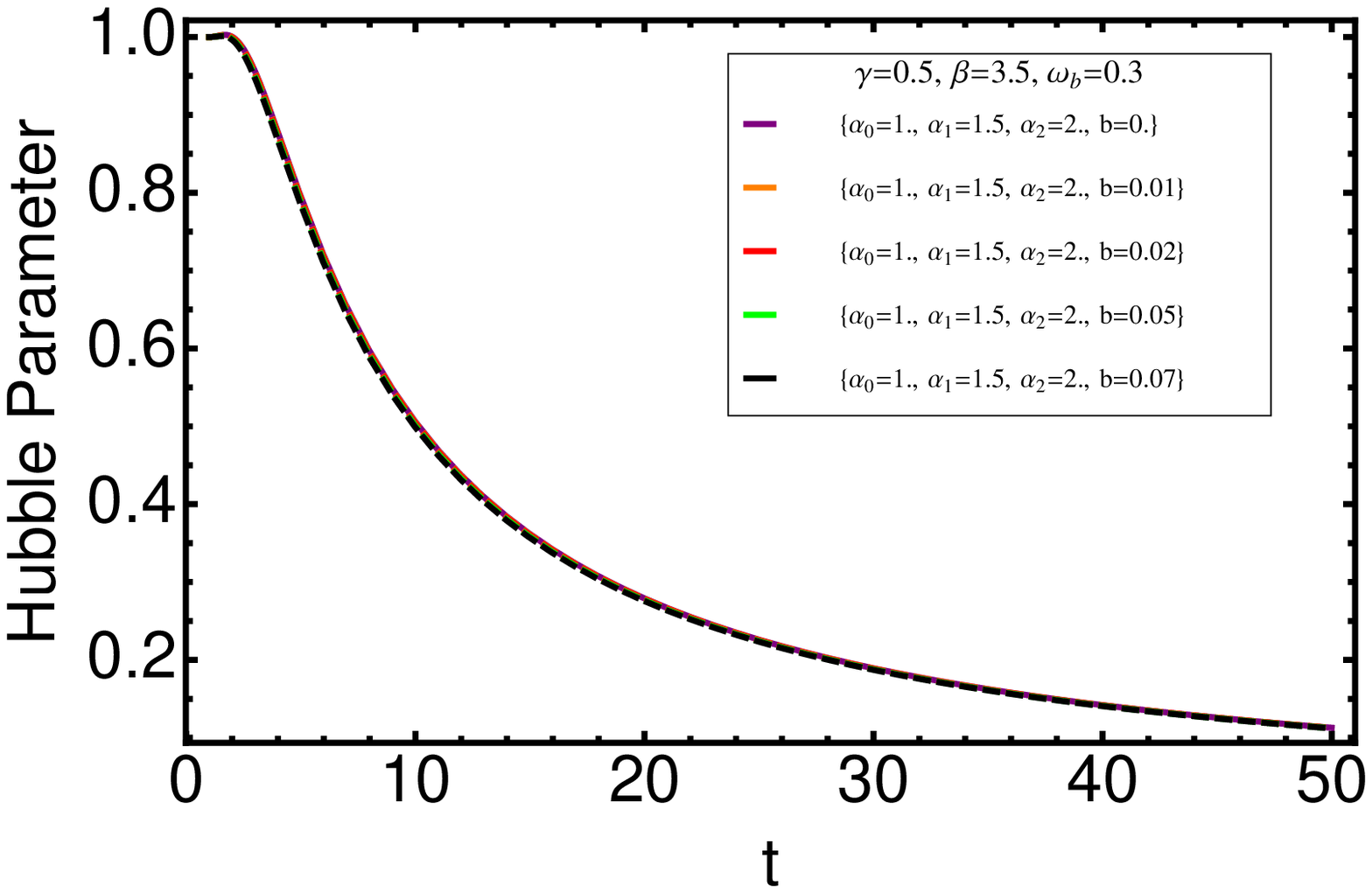}\\
\includegraphics[width=50 mm]{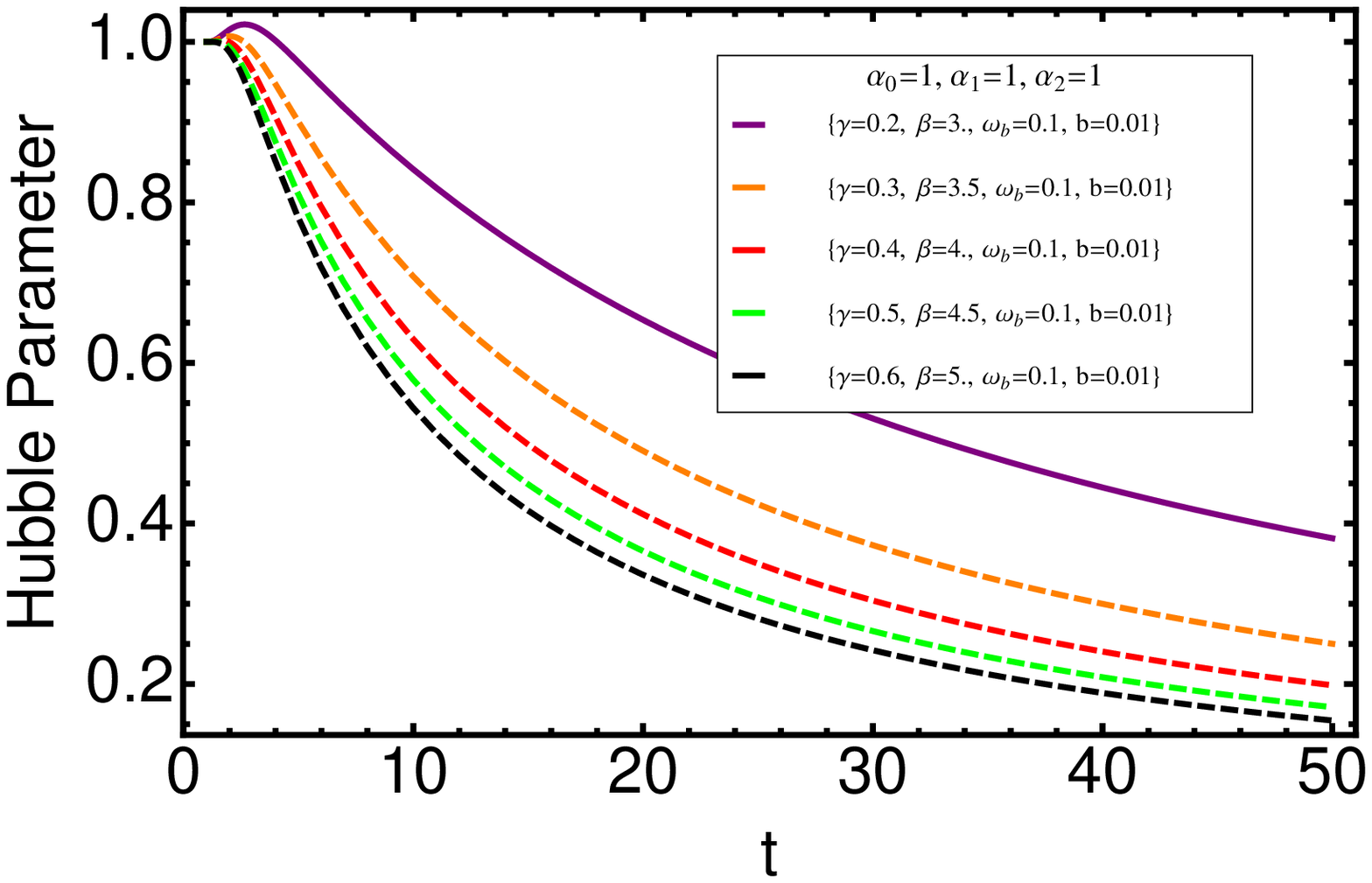} &
\includegraphics[width=50 mm]{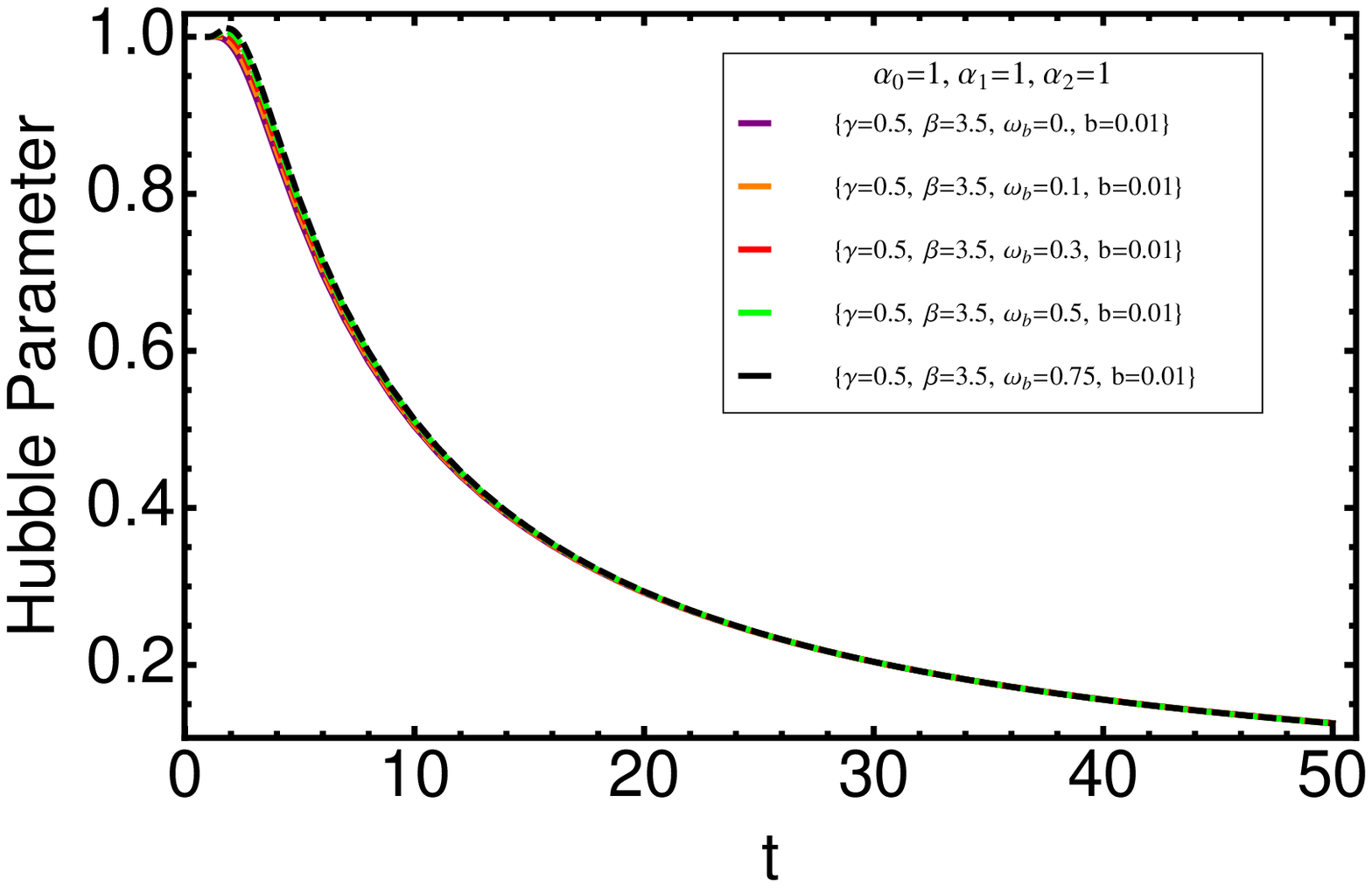}
 \end{array}$
 \end{center}
\caption{Behavior of Hubble parameter $H(t)$ against $t$ for Model 1.}
 \label{fig:10}
\end{figure}
\begin{figure}[h!]
 \begin{center}$
 \begin{array}{cccc}
\includegraphics[width=50 mm]{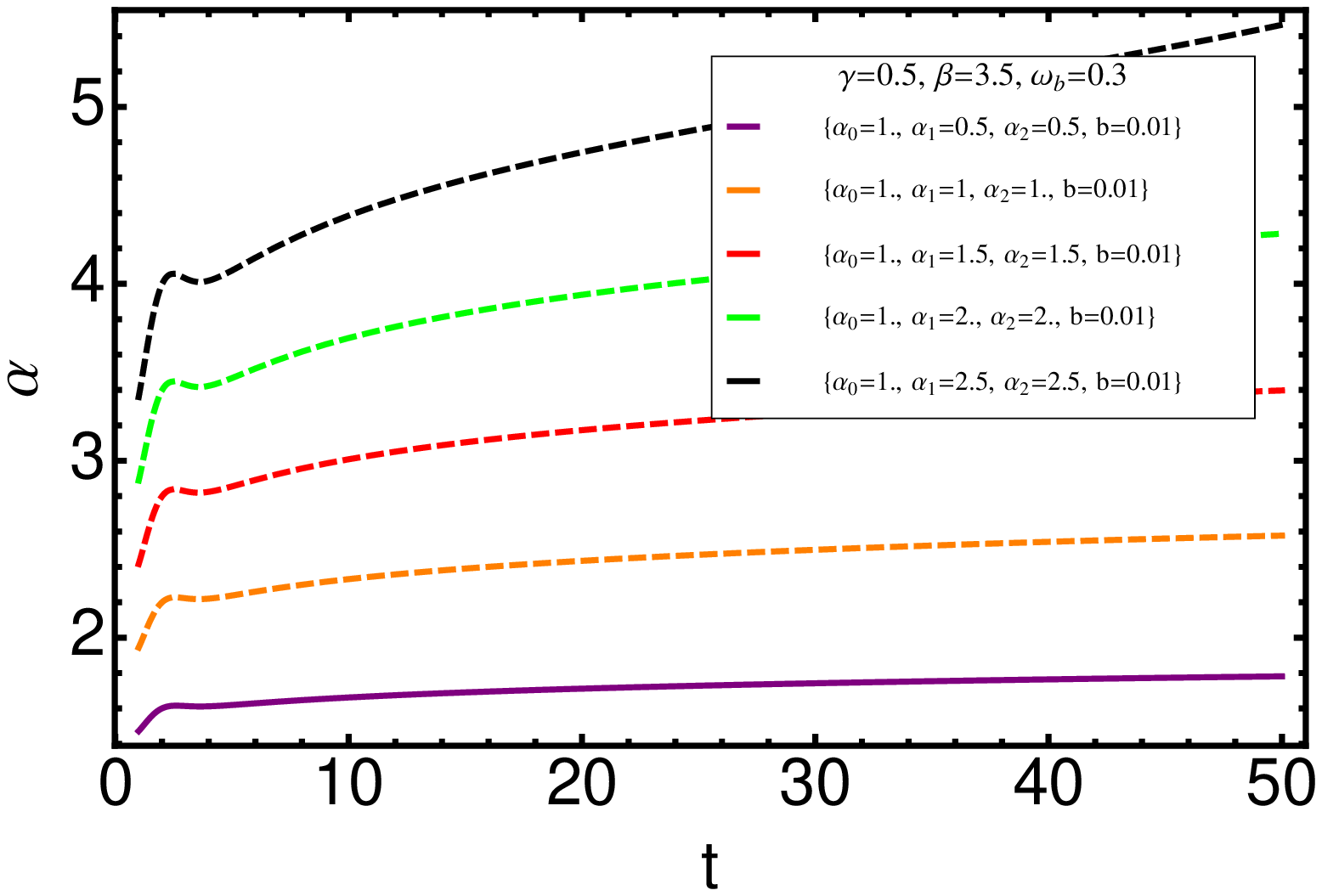} &
\includegraphics[width=50 mm]{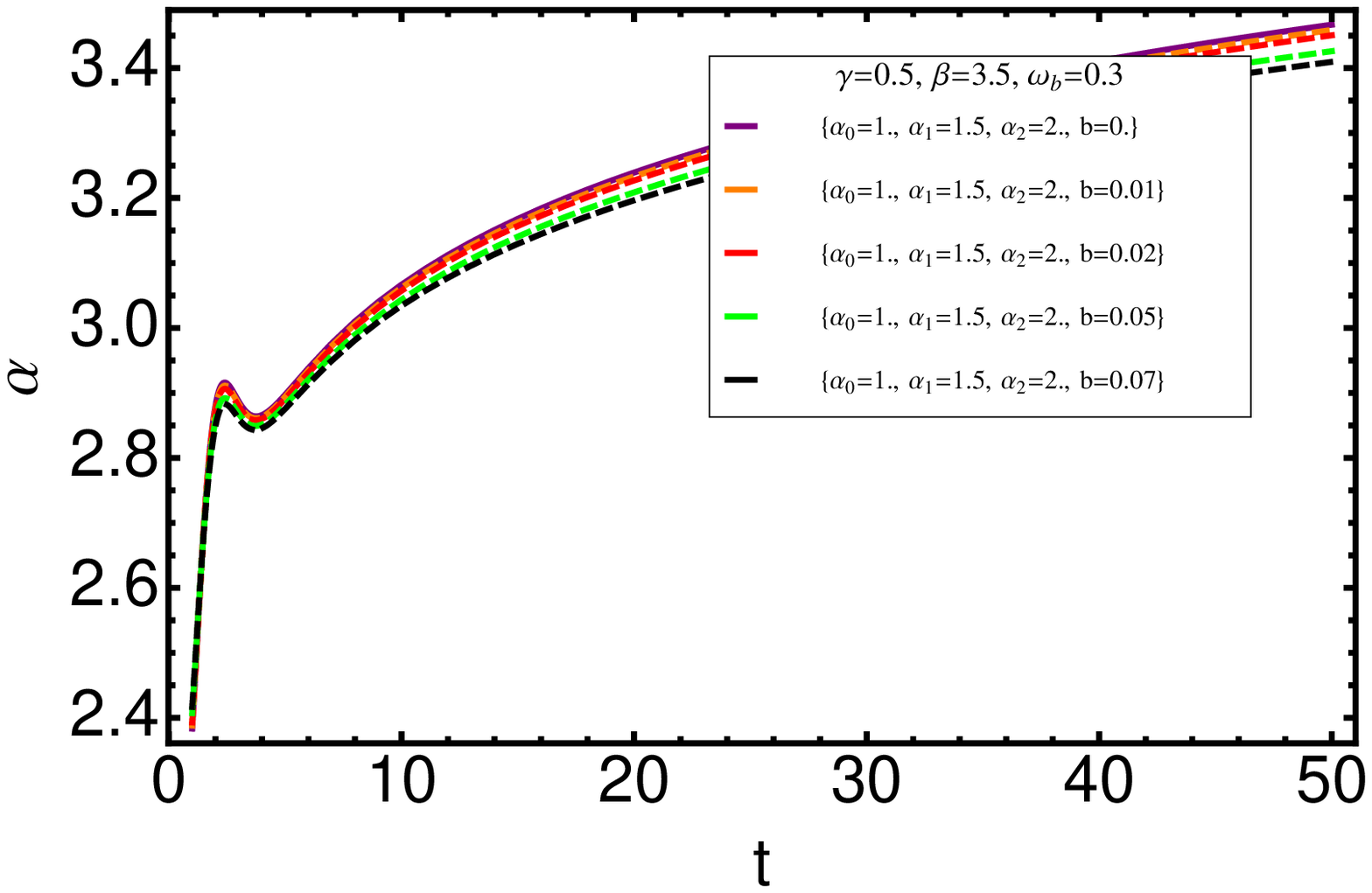}\\
\includegraphics[width=50 mm]{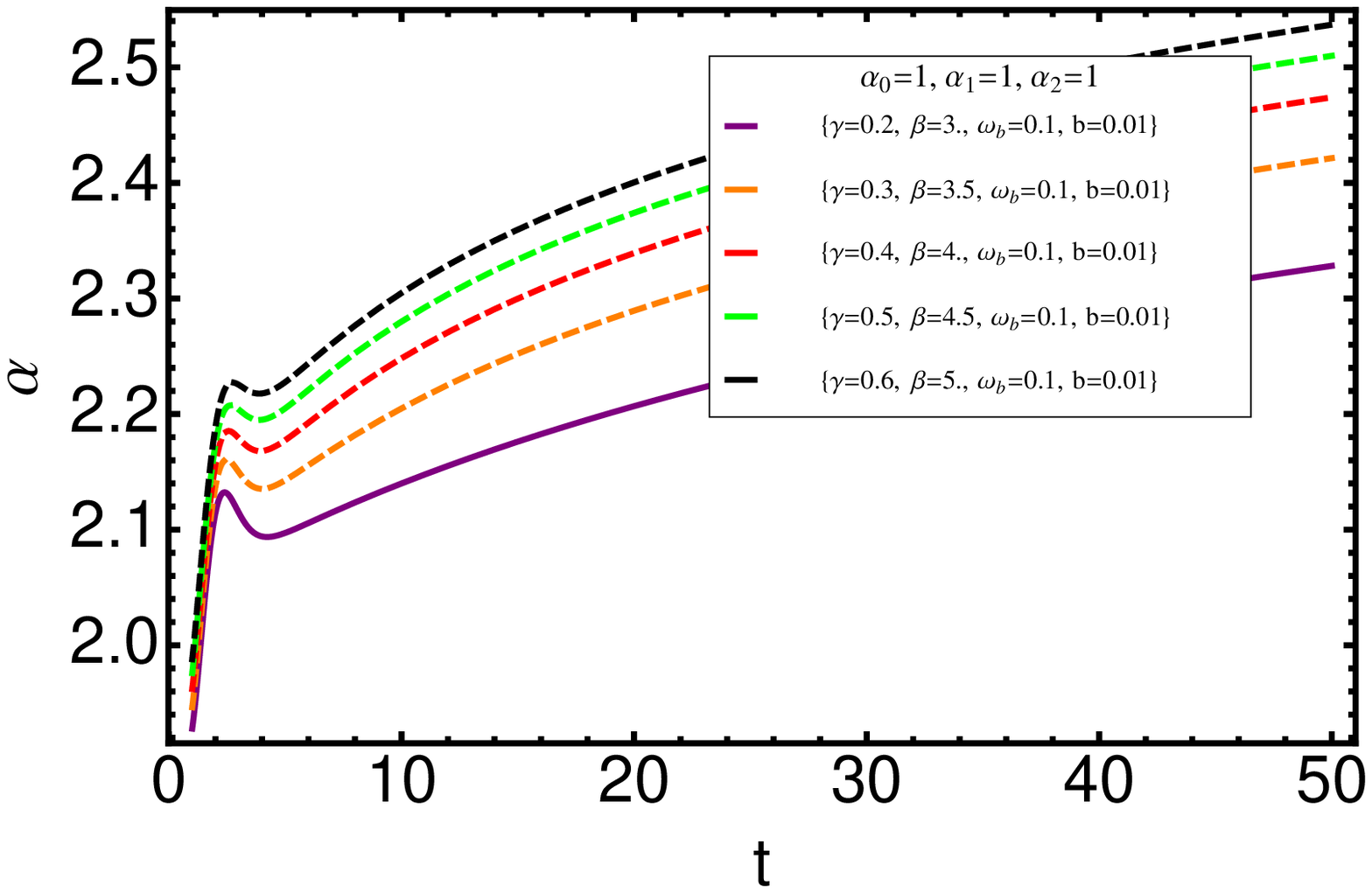} &
\includegraphics[width=50 mm]{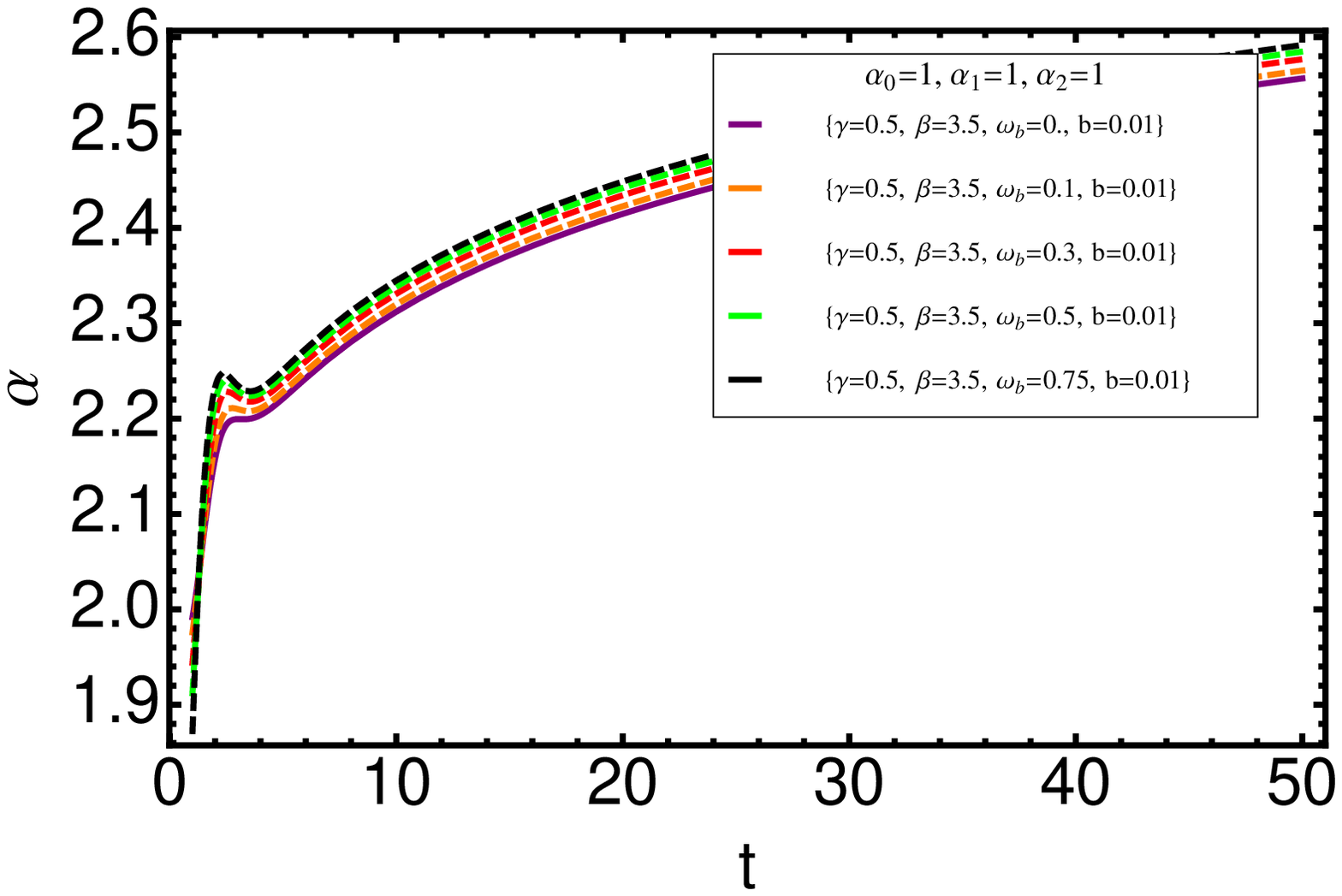}
 \end{array}$
 \end{center}
\caption{Behavior of $\alpha$ against $t$ for Model 1.}
 \label{fig:11}
\end{figure}
\begin{figure}[h!]
 \begin{center}$
 \begin{array}{cccc}
\includegraphics[width=50 mm]{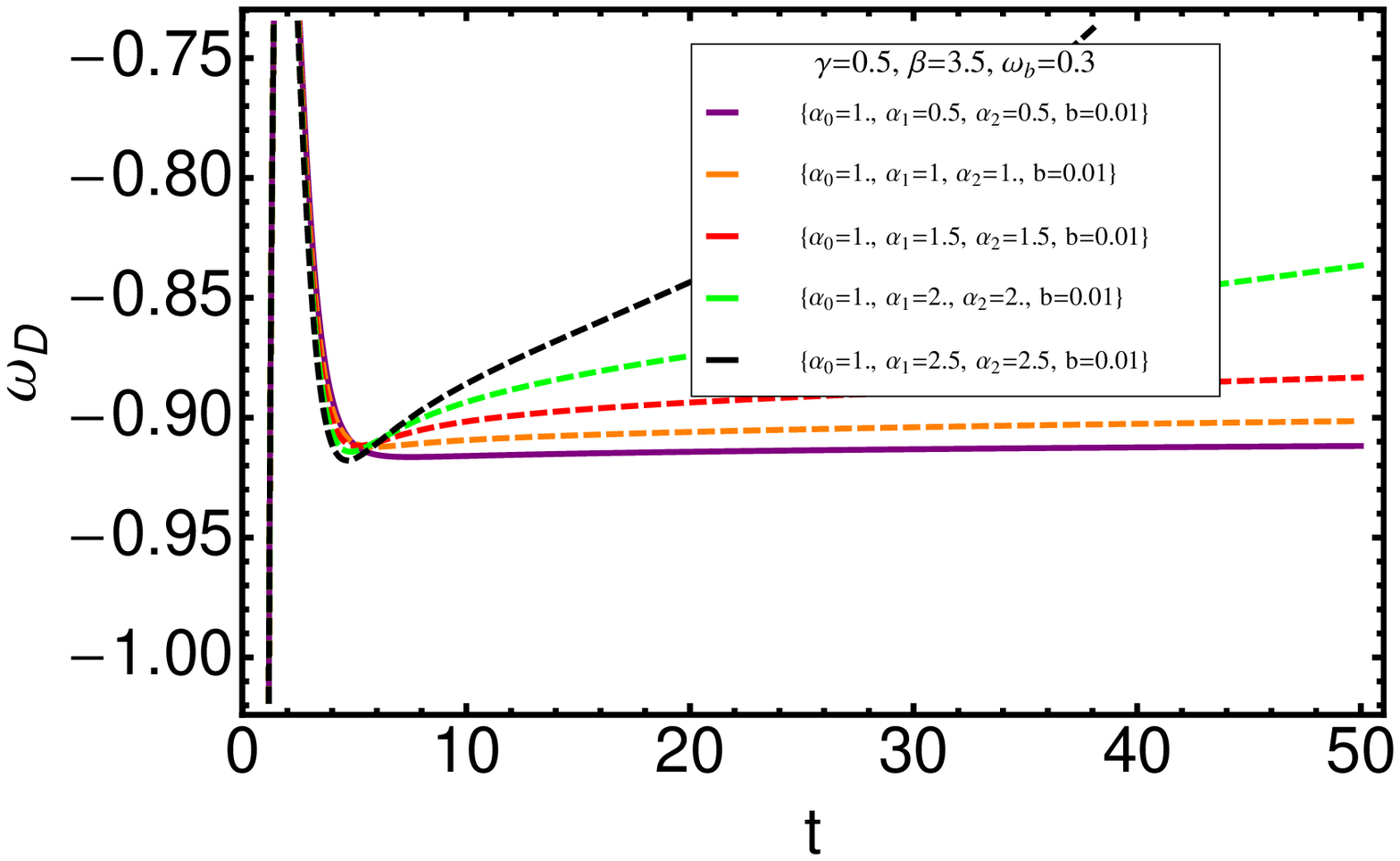} &
\includegraphics[width=50 mm]{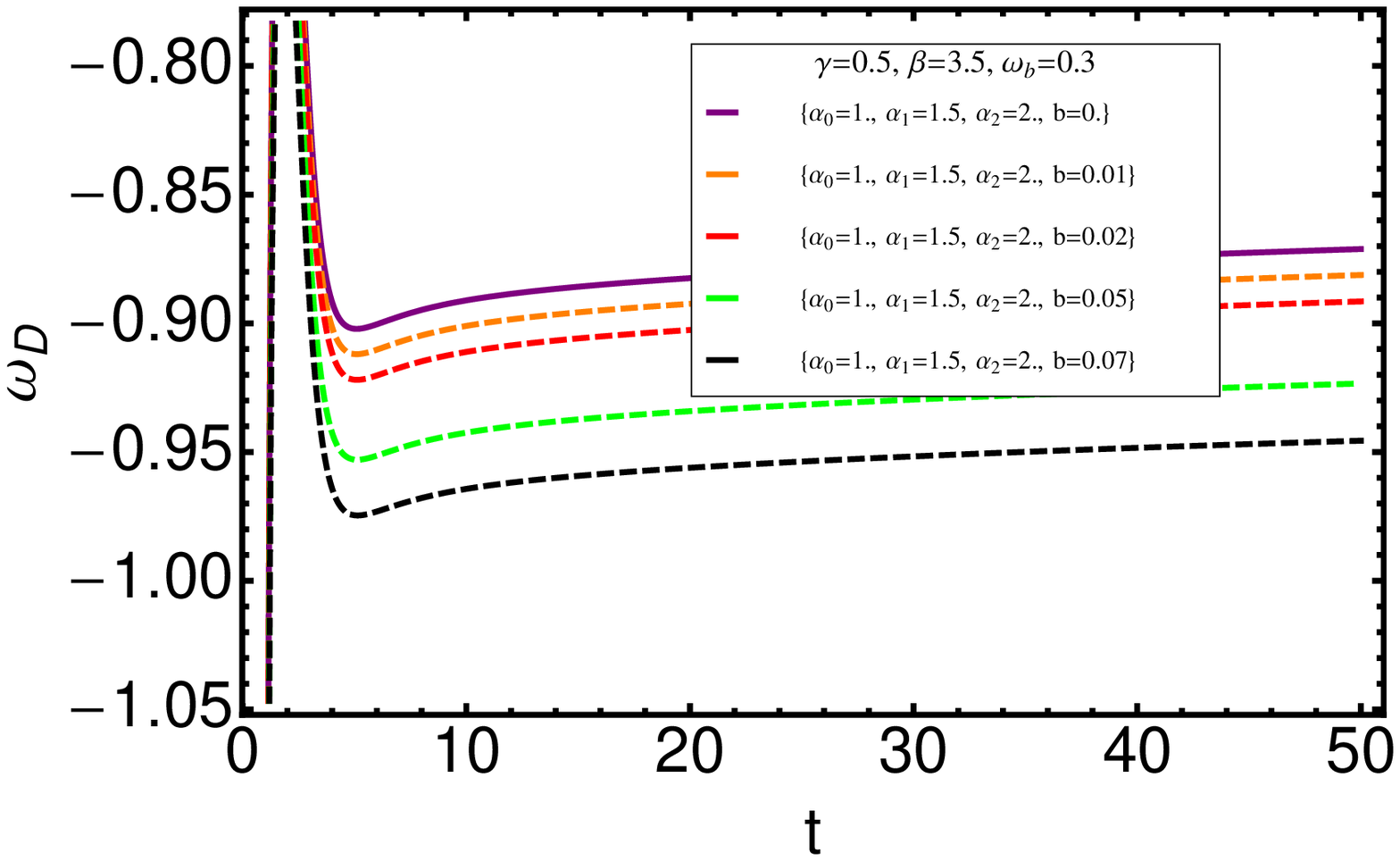}\\
\includegraphics[width=50 mm]{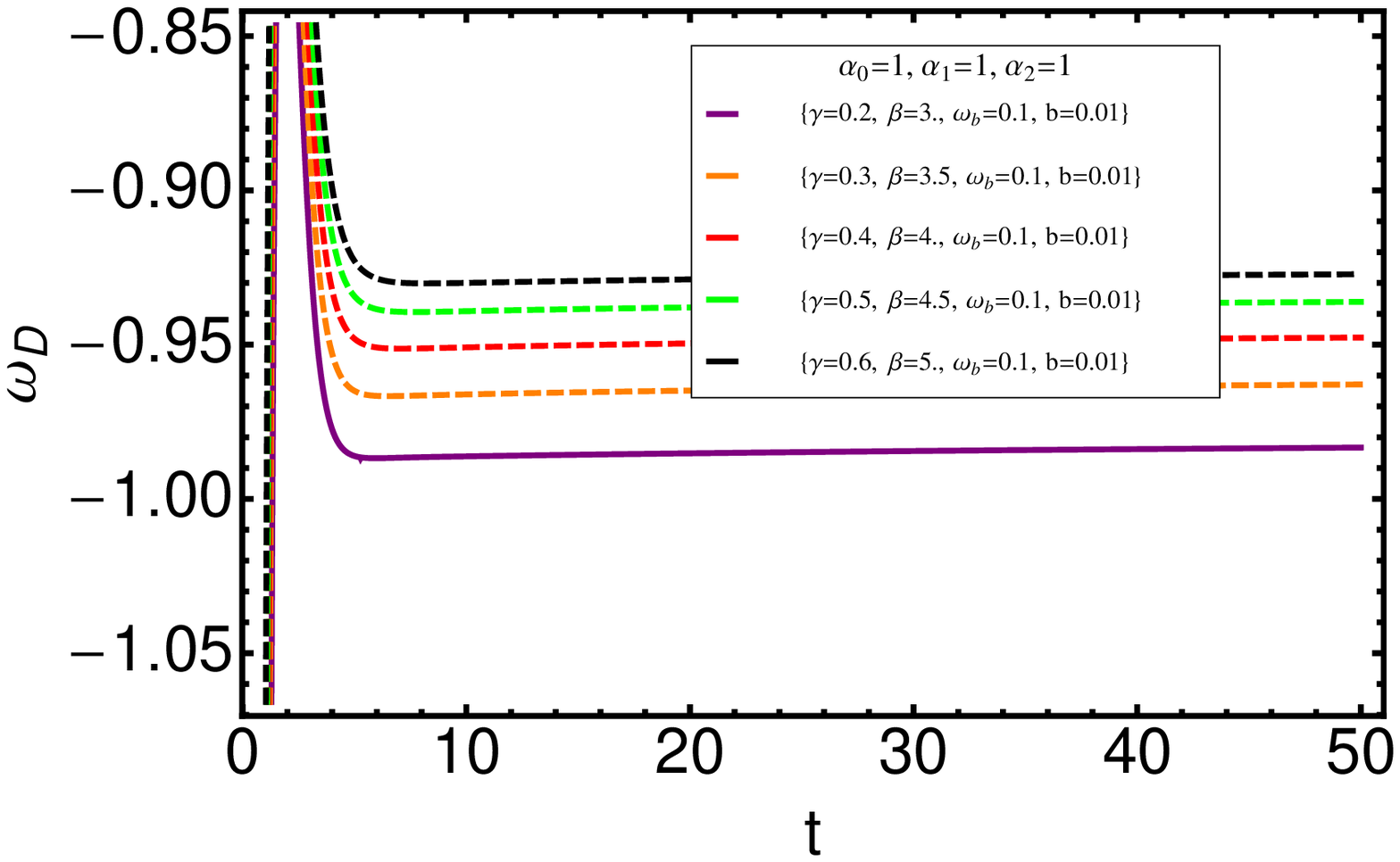} &
\includegraphics[width=50 mm]{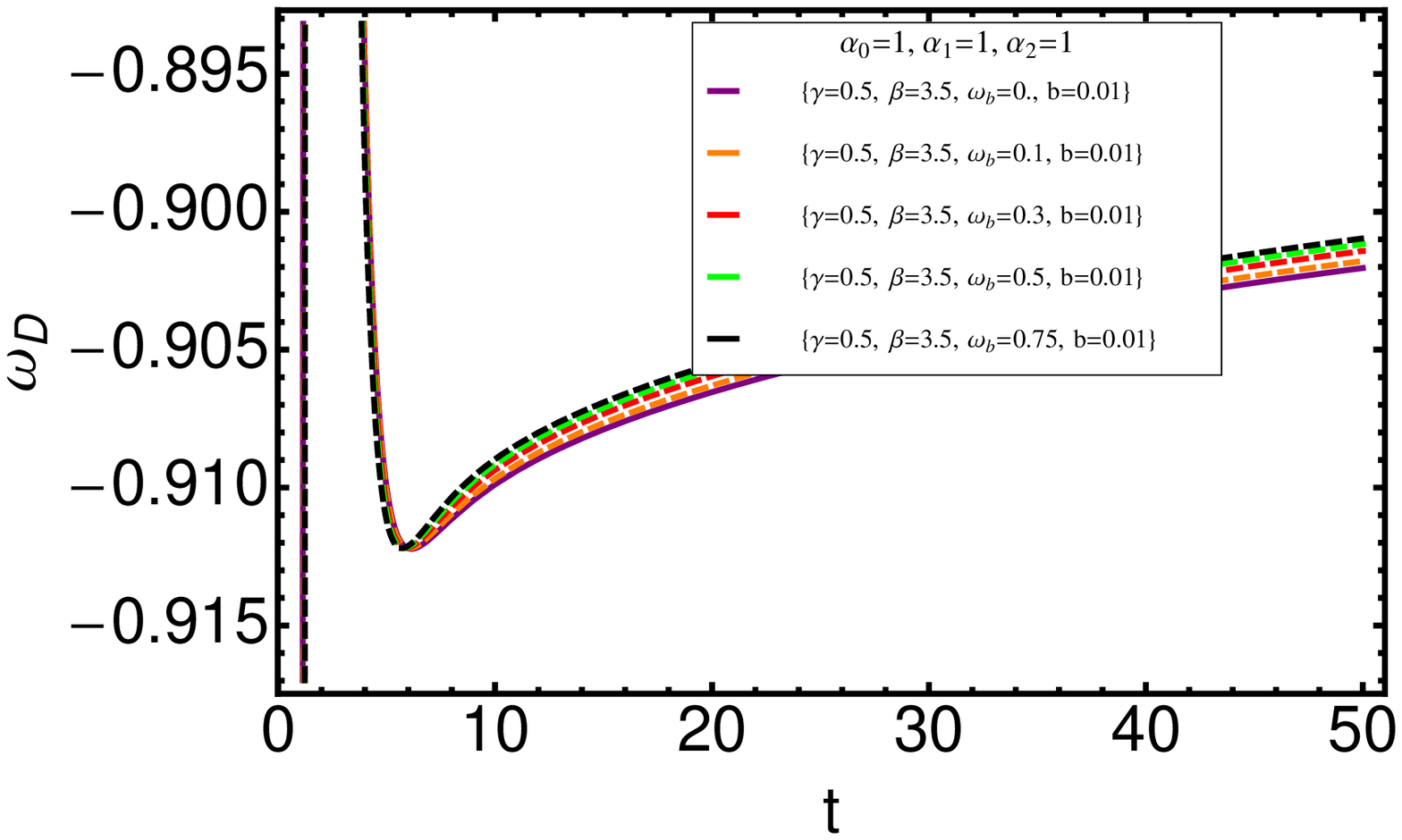}
 \end{array}$
 \end{center}
\caption{Behavior of $\omega_{D}$ against $t$ for Model 1.}
 \label{fig:12}
\end{figure}
\begin{figure}[h!]
 \begin{center}$
 \begin{array}{cccc}
\includegraphics[width=50 mm]{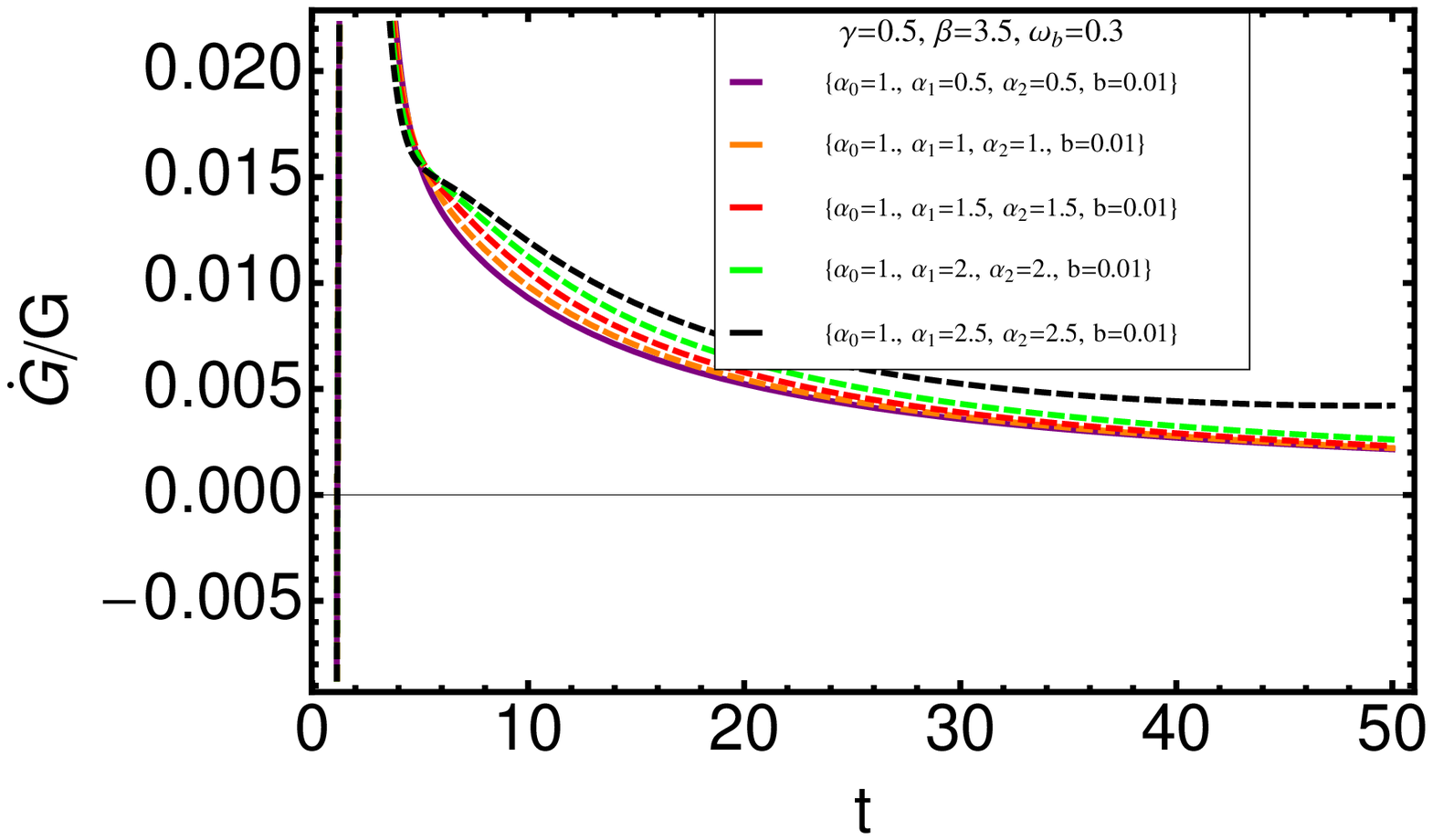} &
\includegraphics[width=50 mm]{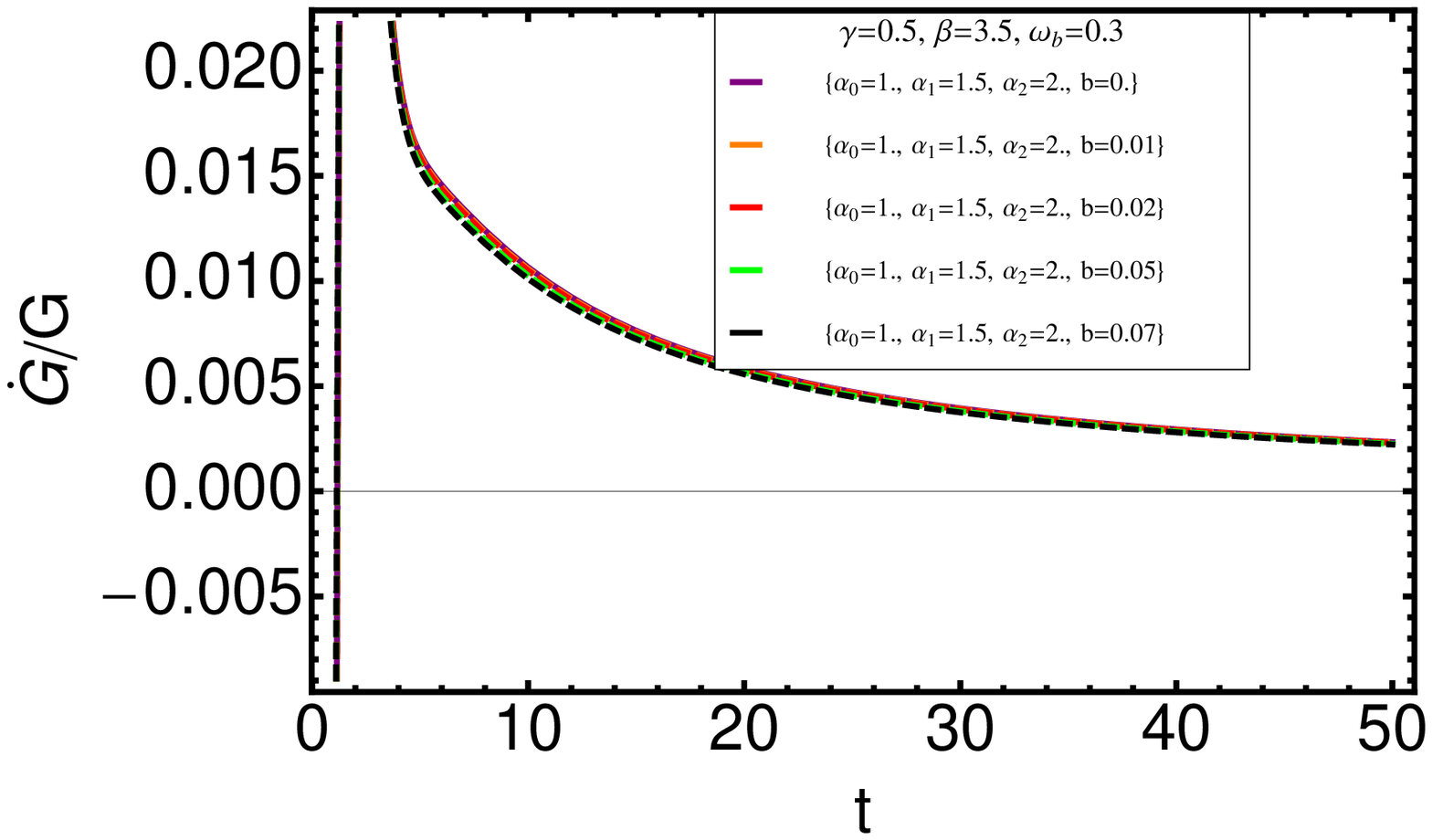}\\
\includegraphics[width=50 mm]{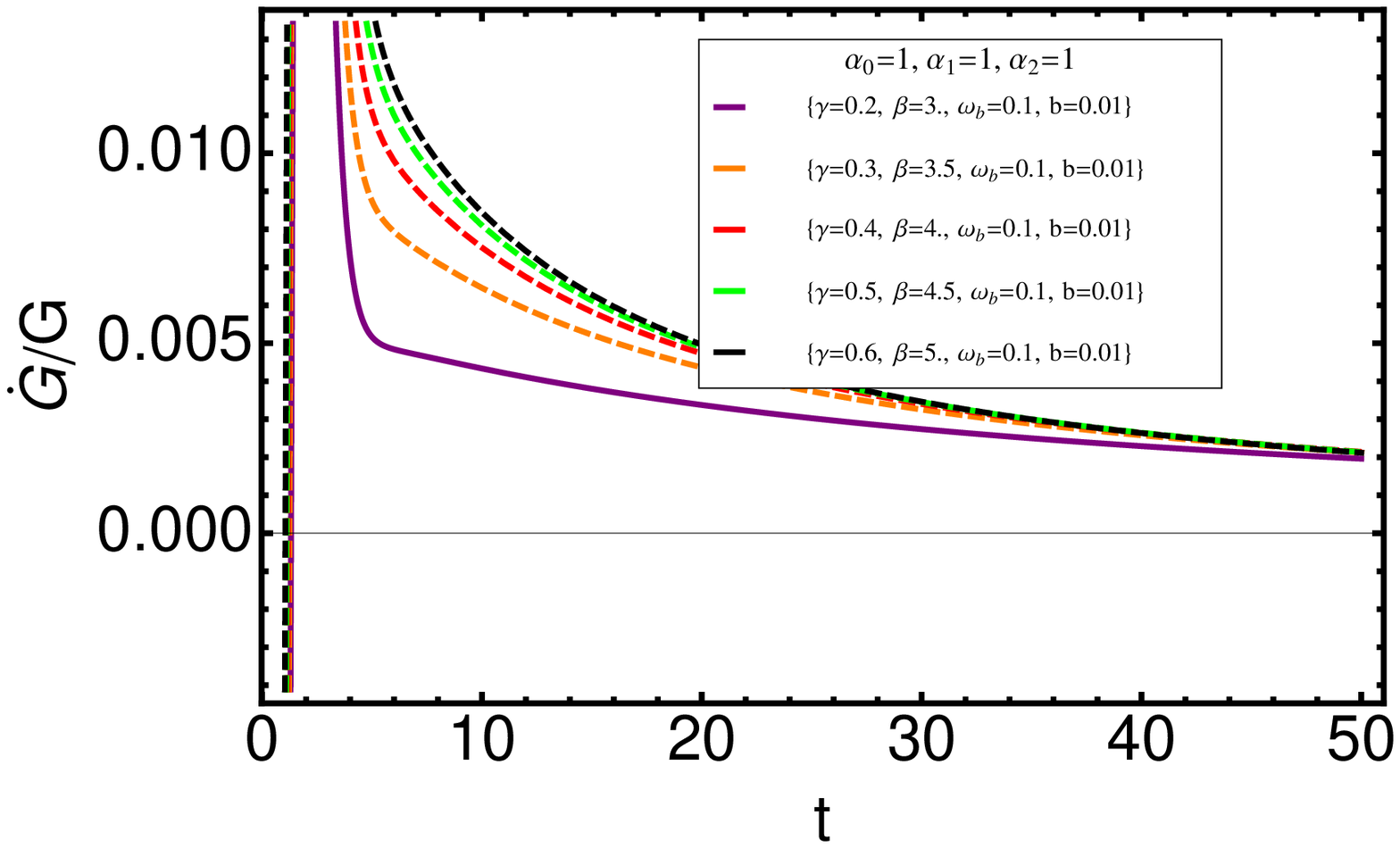} &
\includegraphics[width=50 mm]{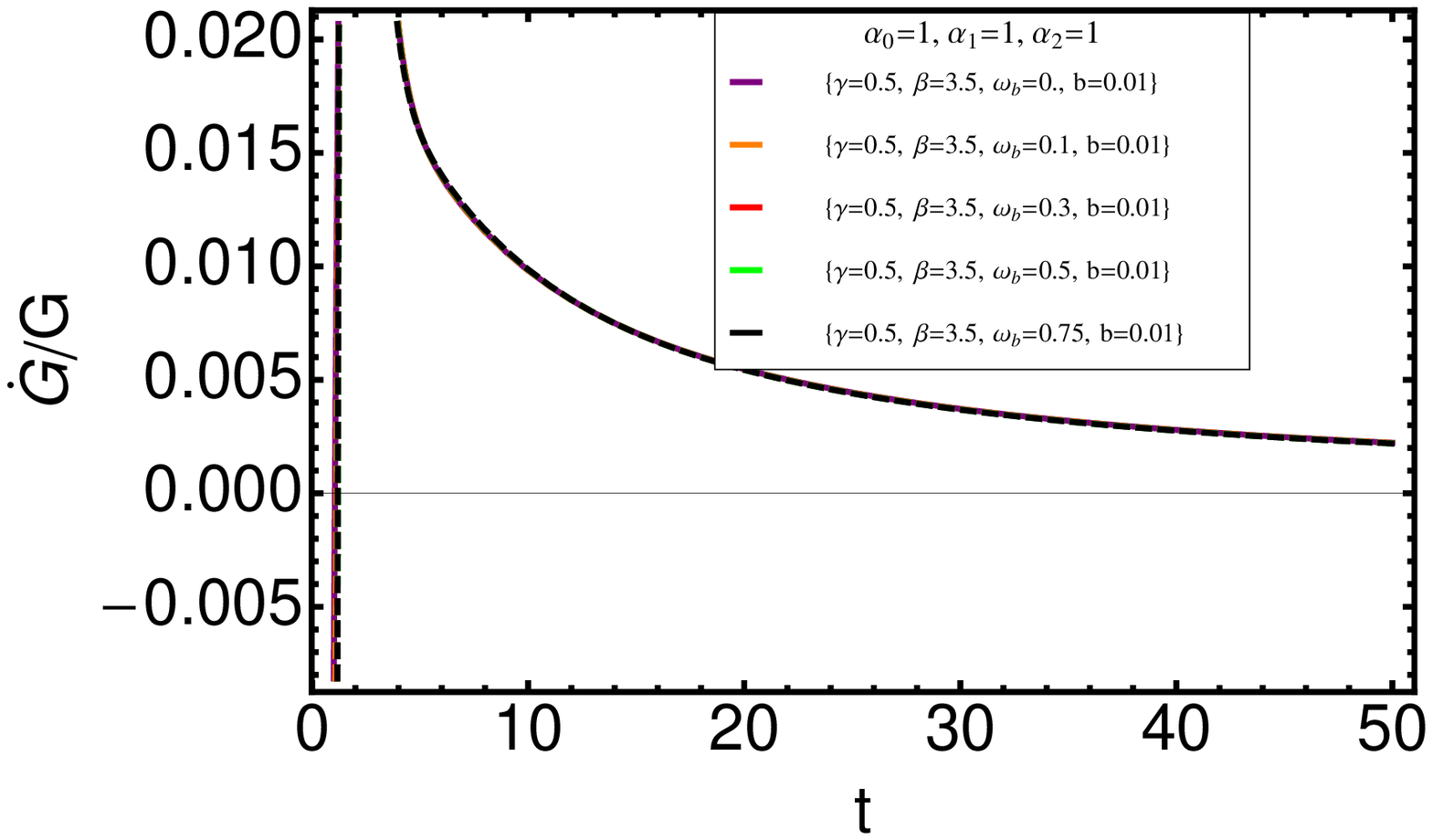}
 \end{array}$
 \end{center}
\caption{Behavior of $\dot{G}(t)/G(t)$ against $t$ for Model 1.}
 \label{fig:13}
\end{figure}

\subsection{Model 2: $\Lambda(t)=\rho_{b}\sin{(tH)}^{3}+\rho_{D}\cos{(tH)}$}
\begin{figure}[h!]
 \begin{center}$
 \begin{array}{cccc}
\includegraphics[width=50 mm]{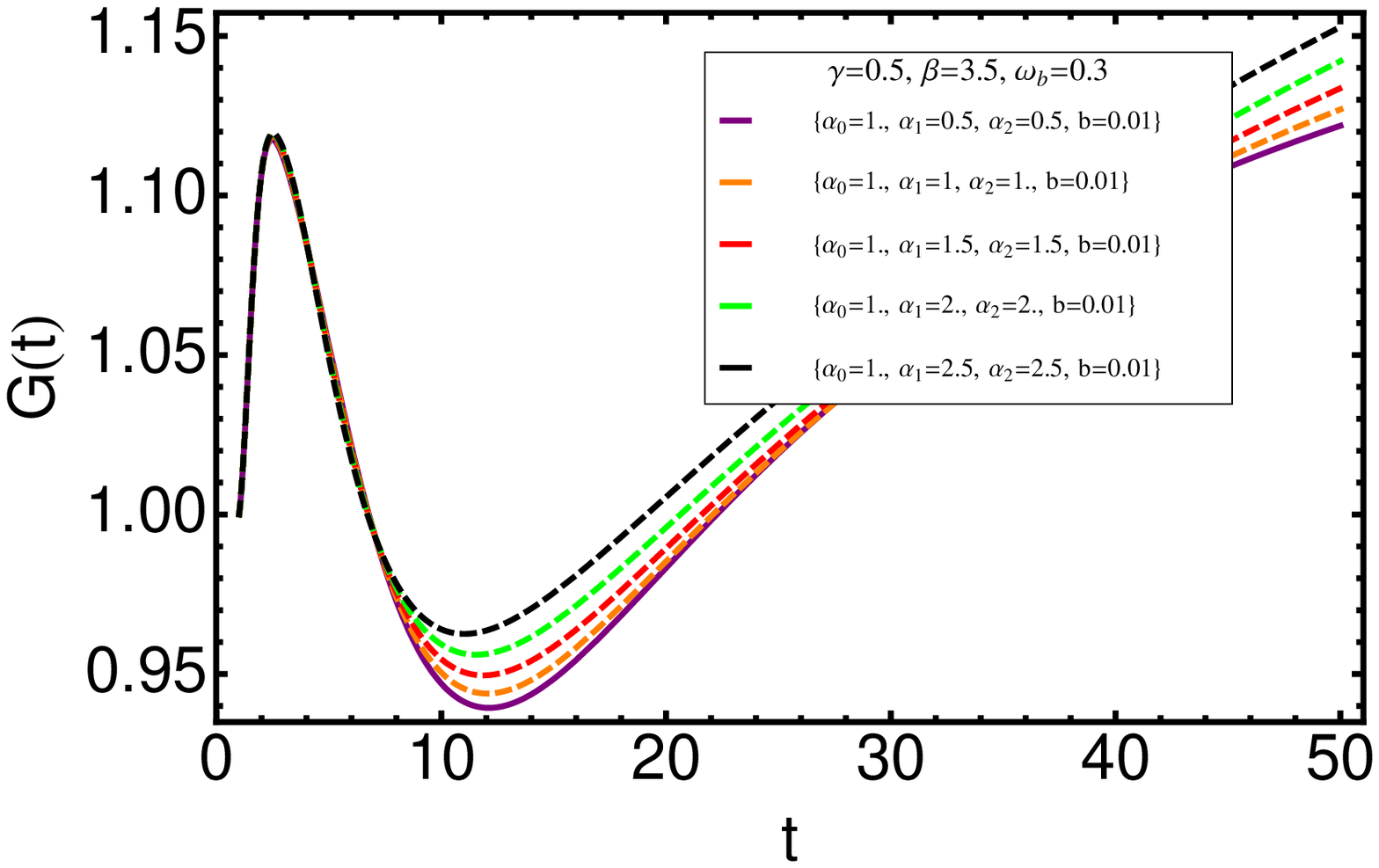} &
\includegraphics[width=50 mm]{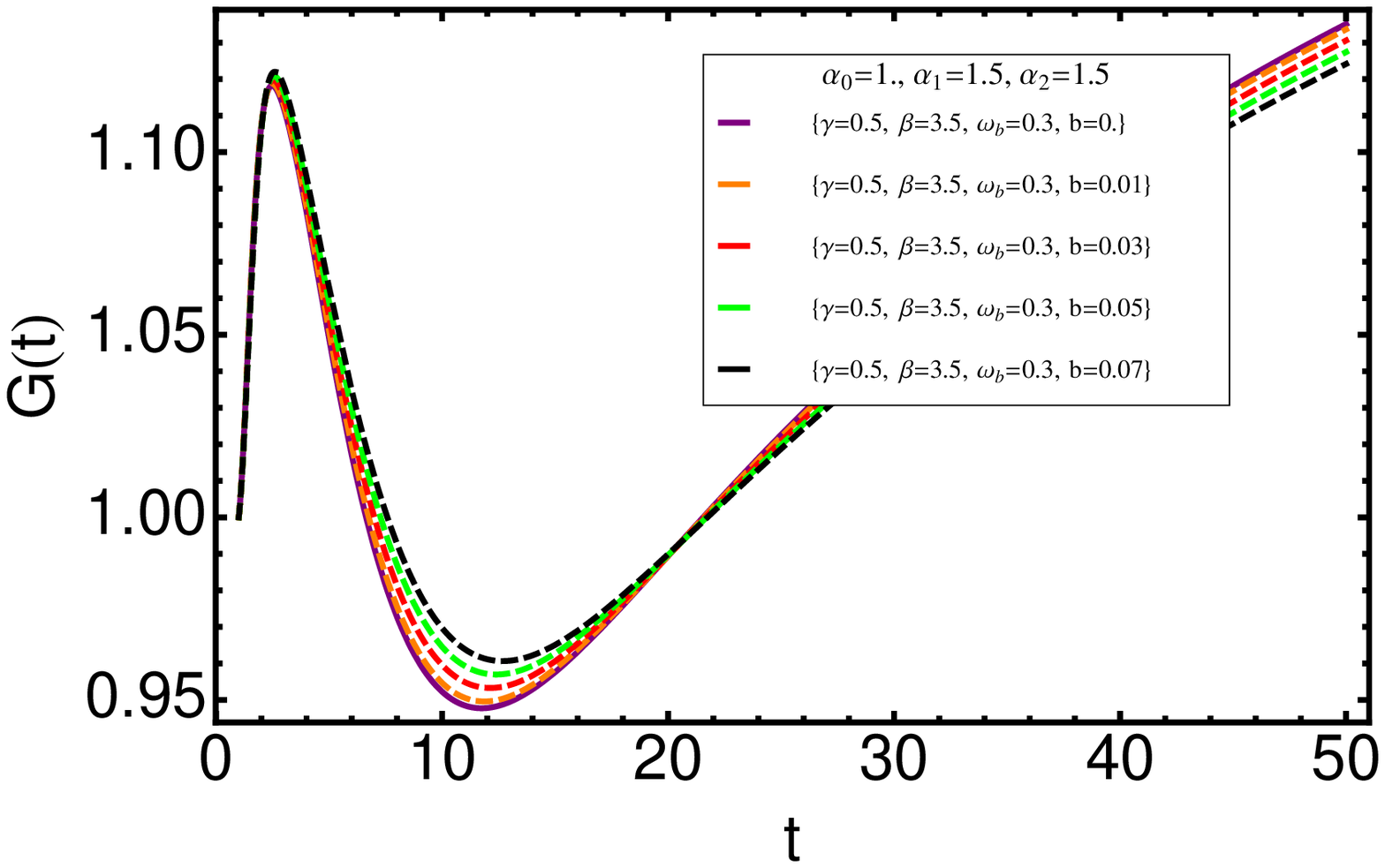}\\
\includegraphics[width=50 mm]{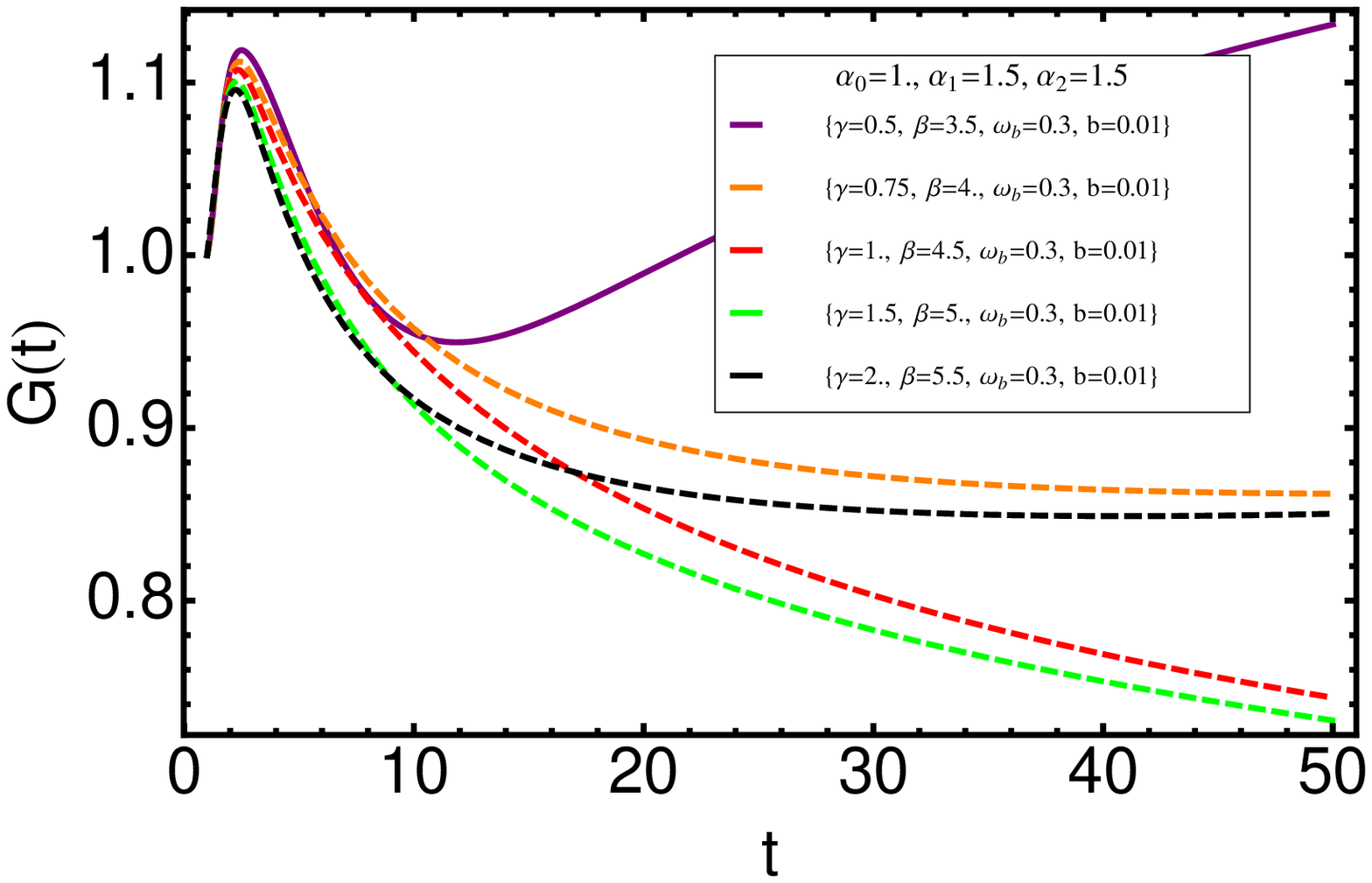} &
\includegraphics[width=50 mm]{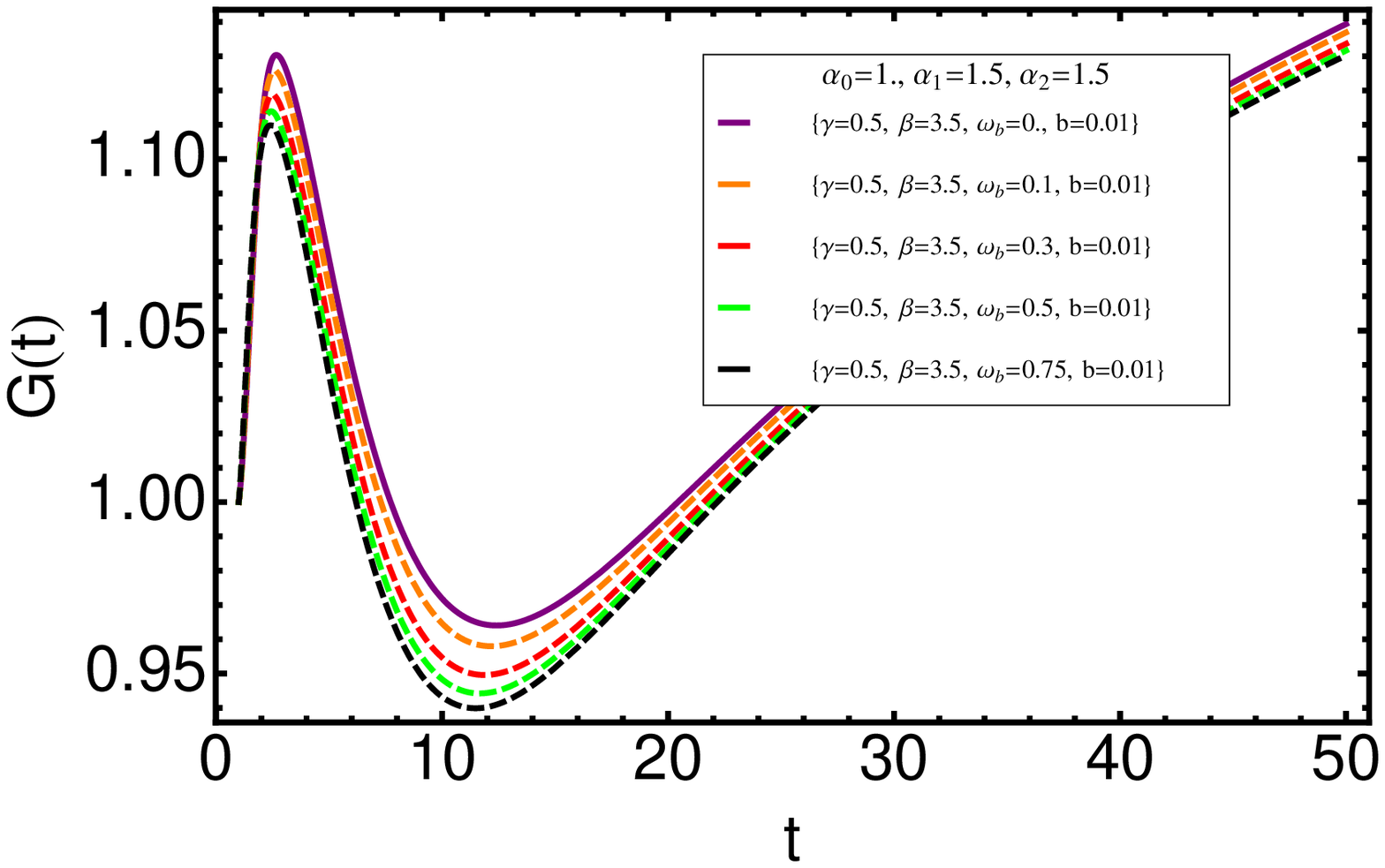}
 \end{array}$
 \end{center}
\caption{Behavior of Gravitational constant $G(t)$ against $t$ Model 2.}
 \label{fig:4}
\end{figure}

\begin{figure}[h!]
 \begin{center}$
 \begin{array}{cccc}
\includegraphics[width=50 mm]{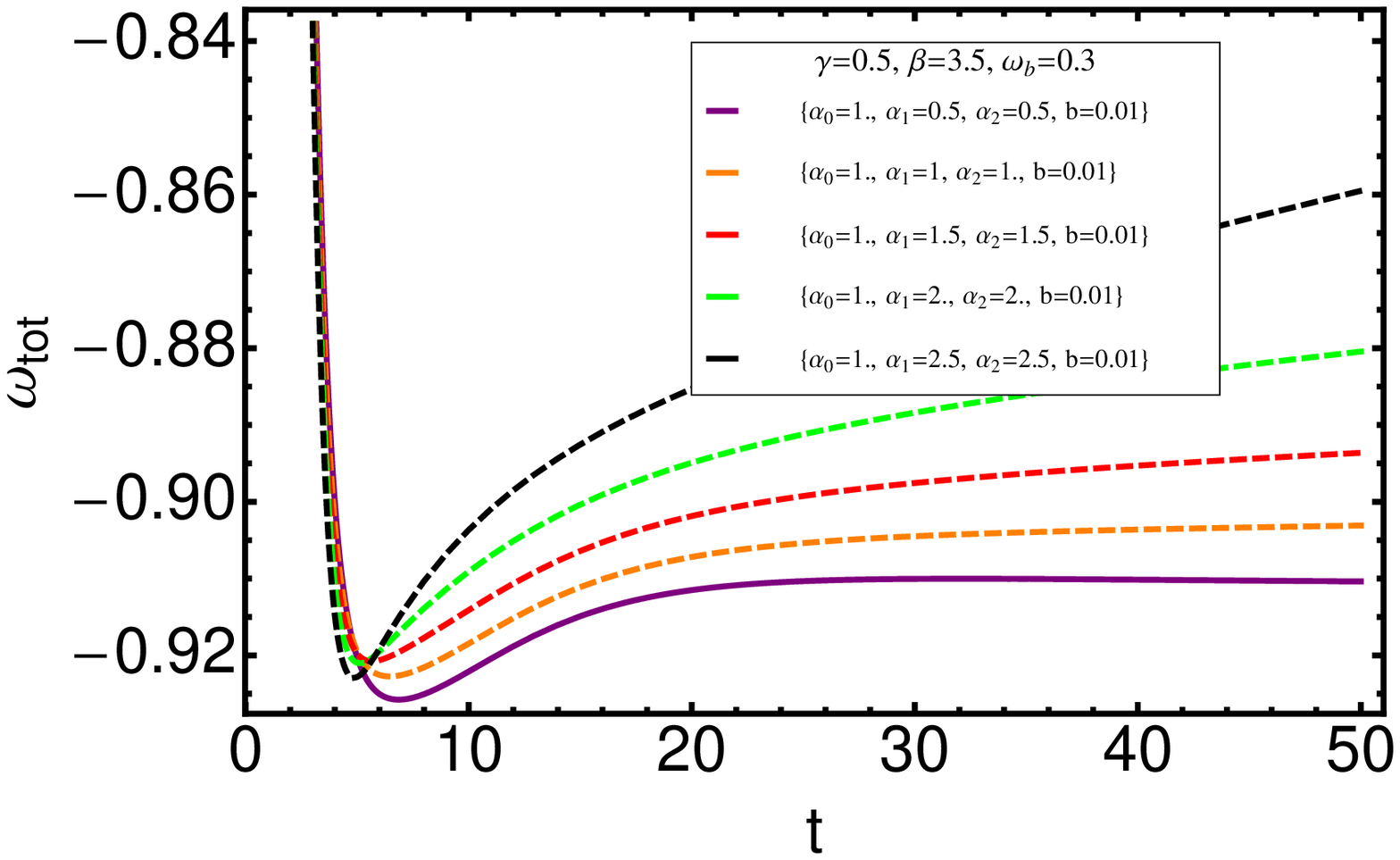} &
\includegraphics[width=50 mm]{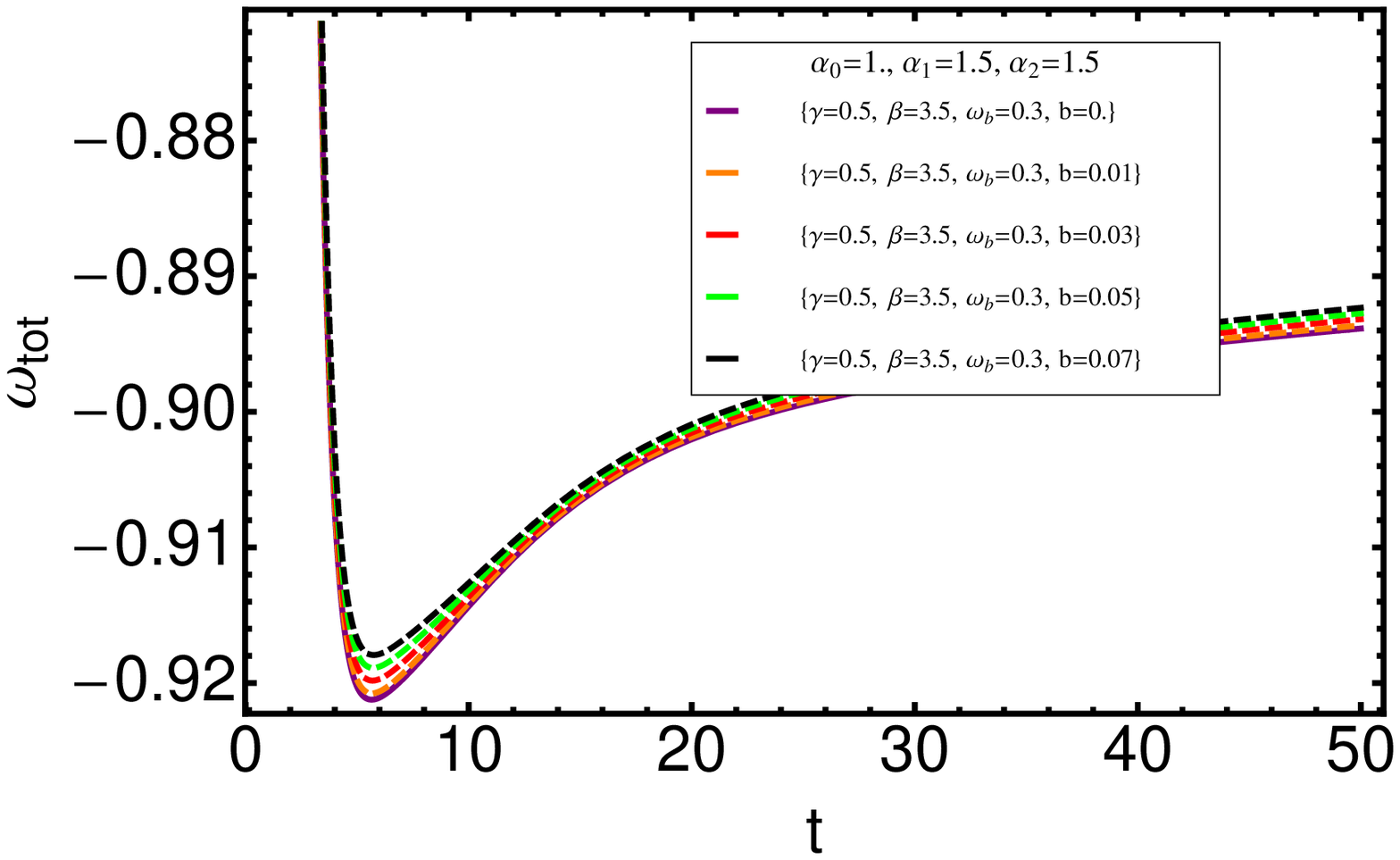}\\
\includegraphics[width=50 mm]{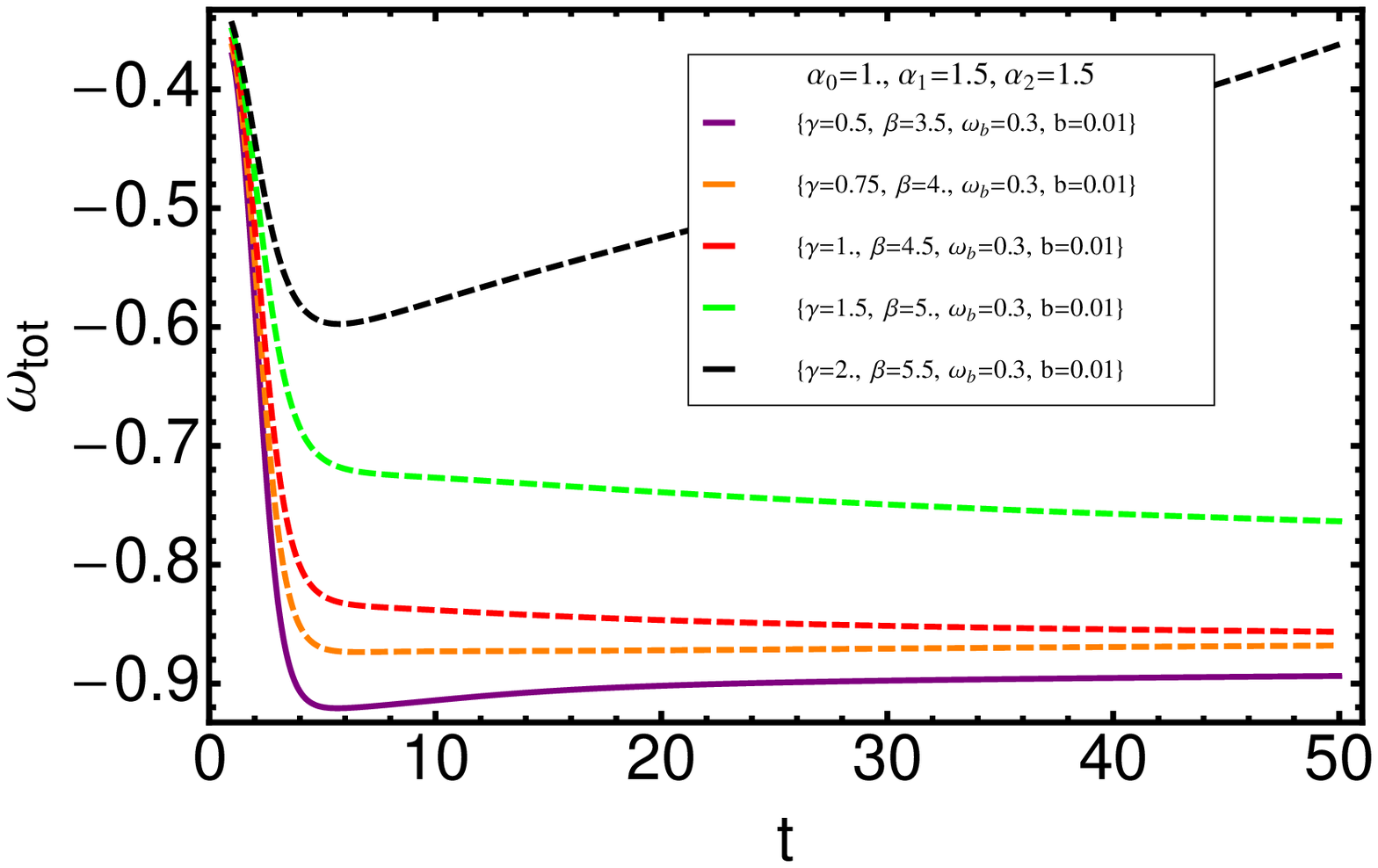} &
\includegraphics[width=50 mm]{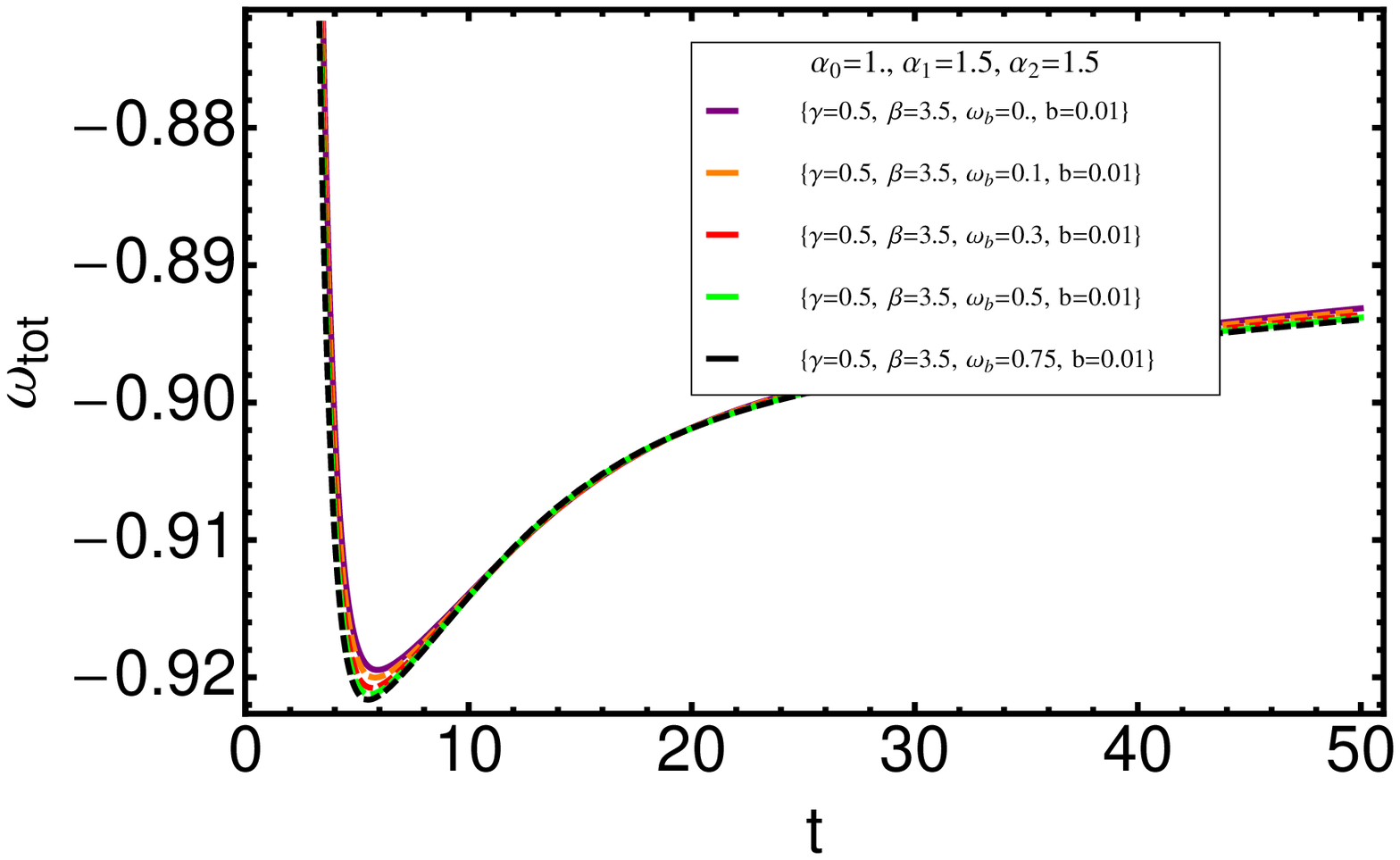}
 \end{array}$
 \end{center}
\caption{Behavior of EoS parameter $\omega_{tot}$ against $t$ for Model 2.}
 \label{fig:5}
\end{figure}

For the second model we will consider the following phenomenological
form of the $\Lambda(t)$
\begin{equation}\label{eq:Lambda2}
\Lambda(t)=\rho_{b}\sin{(tH)}^{3}+\rho_{D}\cos{(tH)}.
\end{equation}
Taking into account (\ref{eq: Fridmman vlambda}) we can write $\Lambda(t)$ in a different form
\begin{equation}\label{eq:Lambda2new}
\Lambda(t)=\left [ 1+\frac{\sin(tH)^{3}}{8 \pi G(t)} \right ]^{-1} \left( \frac{3H^{2}}{8 \pi G(t)}\sin{(tH)}^{3}-\rho_{D}(\sin{(tH)}^{3}-\cos{(tH)})\right ).
\end{equation}
\begin{equation}\label{eq:G2}
\frac{\dot{G}(t)}{G(t)}+\frac{\dot{\Lambda}(t)}{3H^{2}-\Lambda(t)}=0,
\end{equation}
with (\ref{eq:Lambda2new}) will give us the behavior of $G(t)$
Fig(\ref{fig:4}). We see that $G(t)$ is an
increasing-decreasing-increasing function (Top panel and
right-bottom plot). The left-bottom plot gives us an information
about the behavior of $G(t)$ as a function of $\gamma$ and $\beta$
with $\alpha_{0}=1$, $\alpha_{1}=\alpha_{2}=1.5$ and
$\omega_{b}=0.3$, $b=0.01$. We see that with increasing $\gamma$ and
$\beta$ we are able to change the behavior of $G(t)$. For instance,
with $\gamma=0.5$ and $\beta=3.5$ which is a blue line, still
preserves the increasing-decreasing-increasing behavior. While for
higher values of the parameters, we change the behavior of $G(t)$
compared to the other cases within this model and we have
increasing-decreasing behavior. Graphical behavior of $\omega_{tot}$
can be found in Fig.{\ref{fig:5}}. The behavior of the deceleration
parameter $q$ for this model gives us almost the same as for Model
1, where $\Lambda(t)=\rho_{D}$. We also see that with increasing
$\gamma$ and $\beta$ we increase the value of $q$ (left-bottom
plot). The presence of the interaction $Q$ and the barotropic fluid
for which EoS parameter $\omega_{b}<1$ does not leave a serious
impact on the behavior of $q$. This model with this behavior of
$q>-1$ can be comparable with the observational facts.

\begin{figure}[h!]
 \begin{center}$
 \begin{array}{cccc}
\includegraphics[width=50 mm]{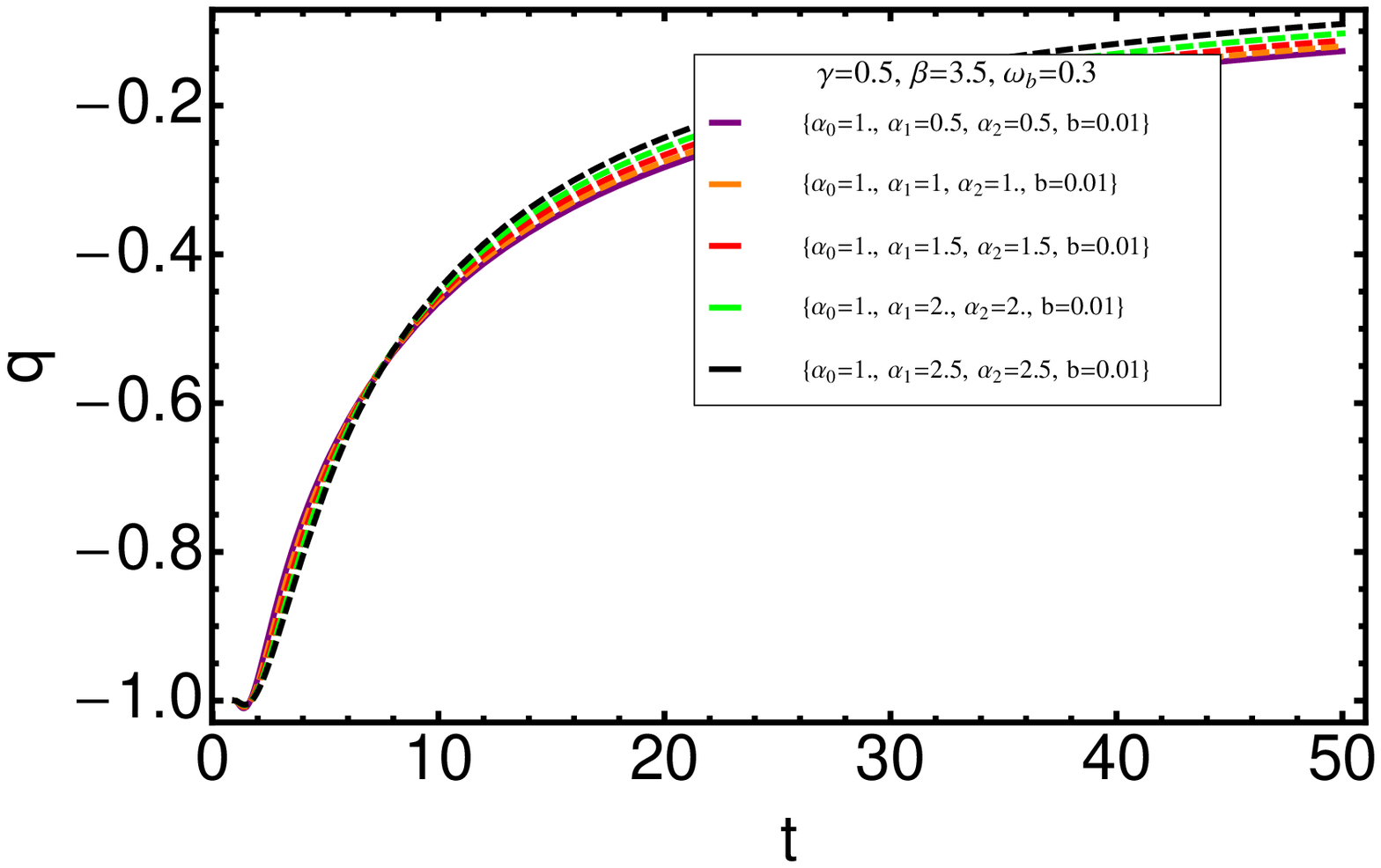} &
\includegraphics[width=50 mm]{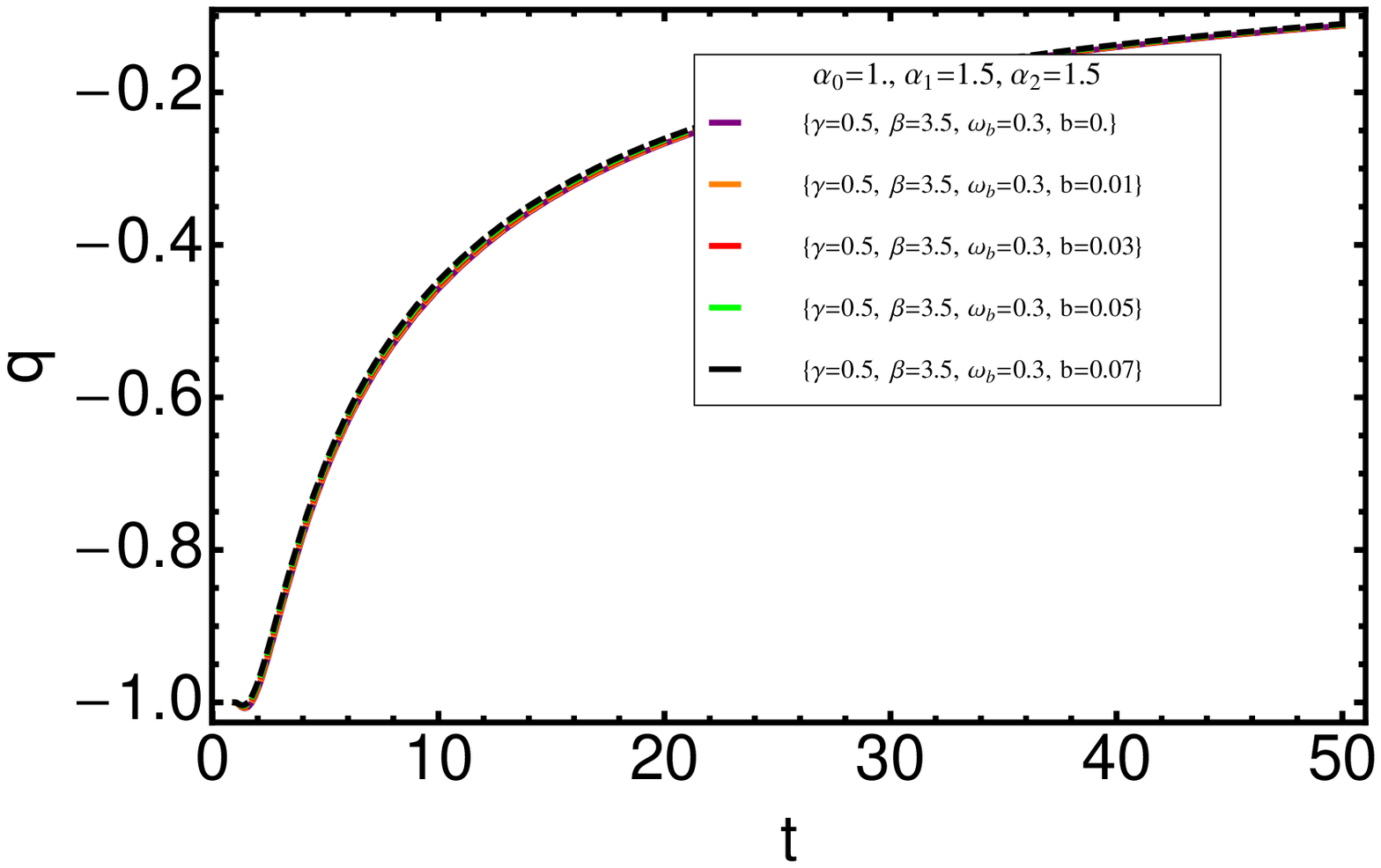}\\
\includegraphics[width=50 mm]{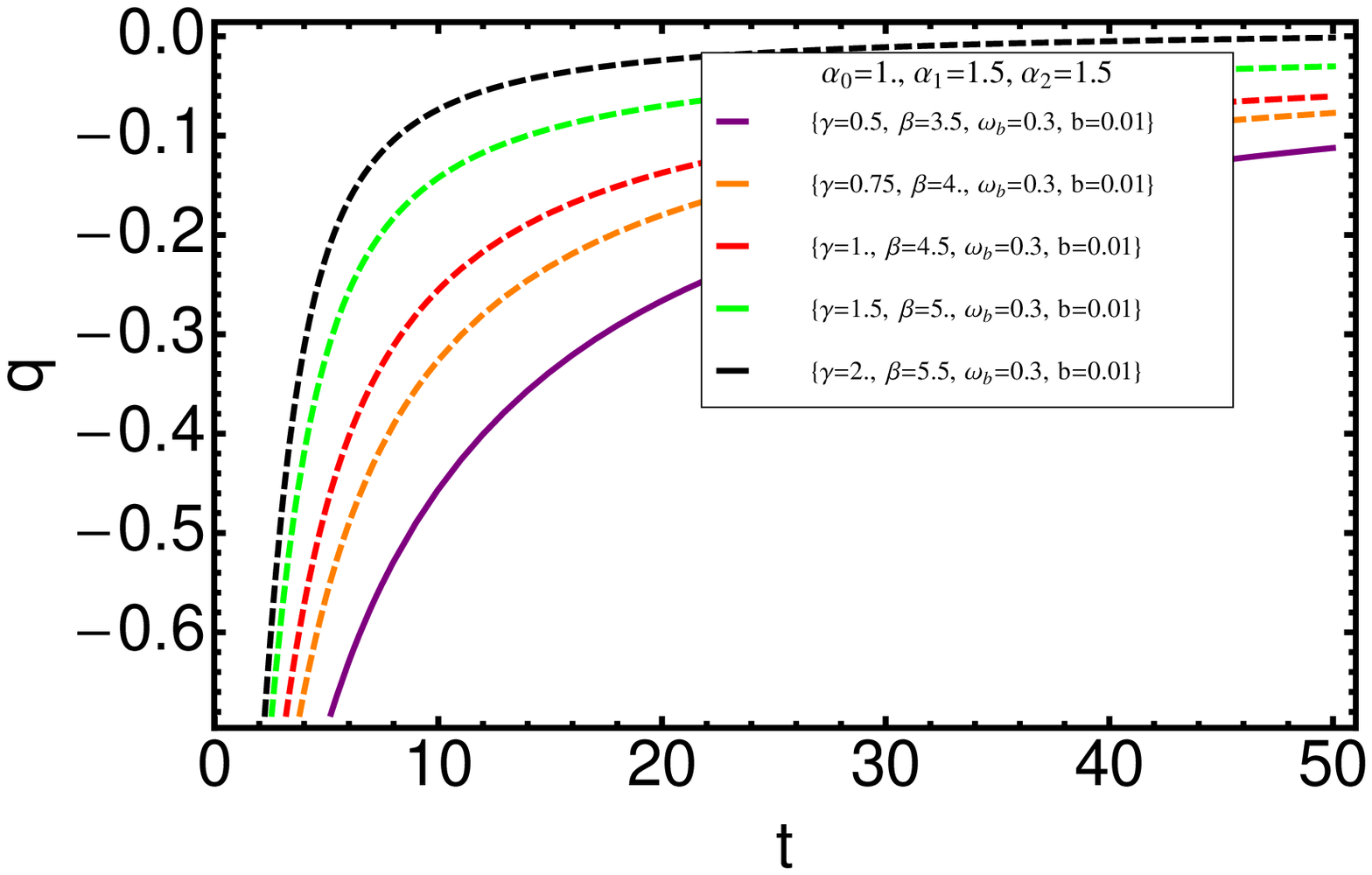} &
\includegraphics[width=50 mm]{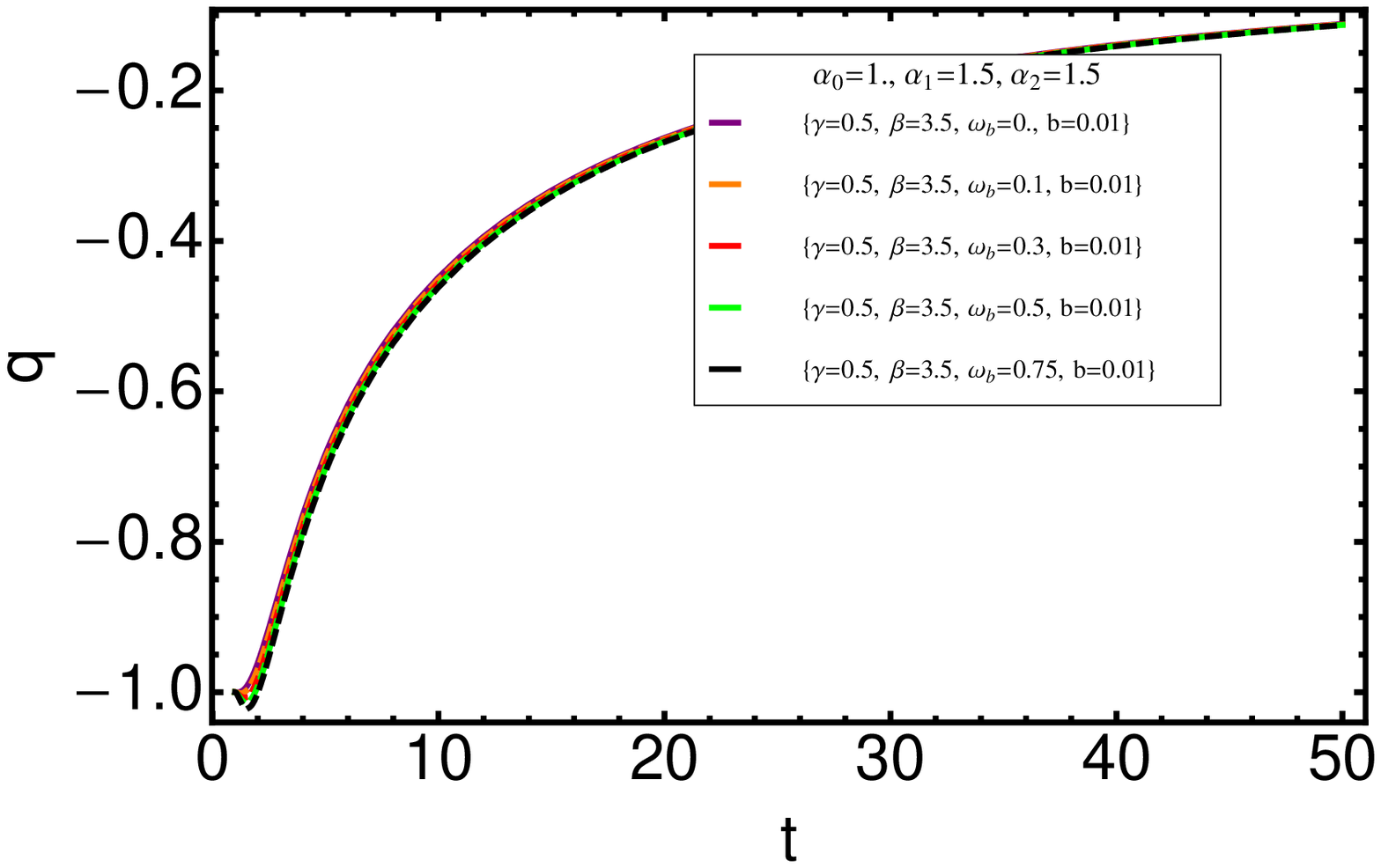}
 \end{array}$
 \end{center}
\caption{Behavior of deceleration parameter $q$ against $t$ for Model 2.}
 \label{fig:6}
\end{figure}
\begin{figure}[h!]
 \begin{center}$
 \begin{array}{cccc}
\includegraphics[width=50 mm]{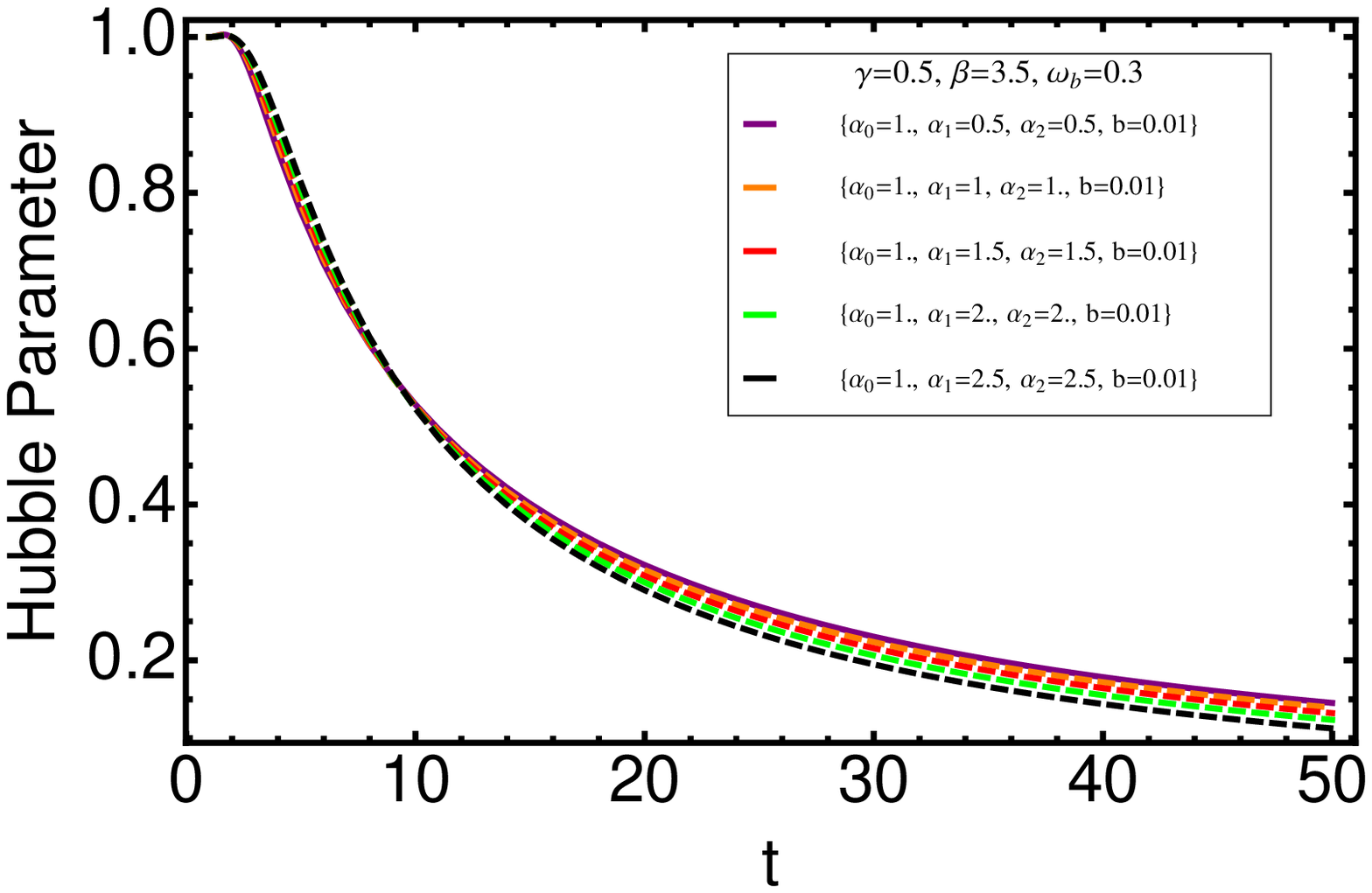} &
\includegraphics[width=50 mm]{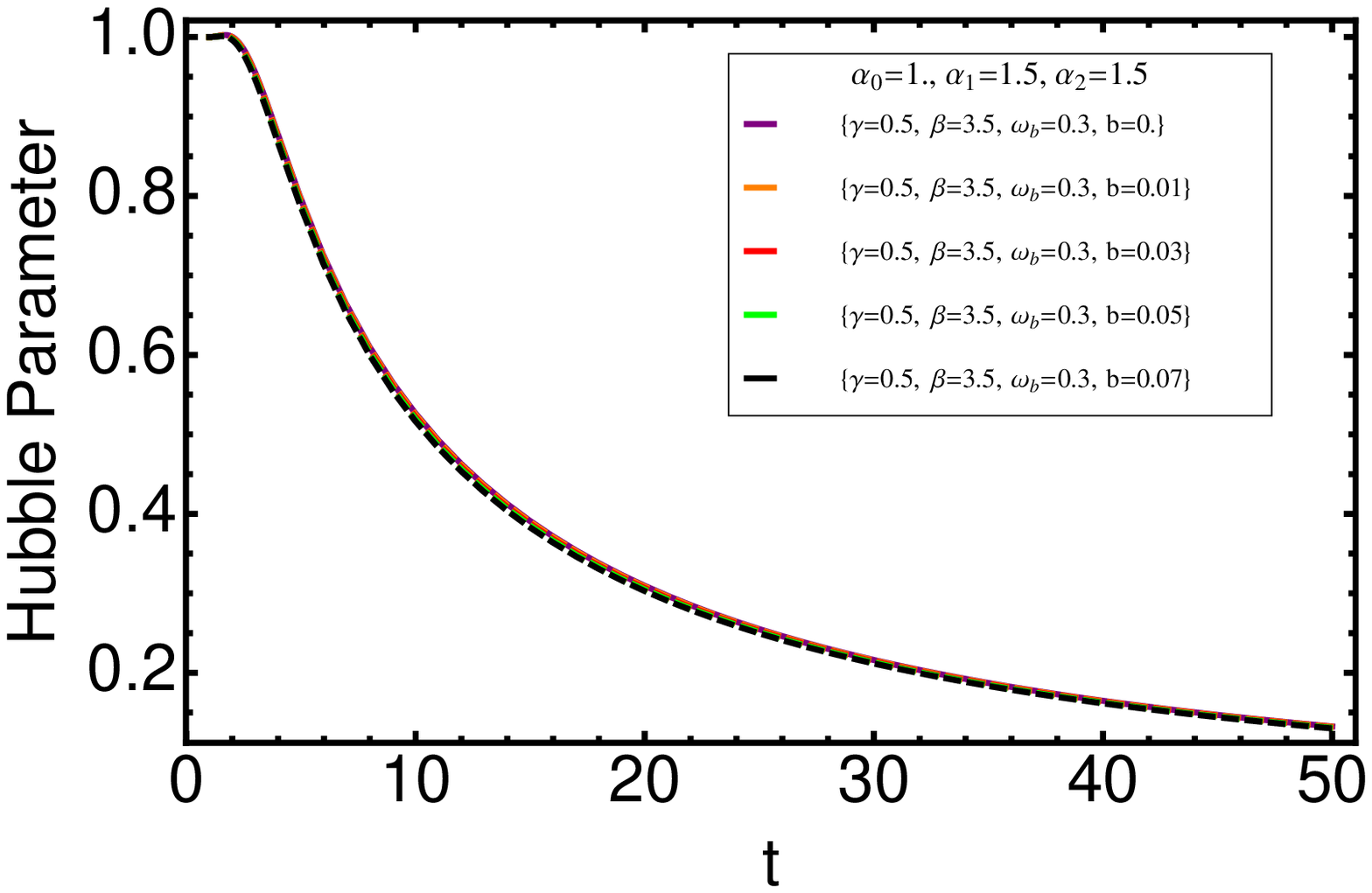}\\
\includegraphics[width=50 mm]{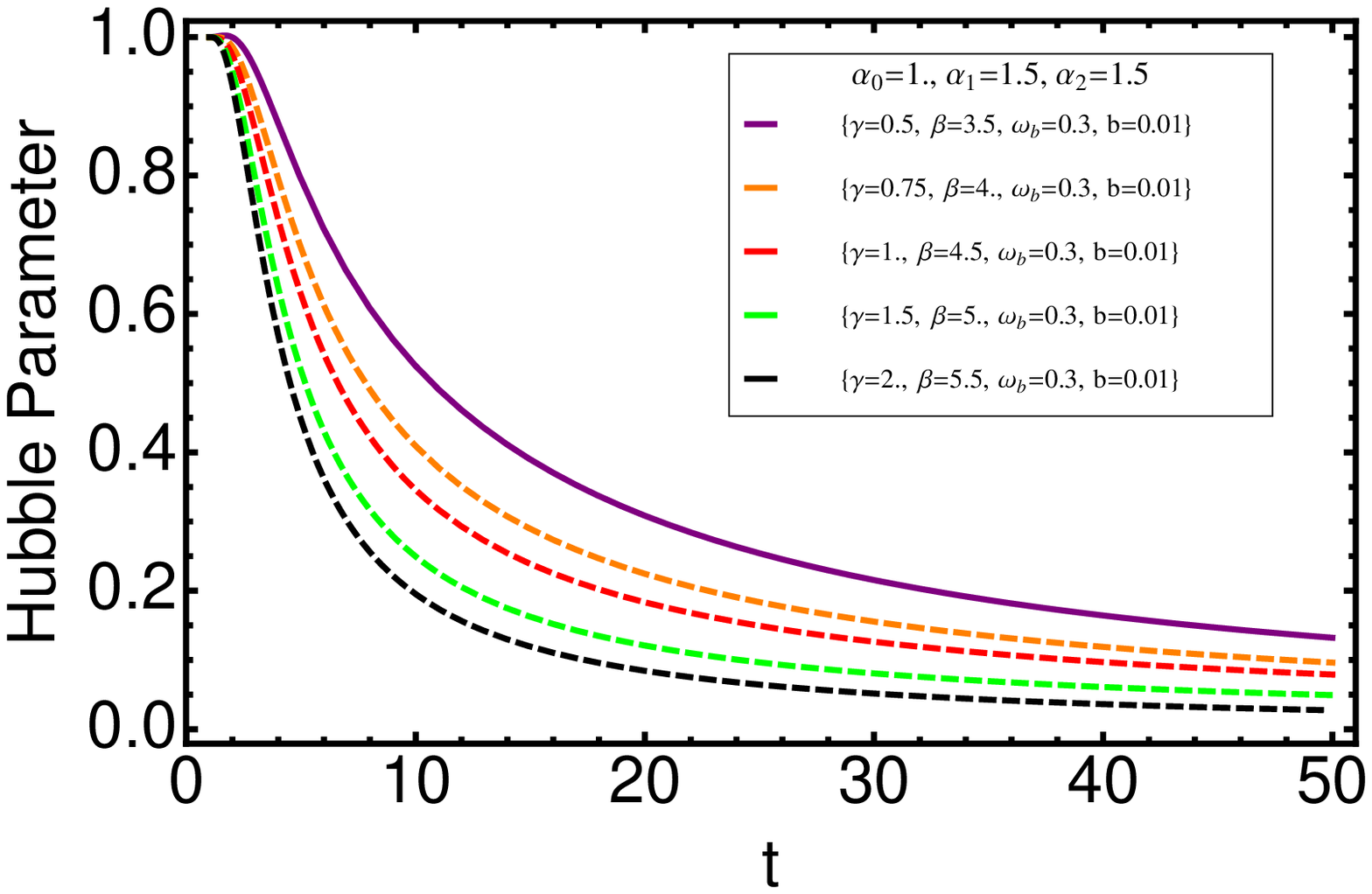} &
\includegraphics[width=50 mm]{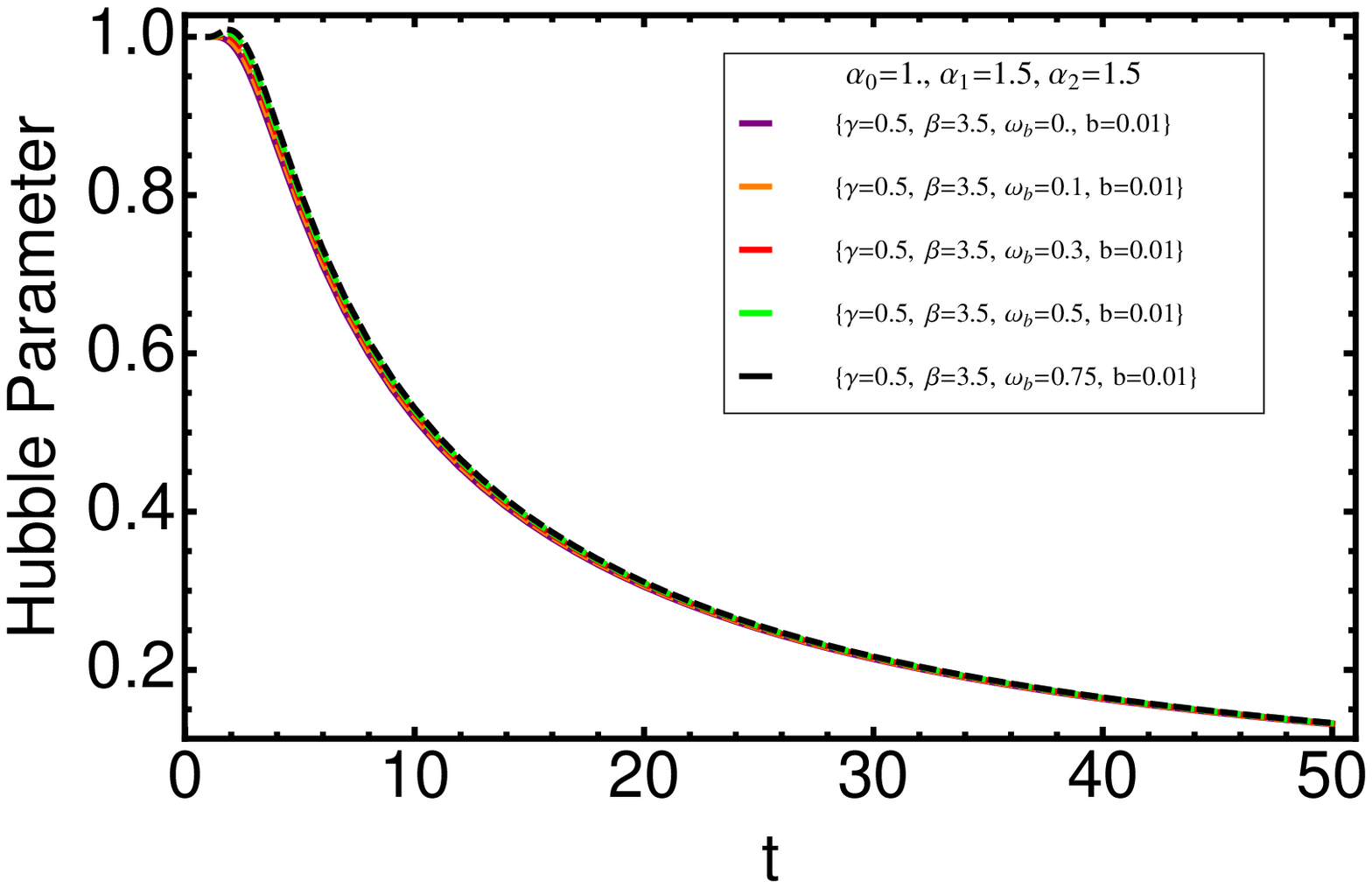}
 \end{array}$
 \end{center}
\caption{Behavior of Hubble parameter $H(t)$ versus $t$ for Model 2.}
 \label{fig:14}
\end{figure}

\begin{figure}[h!]
 \begin{center}$
 \begin{array}{cccc}
\includegraphics[width=50 mm]{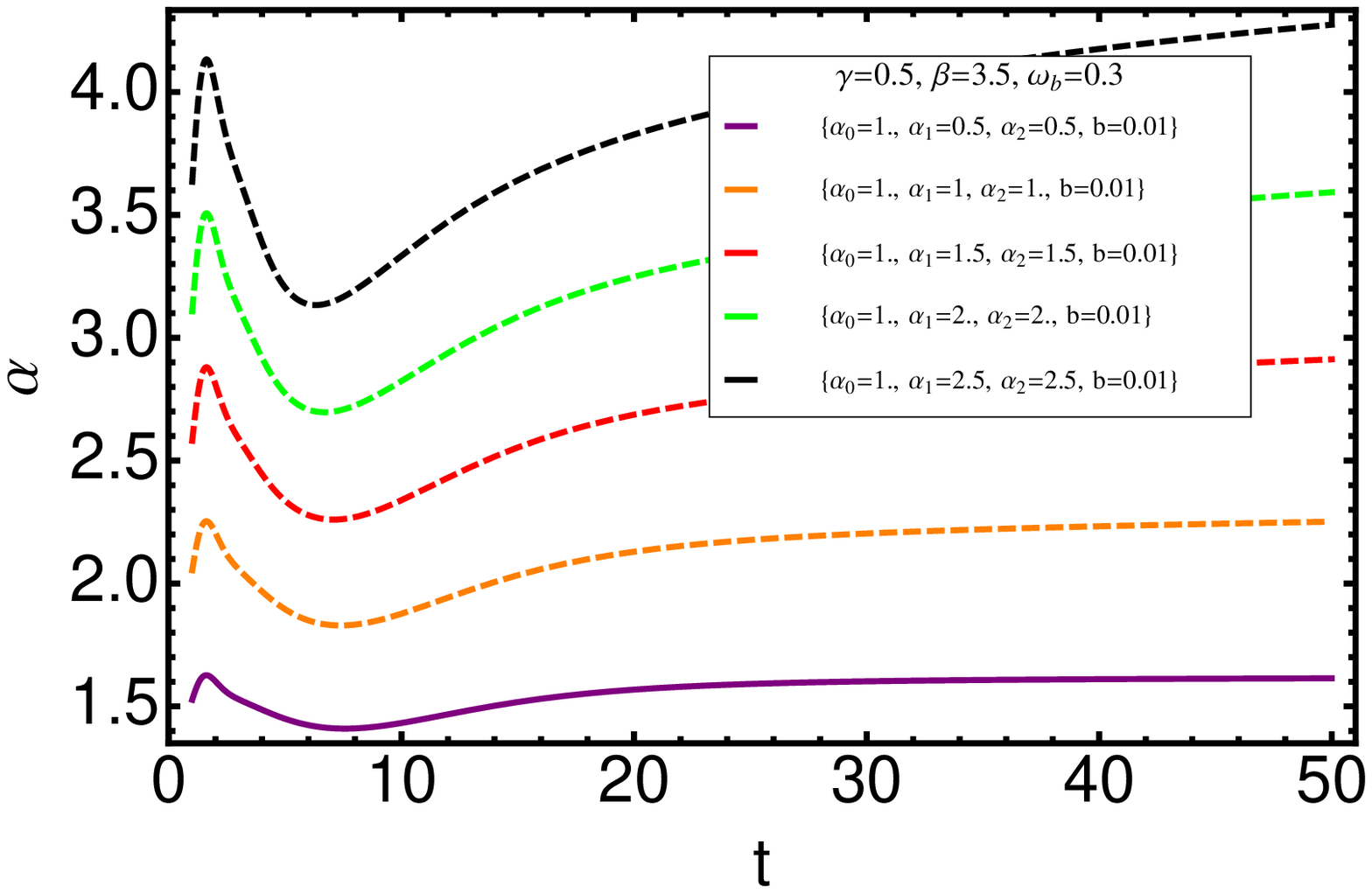} &
\includegraphics[width=50 mm]{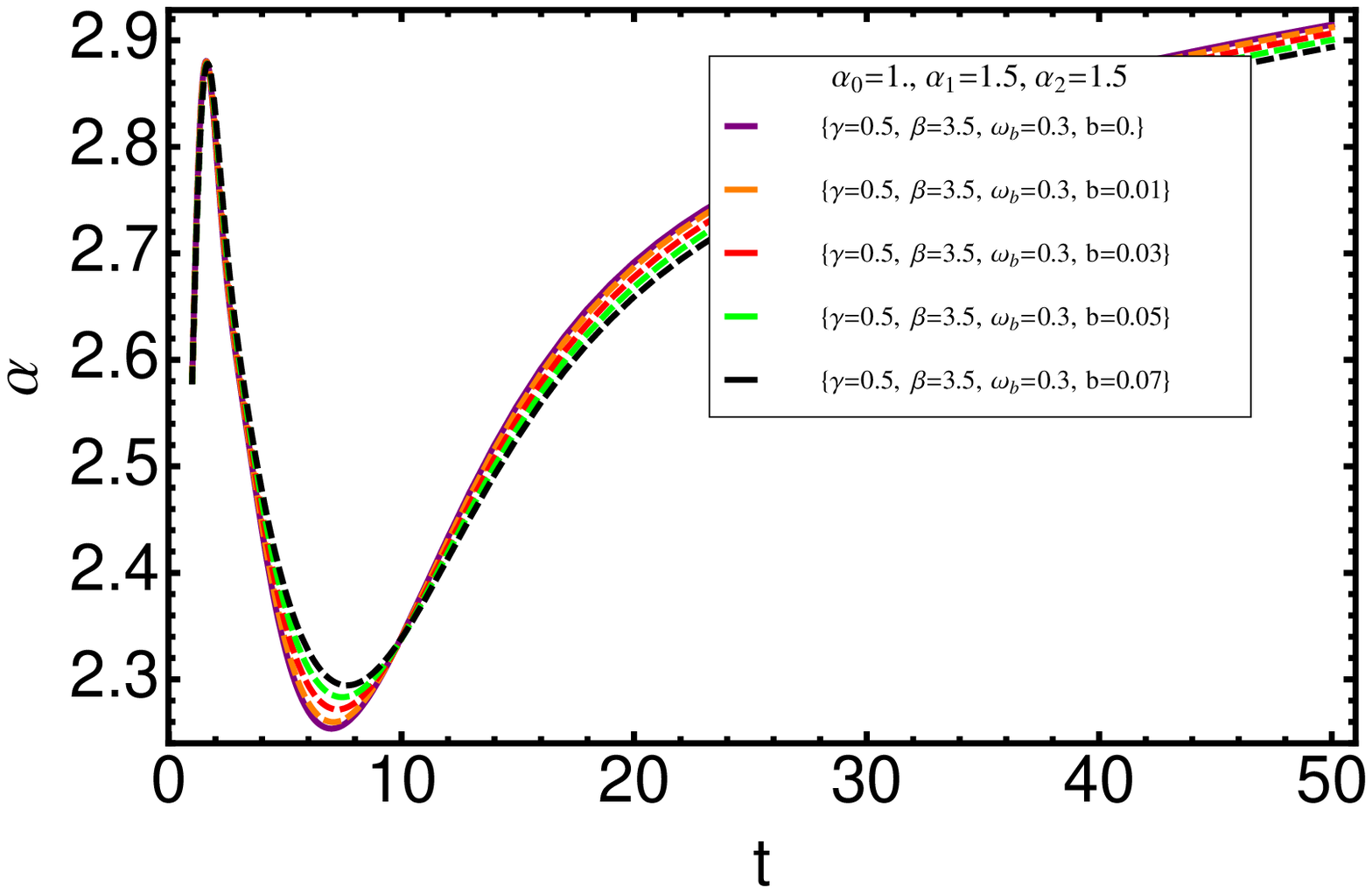}\\
\includegraphics[width=50 mm]{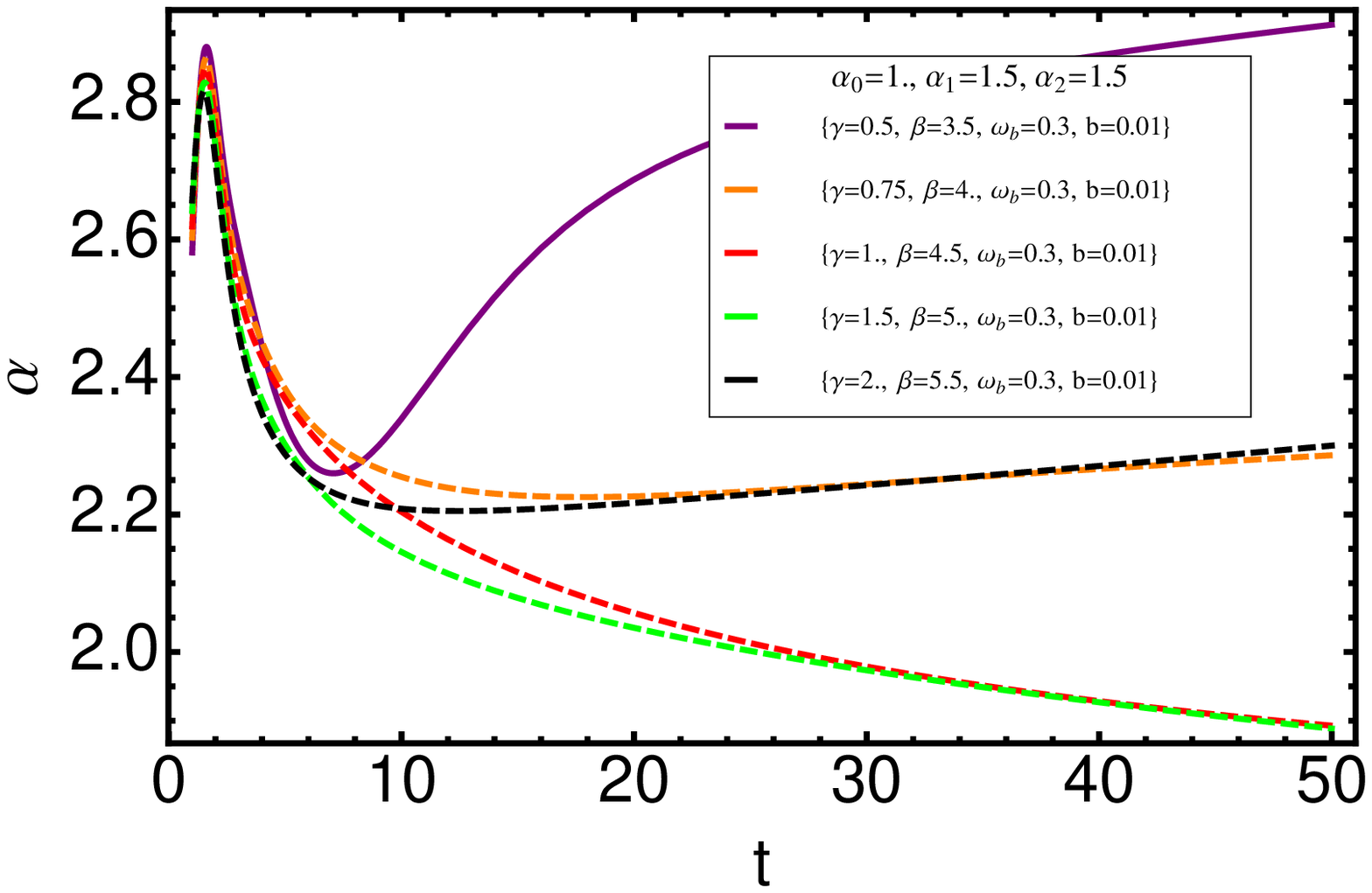} &
\includegraphics[width=50 mm]{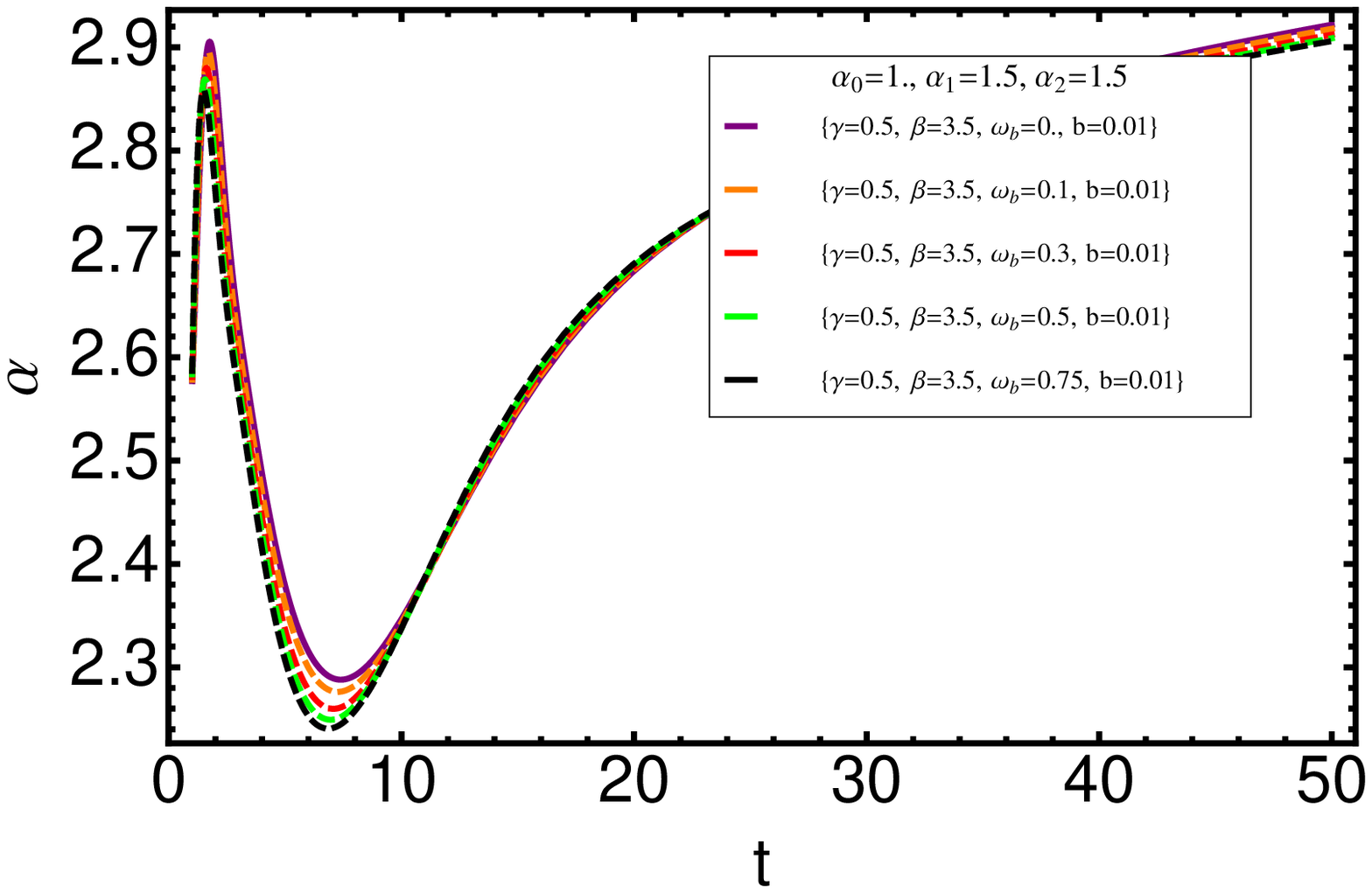}
 \end{array}$
 \end{center}
\caption{Behavior of $\alpha$ versus $t$ for Model 2.}
 \label{fig:15}
\end{figure}
\begin{figure}[h!]
 \begin{center}$
 \begin{array}{cccc}
\includegraphics[width=50 mm]{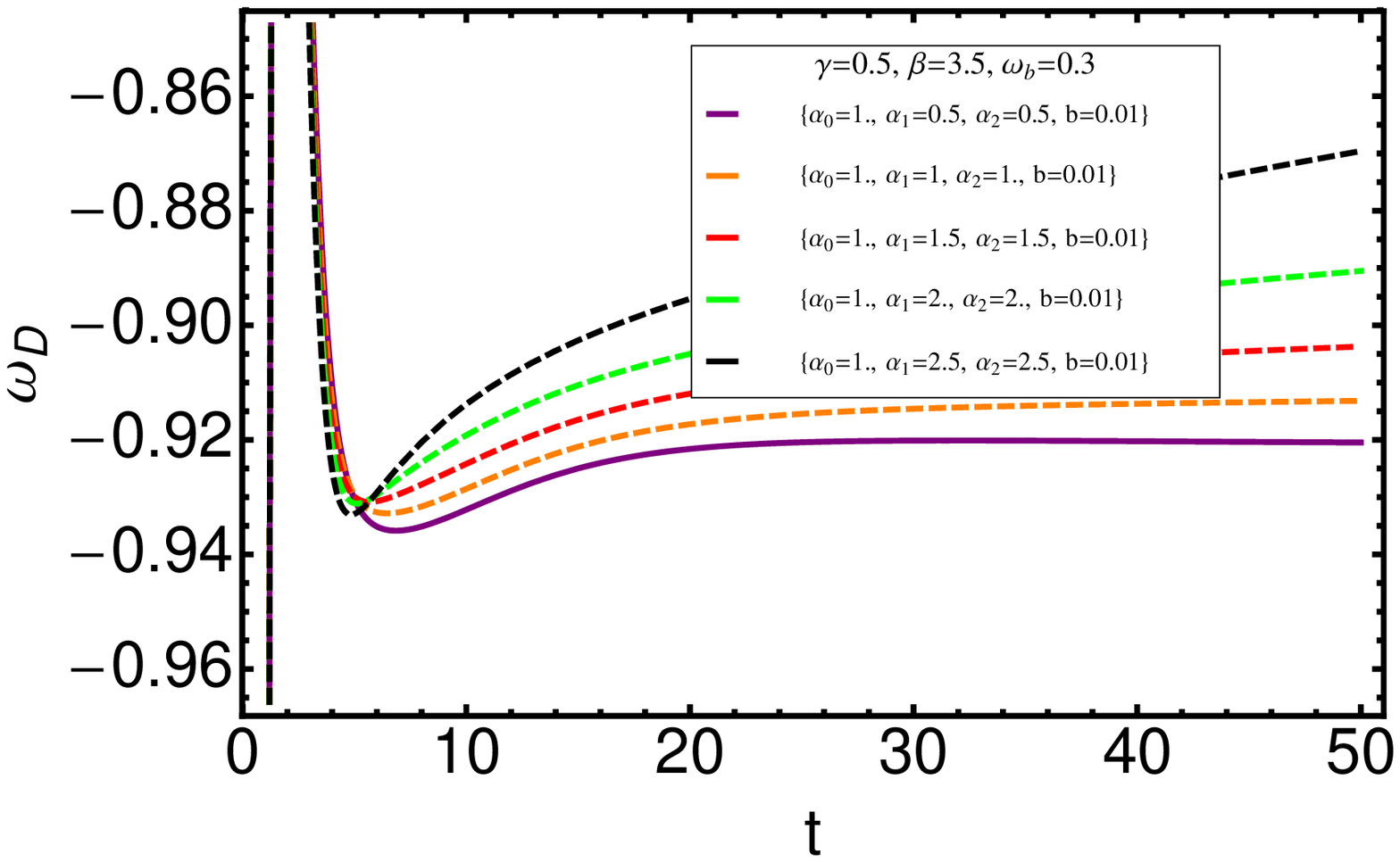} &
\includegraphics[width=50 mm]{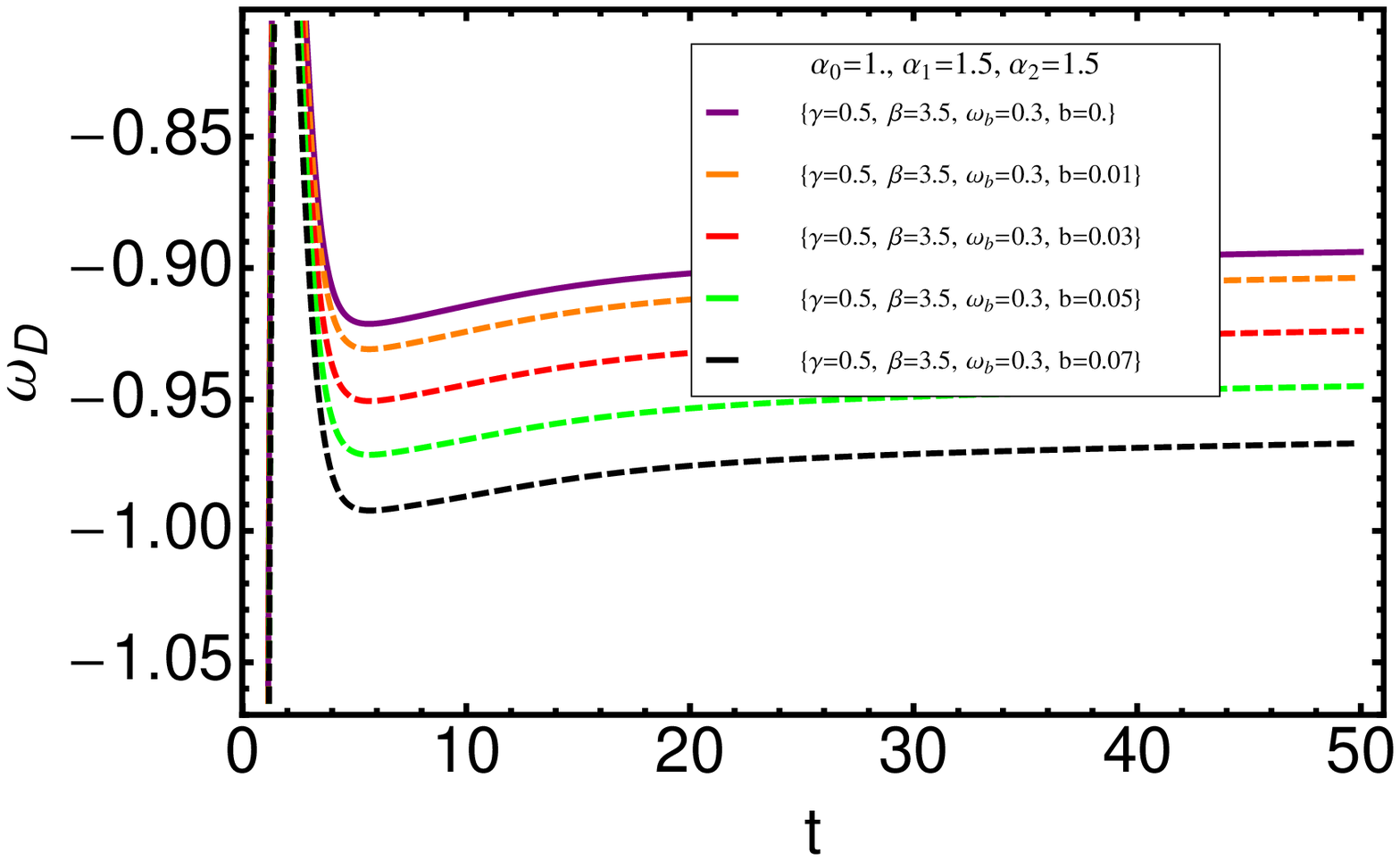}\\
\includegraphics[width=50 mm]{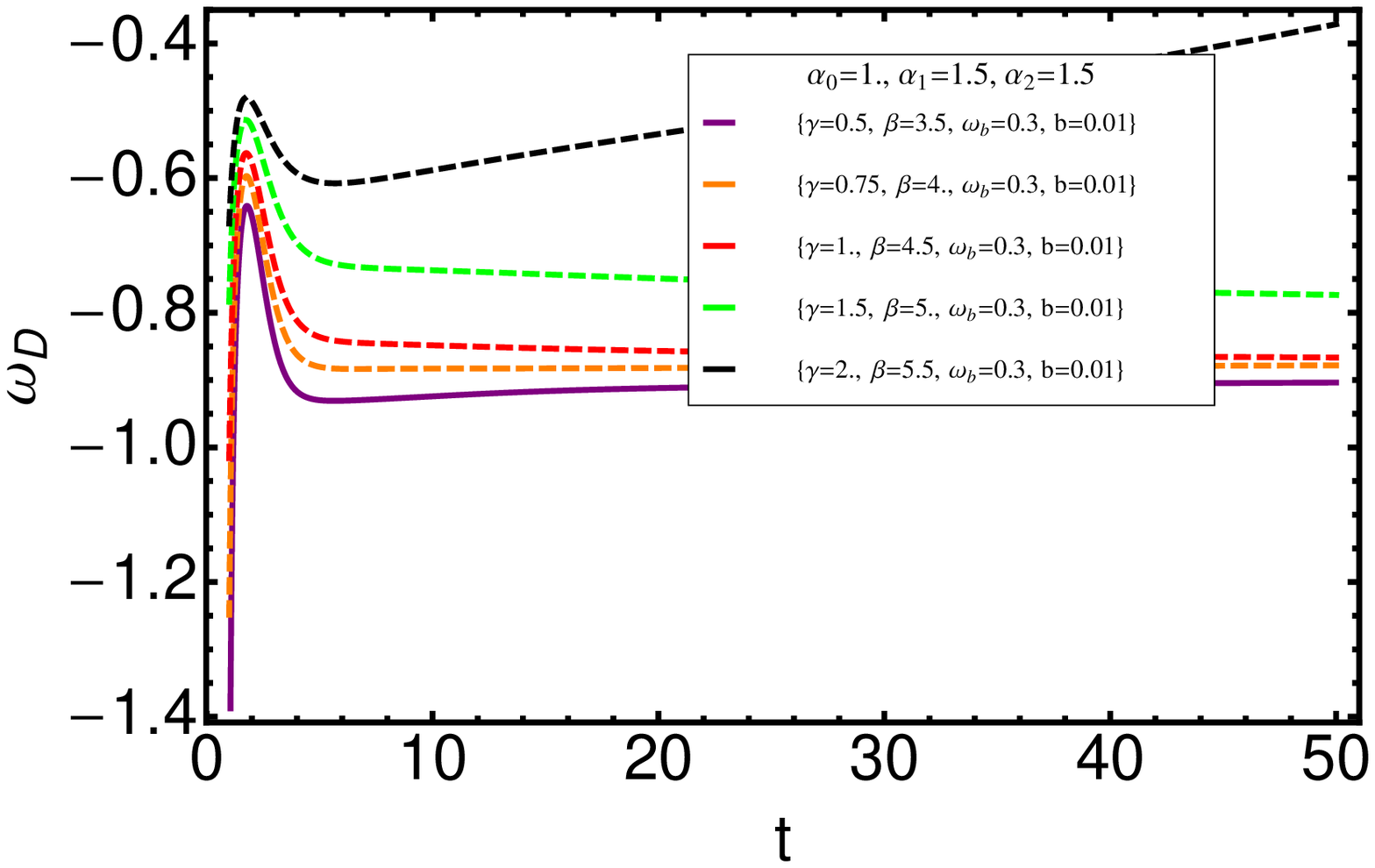} &
\includegraphics[width=50 mm]{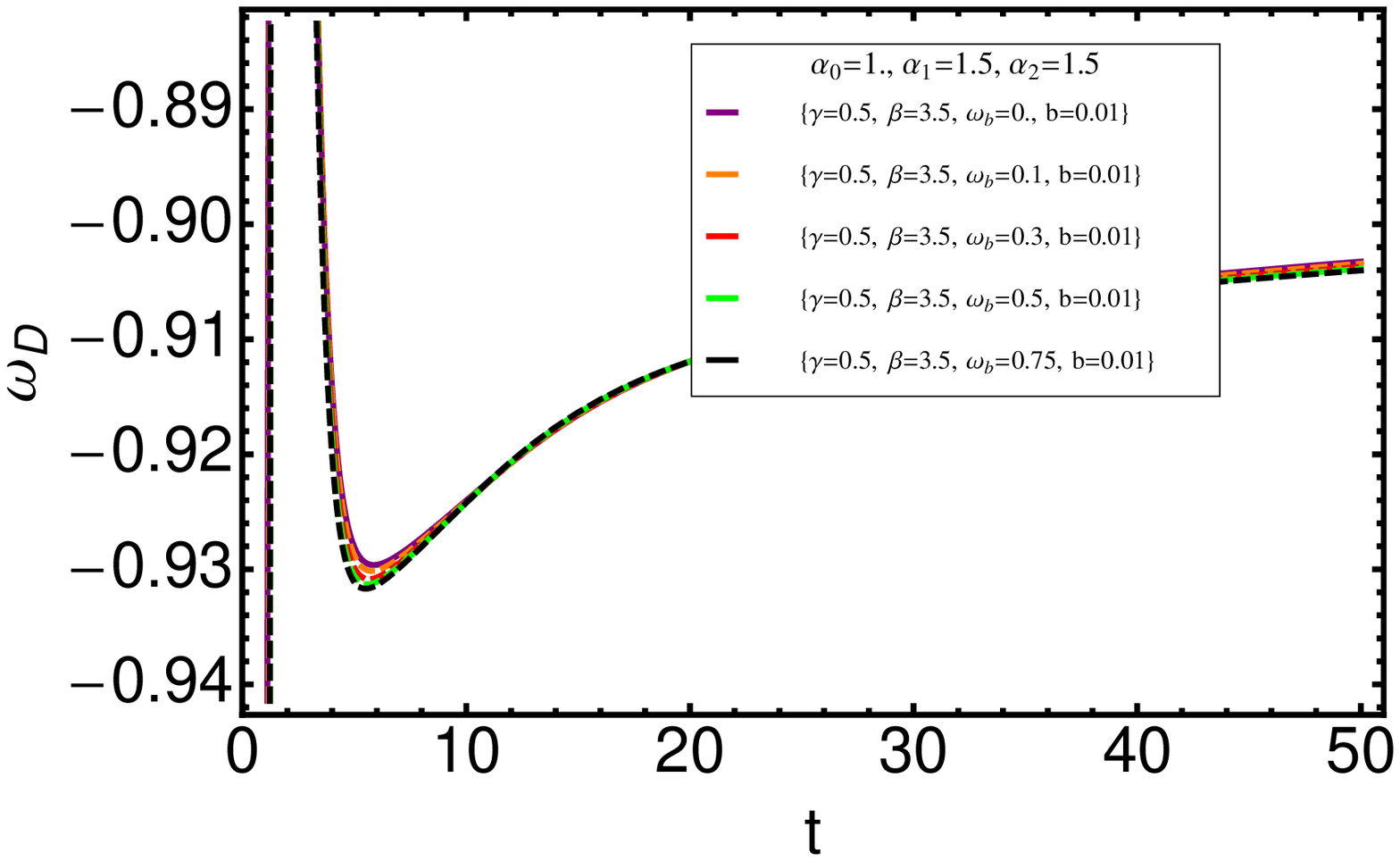}
 \end{array}$
 \end{center}
\caption{Behavior of $\omega_{D}$ against $t$ for Model 2.}
 \label{fig:16}
\end{figure}
\begin{figure}[h!]
 \begin{center}$
 \begin{array}{cccc}
\includegraphics[width=50 mm]{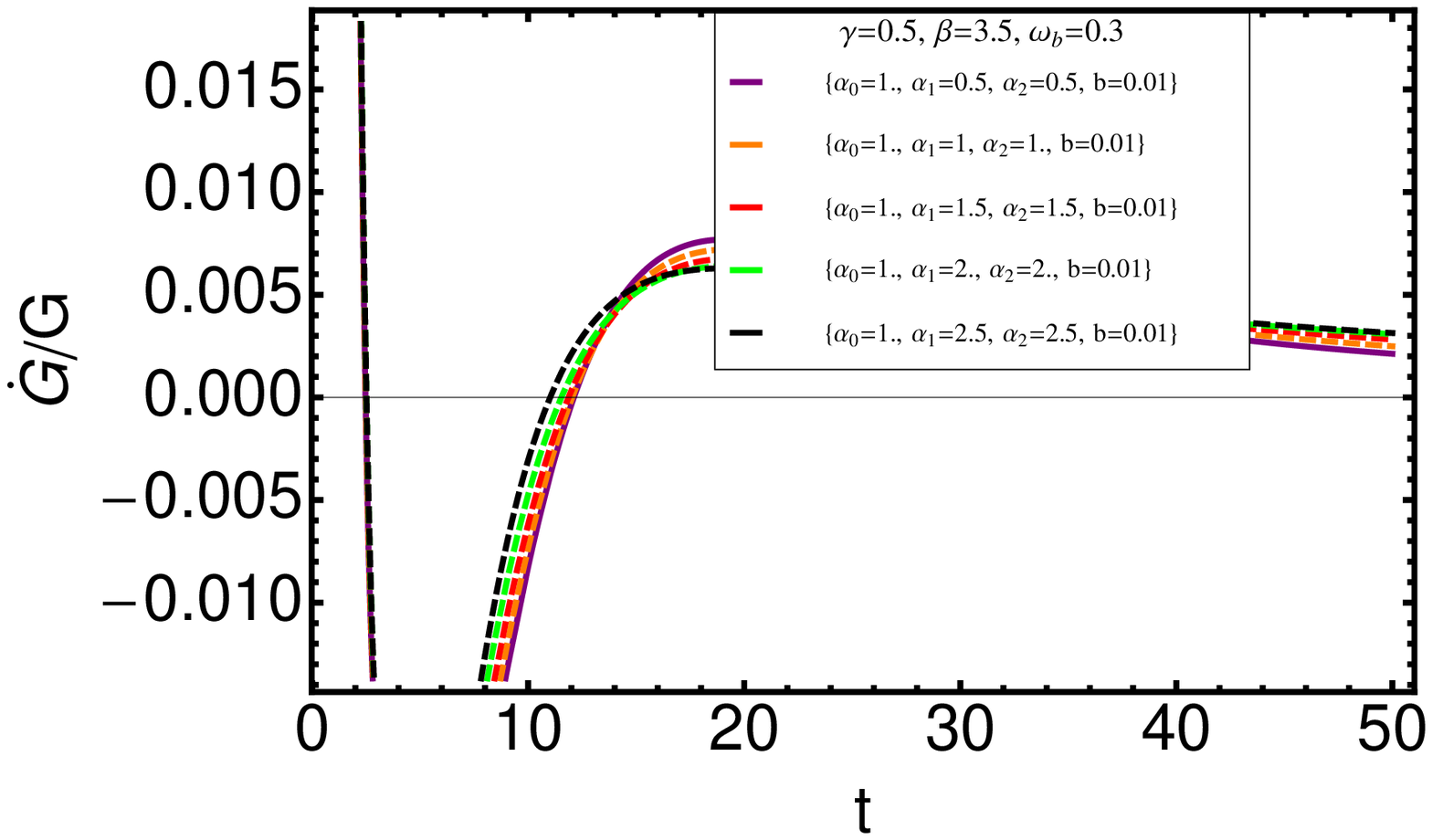} &
\includegraphics[width=50 mm]{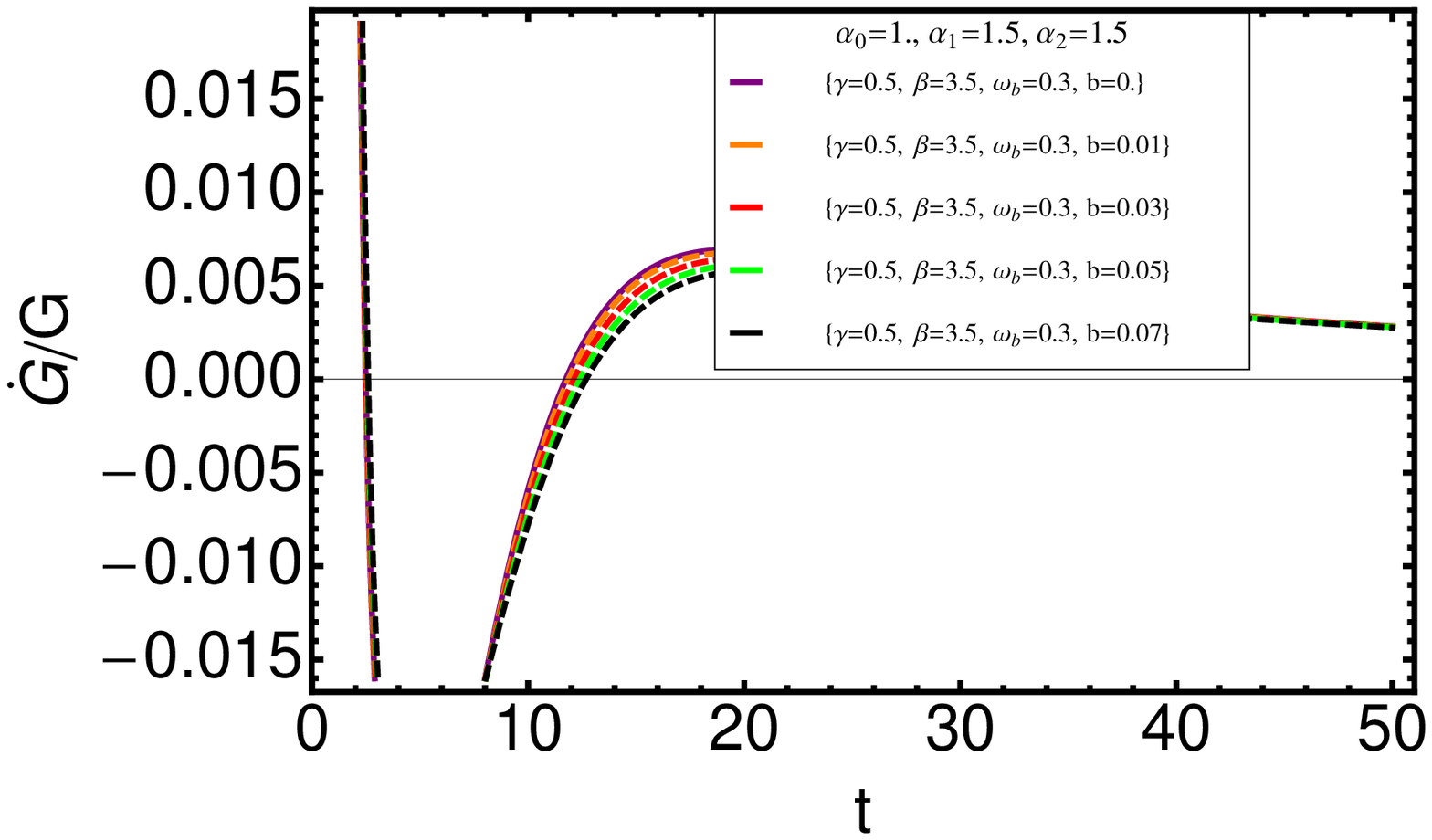}\\
\includegraphics[width=50 mm]{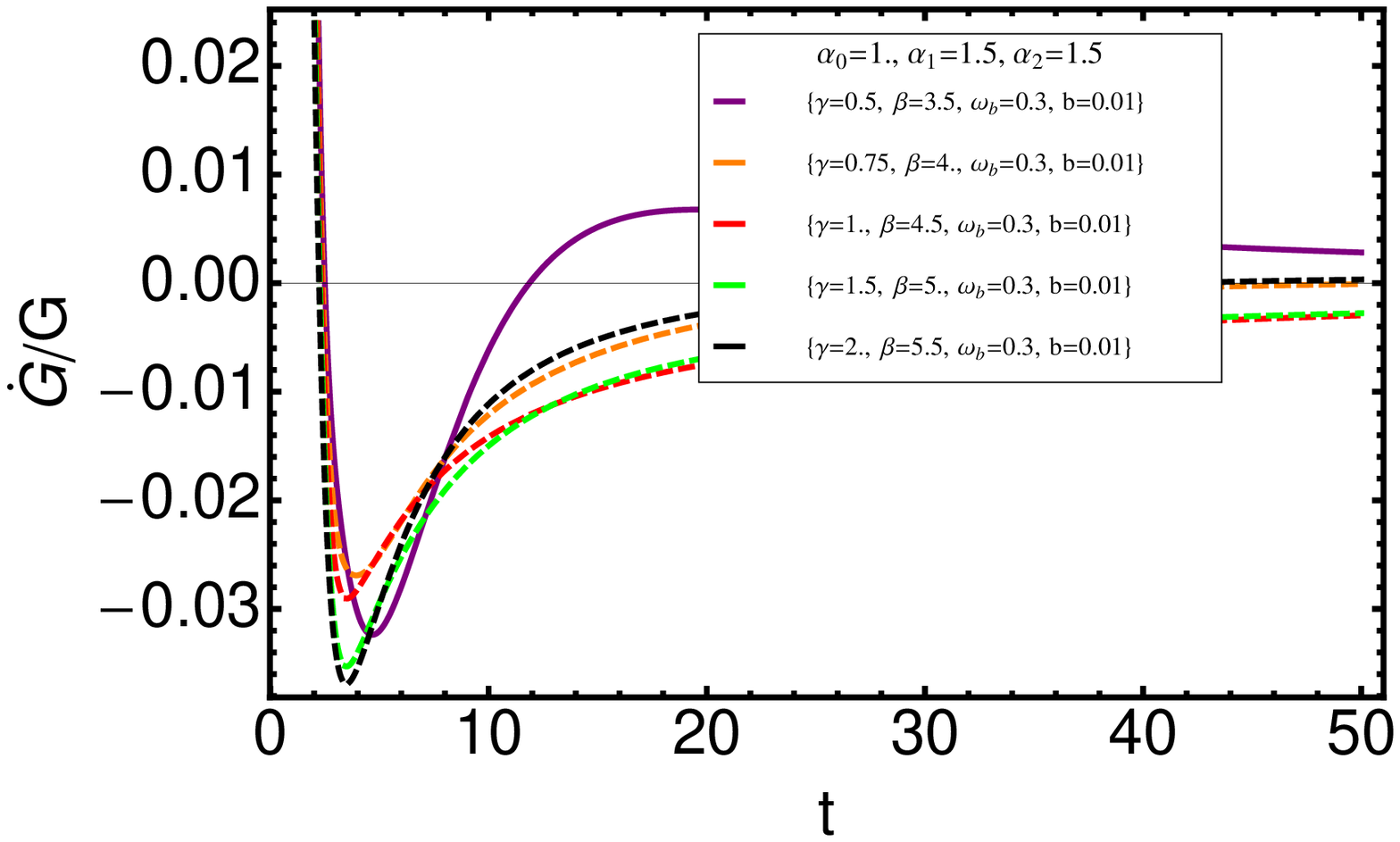} &
\includegraphics[width=50 mm]{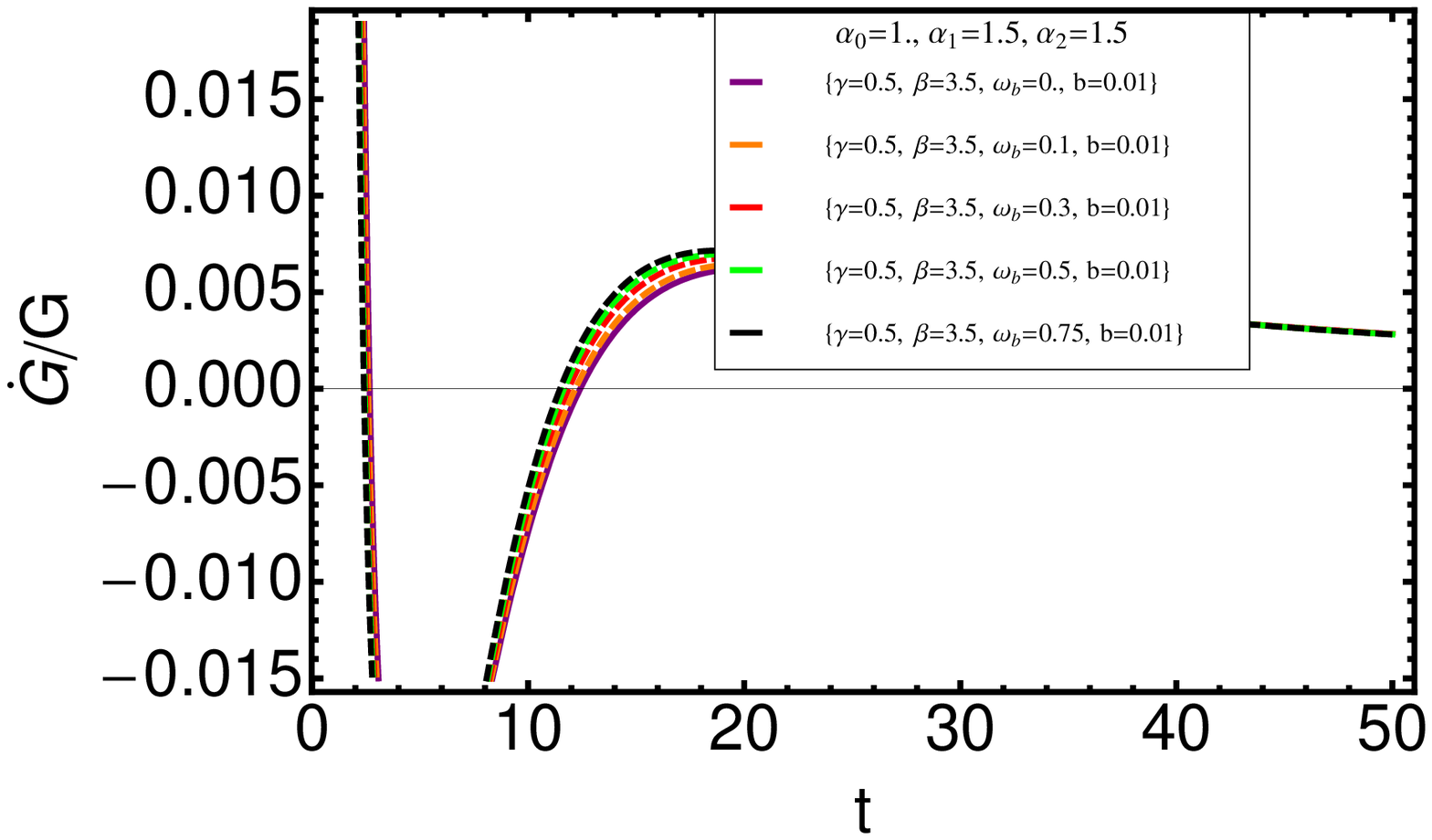}
 \end{array}$
 \end{center}
\caption{Behavior of $\dot{G}(t)/G(t)$ against $t$ for Model 2.}
 \label{fig:17}
\end{figure}

\subsection{Model 3: $\Lambda(t)=\rho_{b}\ln{(tH)}+\rho_{D}\sin{\left (t\frac{\dot{G}(t)}{G(t)}\right) }$}
\begin{figure}[h!]
 \begin{center}$
 \begin{array}{cccc}
\includegraphics[width=50 mm]{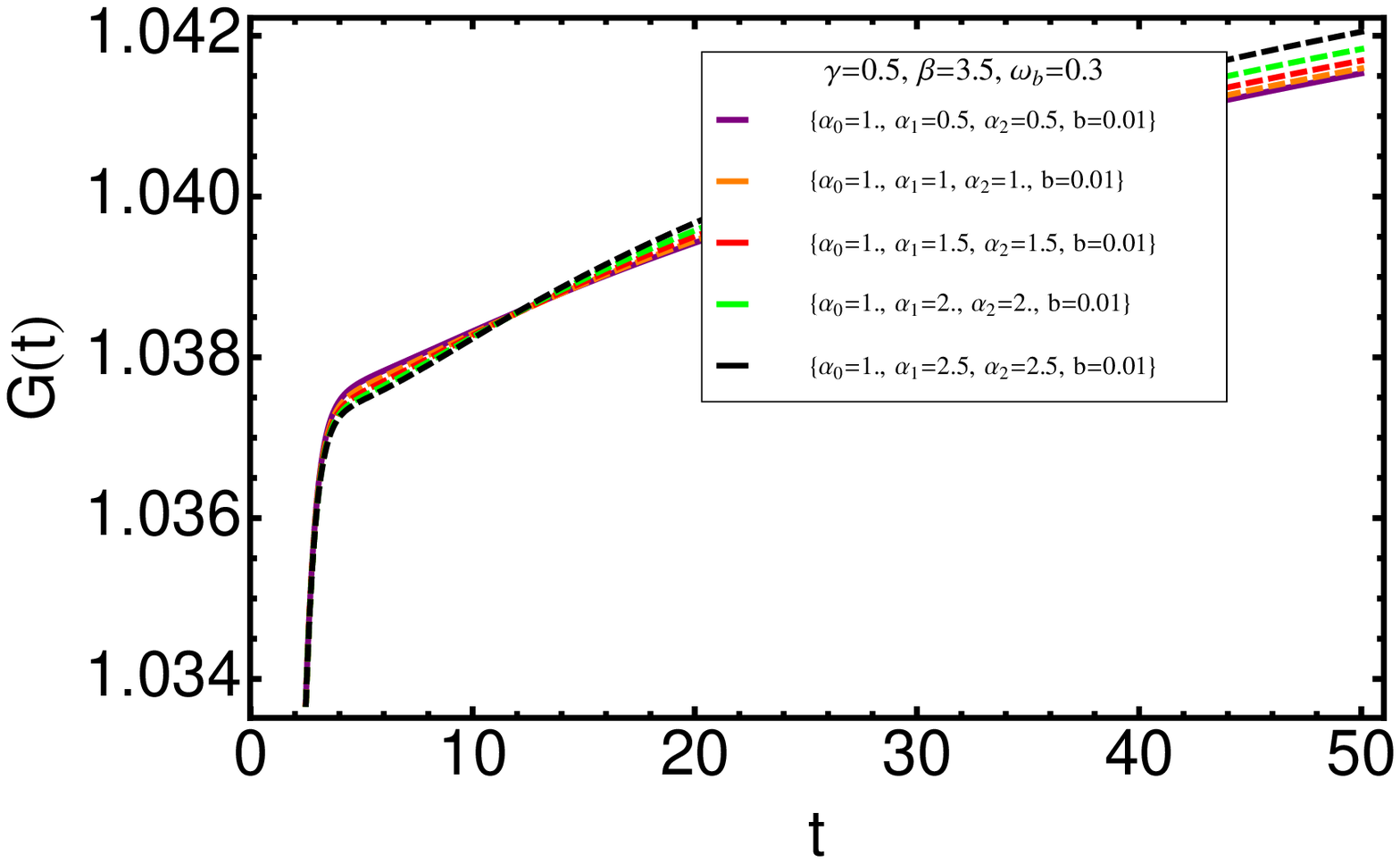} &
\includegraphics[width=50 mm]{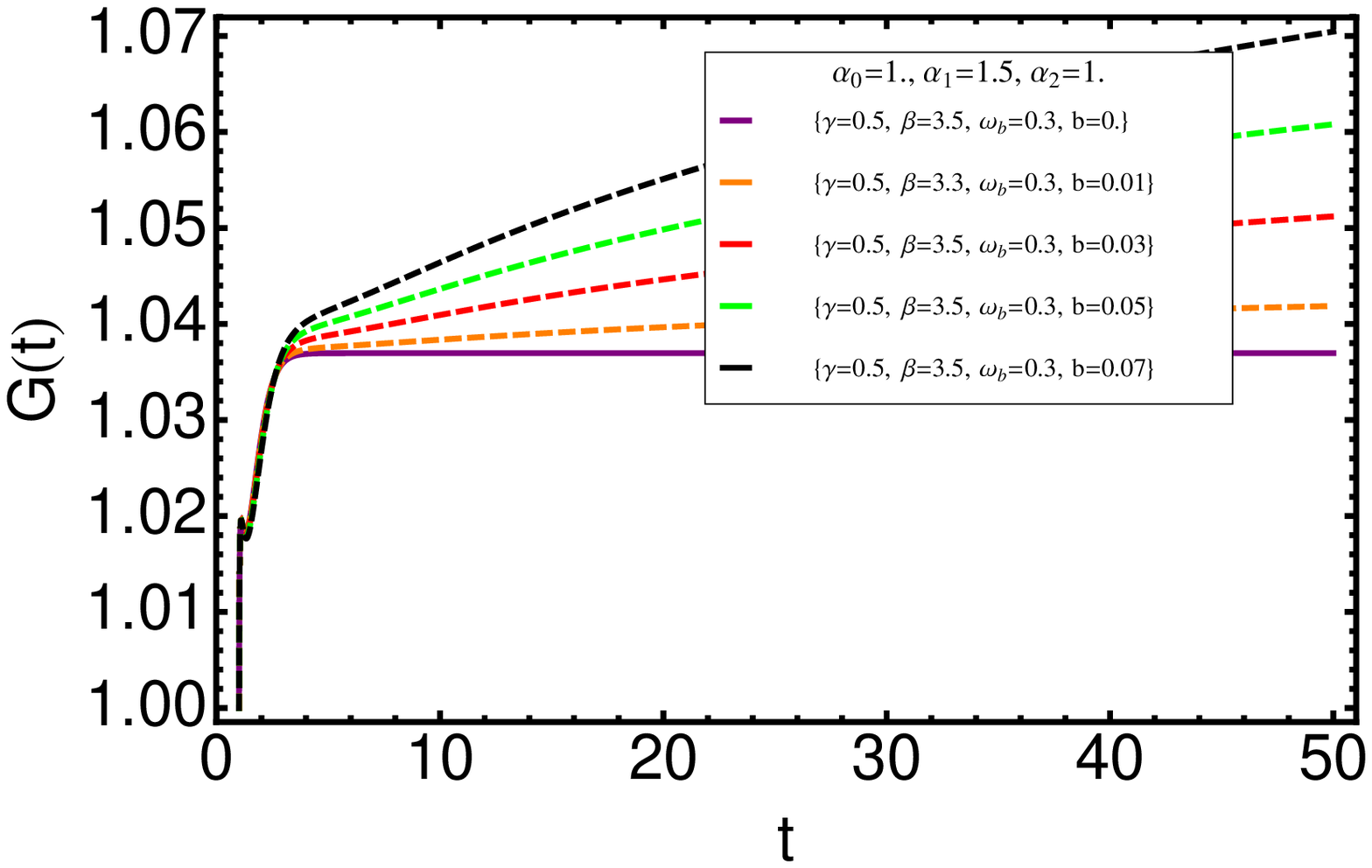}\\
\includegraphics[width=50 mm]{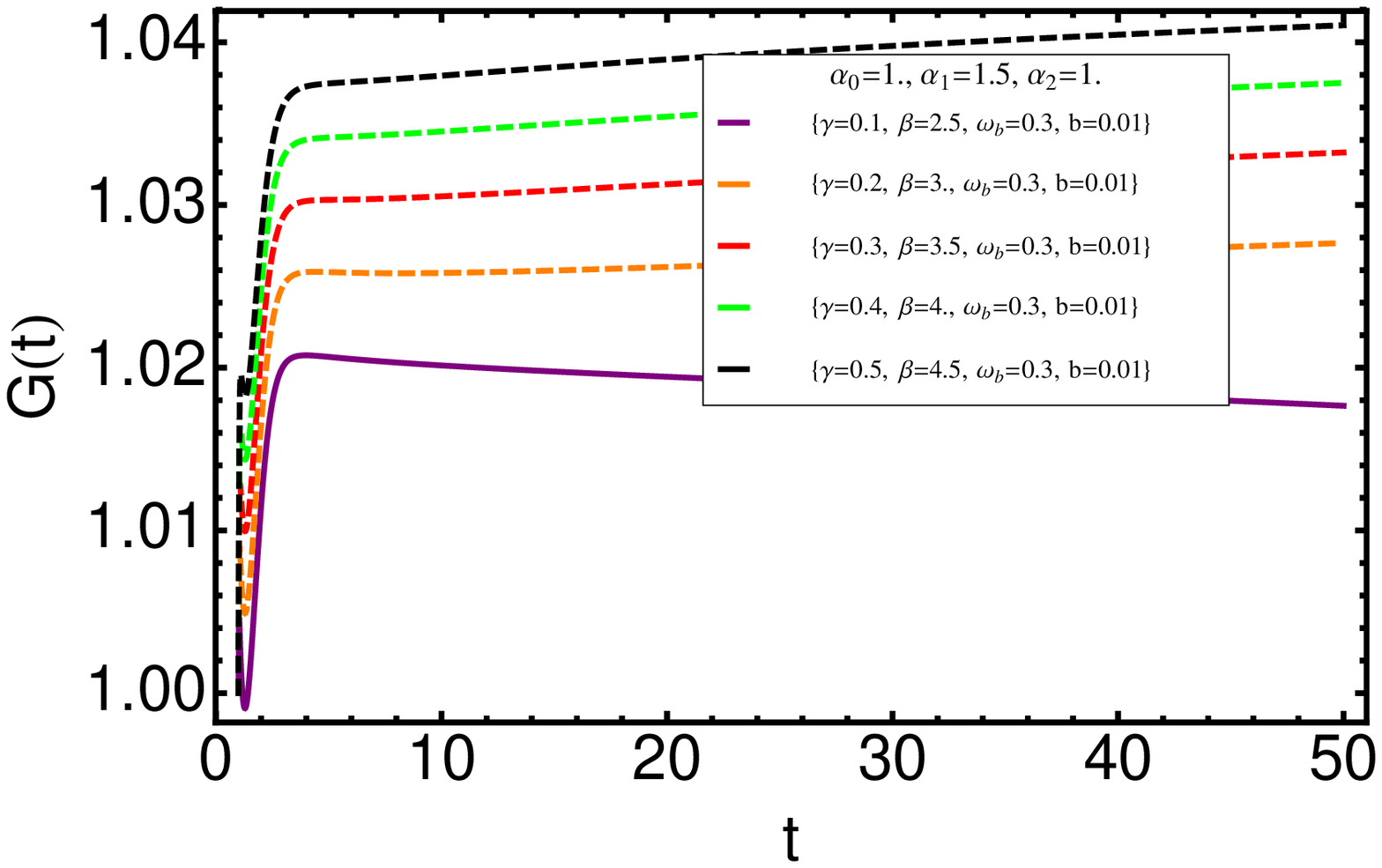} &
\includegraphics[width=50 mm]{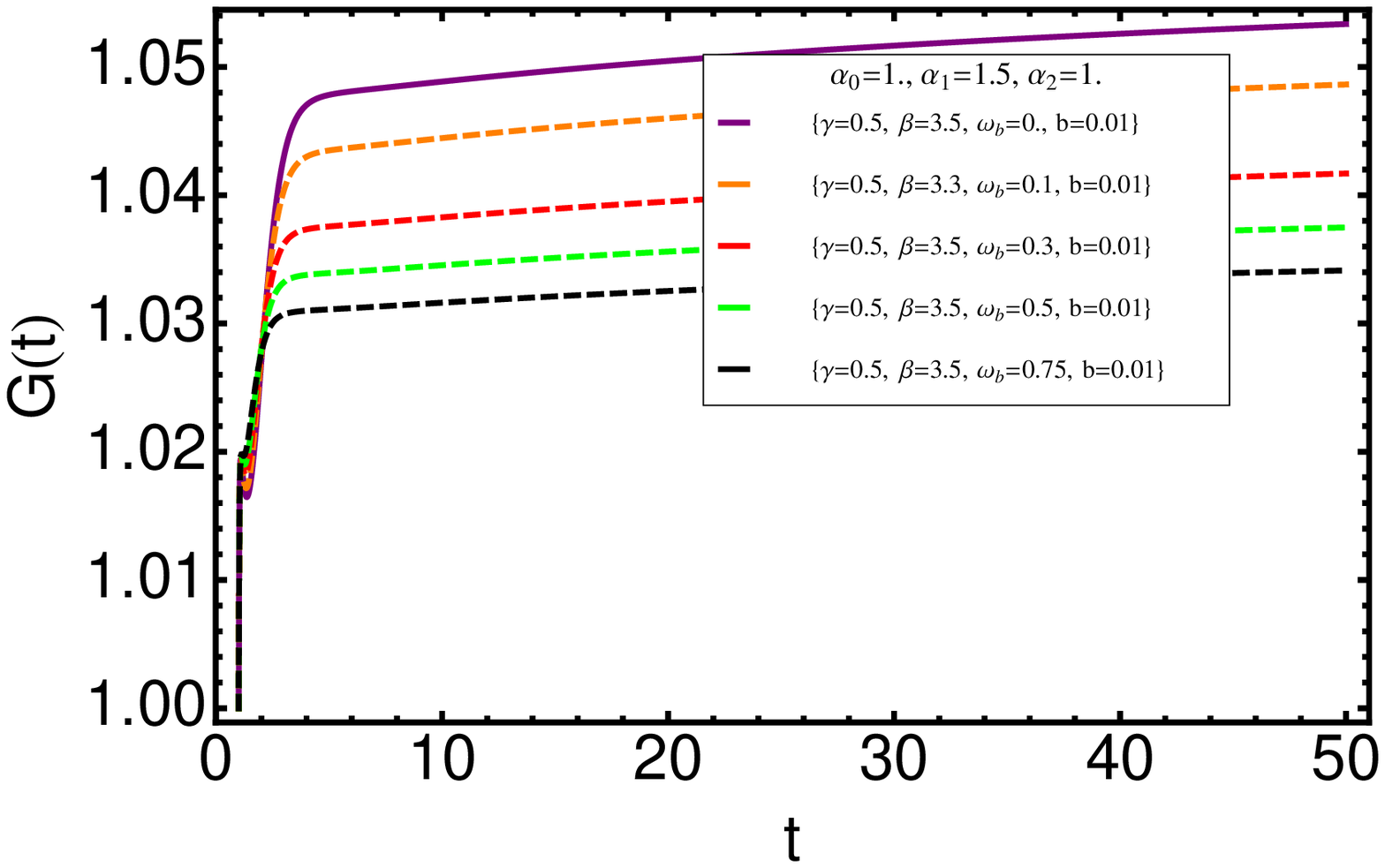}
 \end{array}$
 \end{center}
\caption{Behavior of Gravitational constant $G(t)$ against $t$ Model 3.}
 \label{fig:7}
\end{figure}

\begin{figure}[h!]
 \begin{center}$
 \begin{array}{cccc}
\includegraphics[width=50 mm]{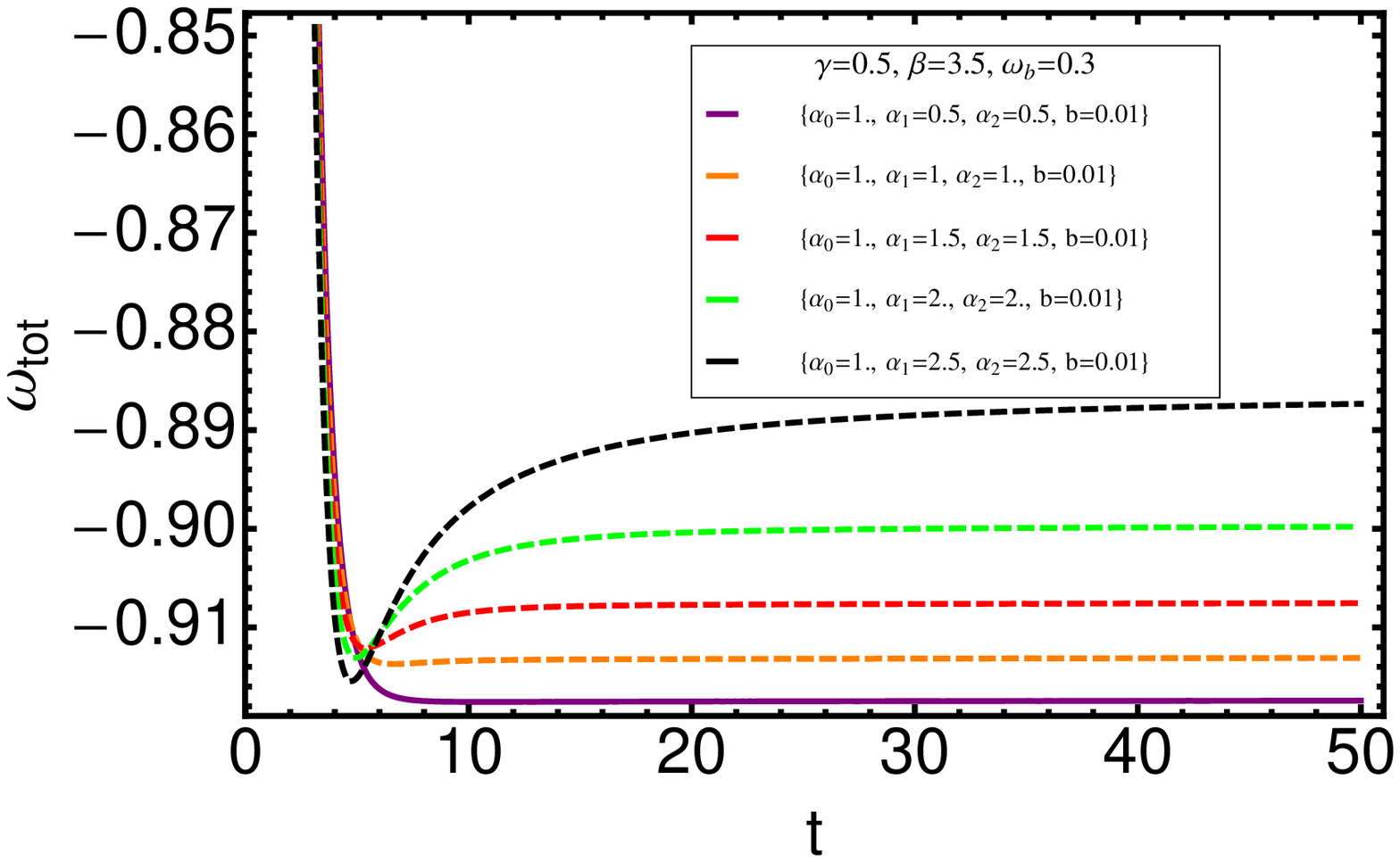} &
\includegraphics[width=50 mm]{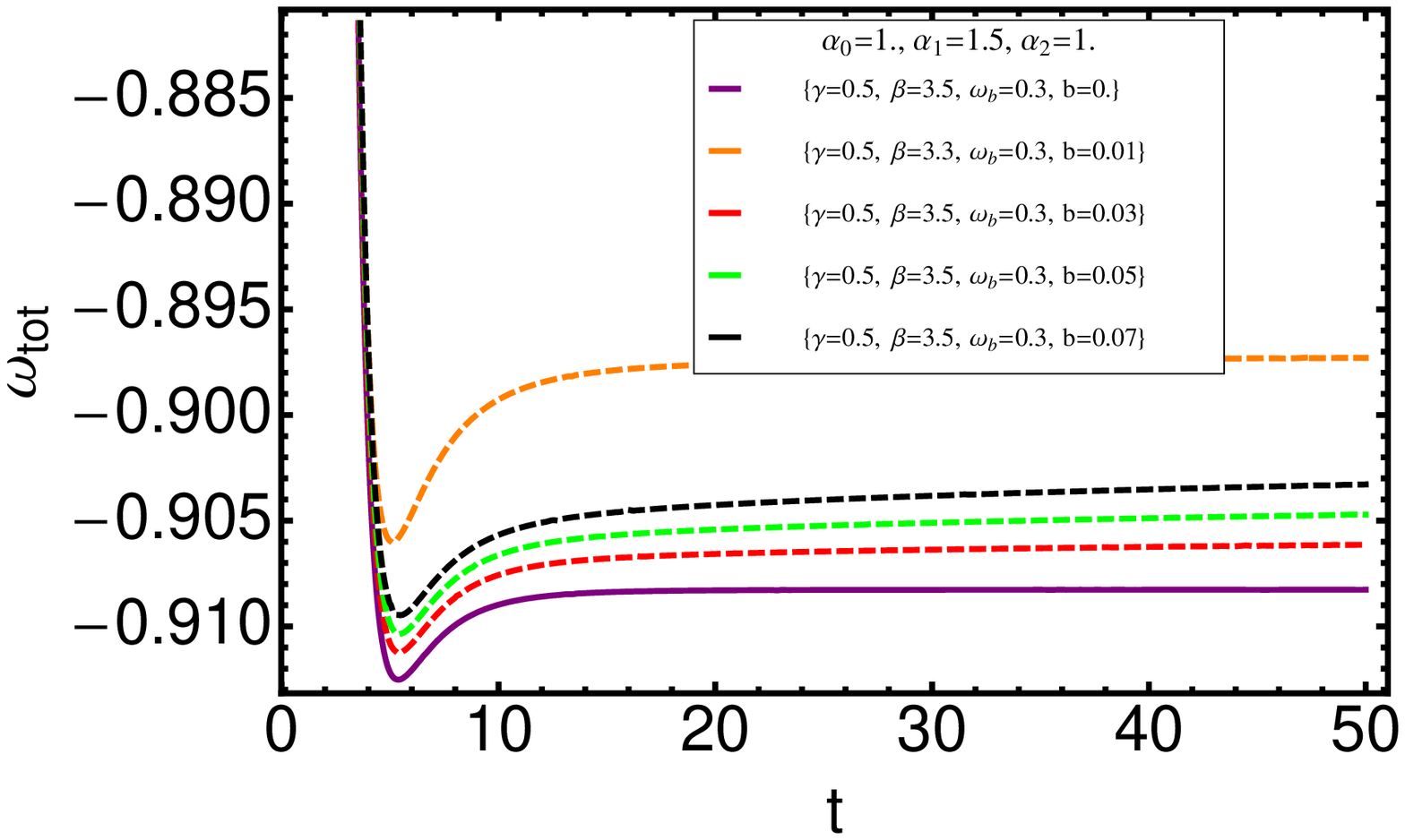}\\
\includegraphics[width=50 mm]{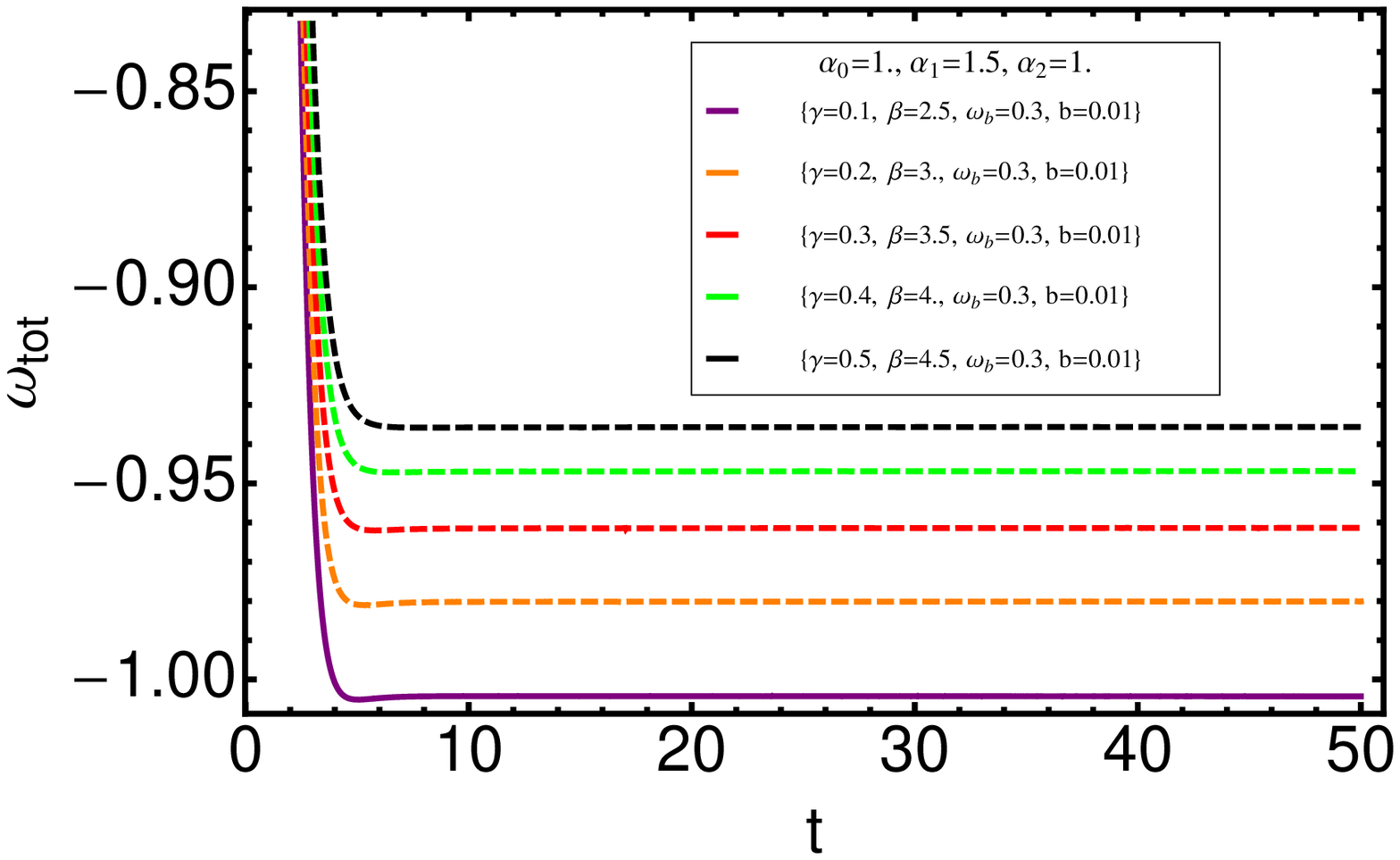} &
\includegraphics[width=50 mm]{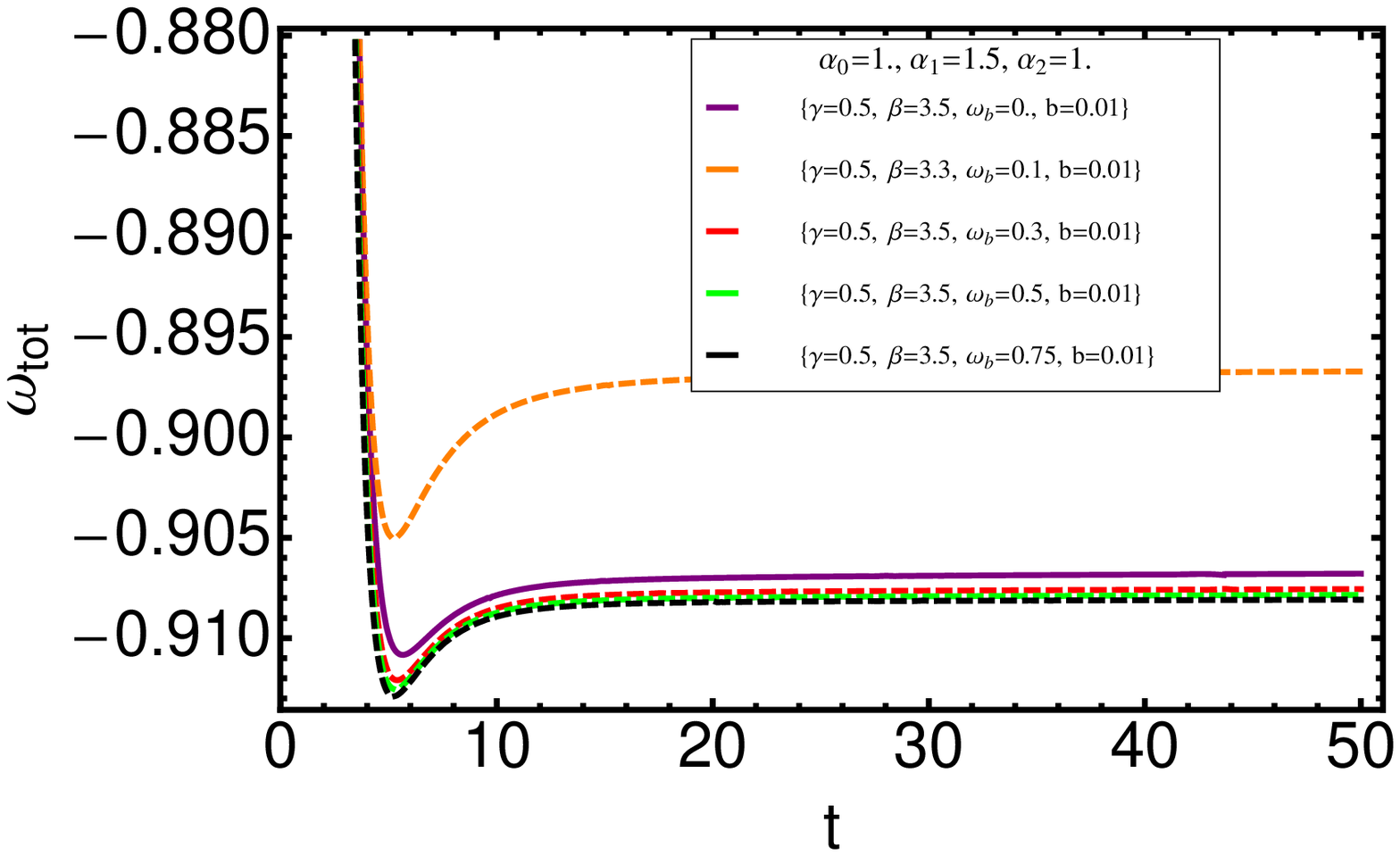}
 \end{array}$
 \end{center}
\caption{Behavior of EoS parameter $\omega_{tot}$ against $t$ for Model 3.}
 \label{fig:8}
\end{figure}

\begin{figure}[h!]
 \begin{center}$
 \begin{array}{cccc}
\includegraphics[width=50 mm]{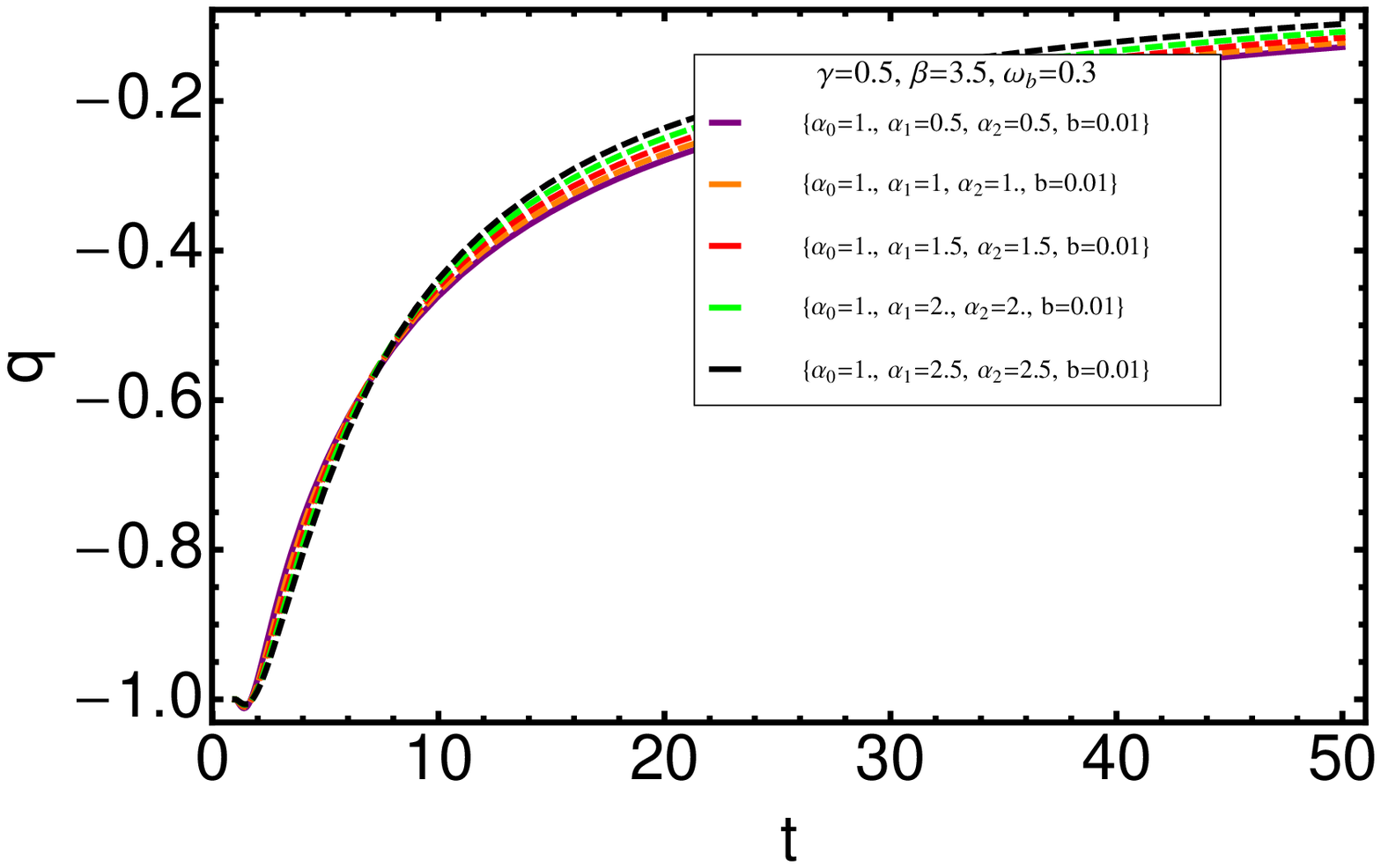} &
\includegraphics[width=50 mm]{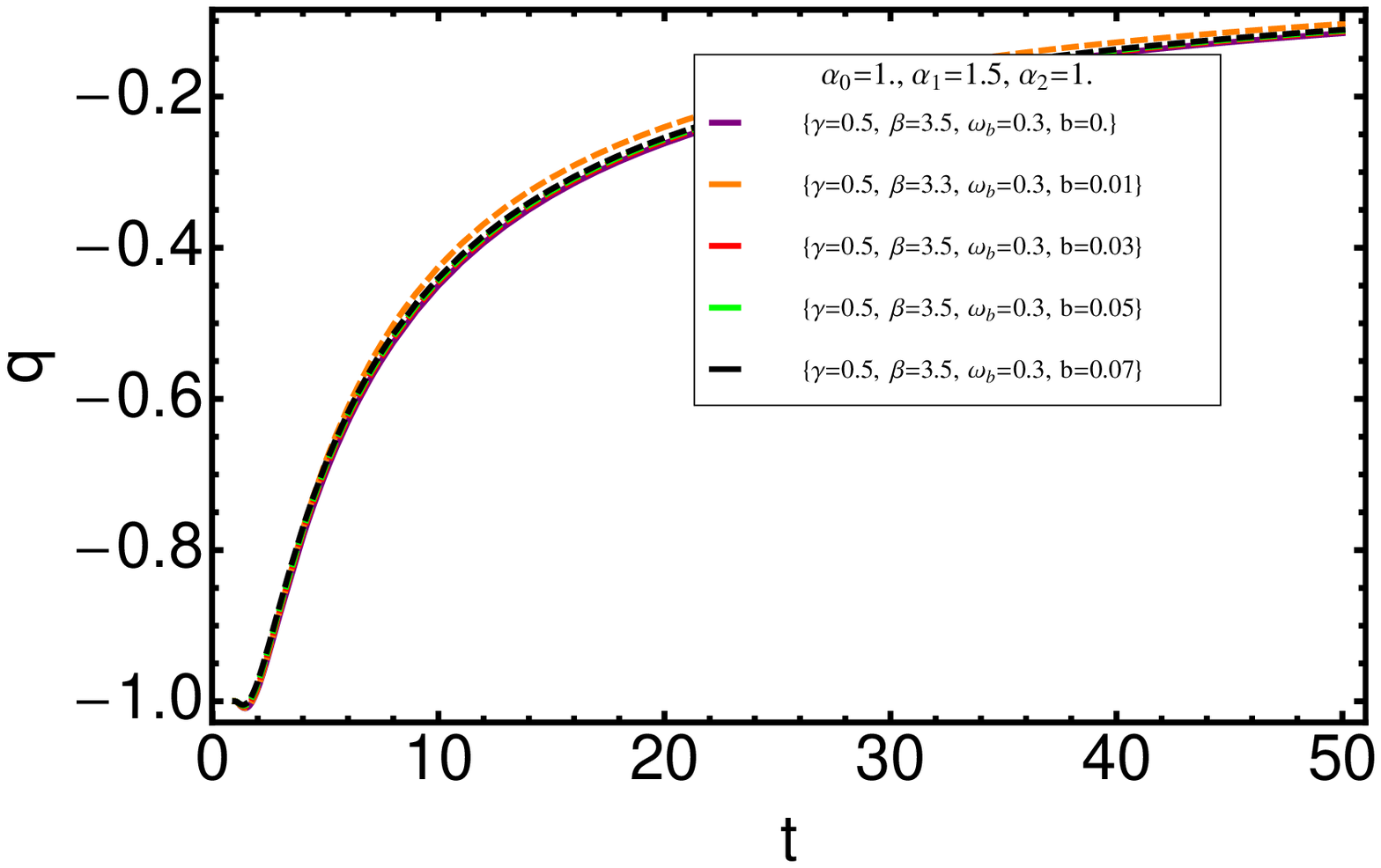}\\
\includegraphics[width=50 mm]{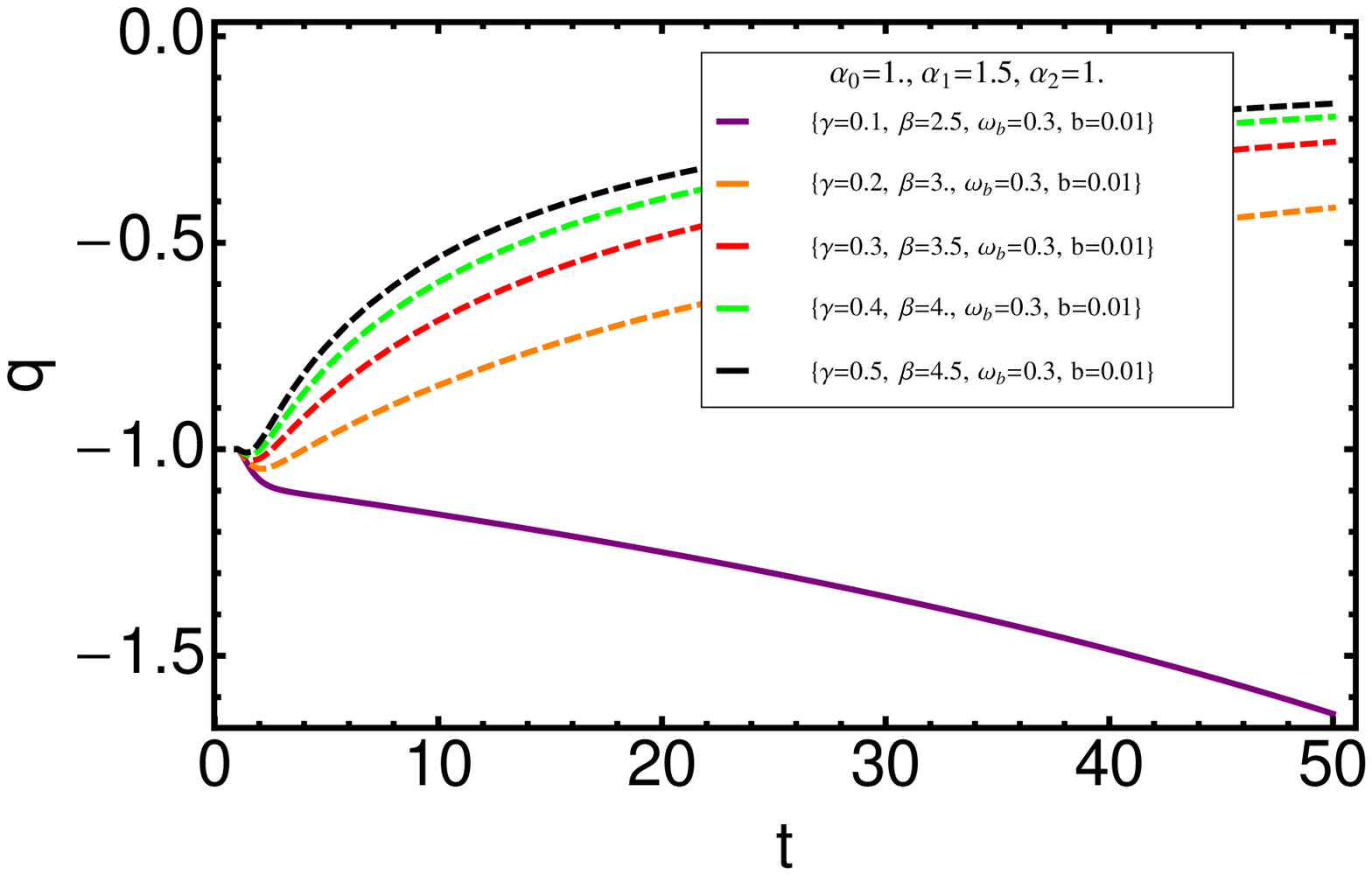} &
\includegraphics[width=50 mm]{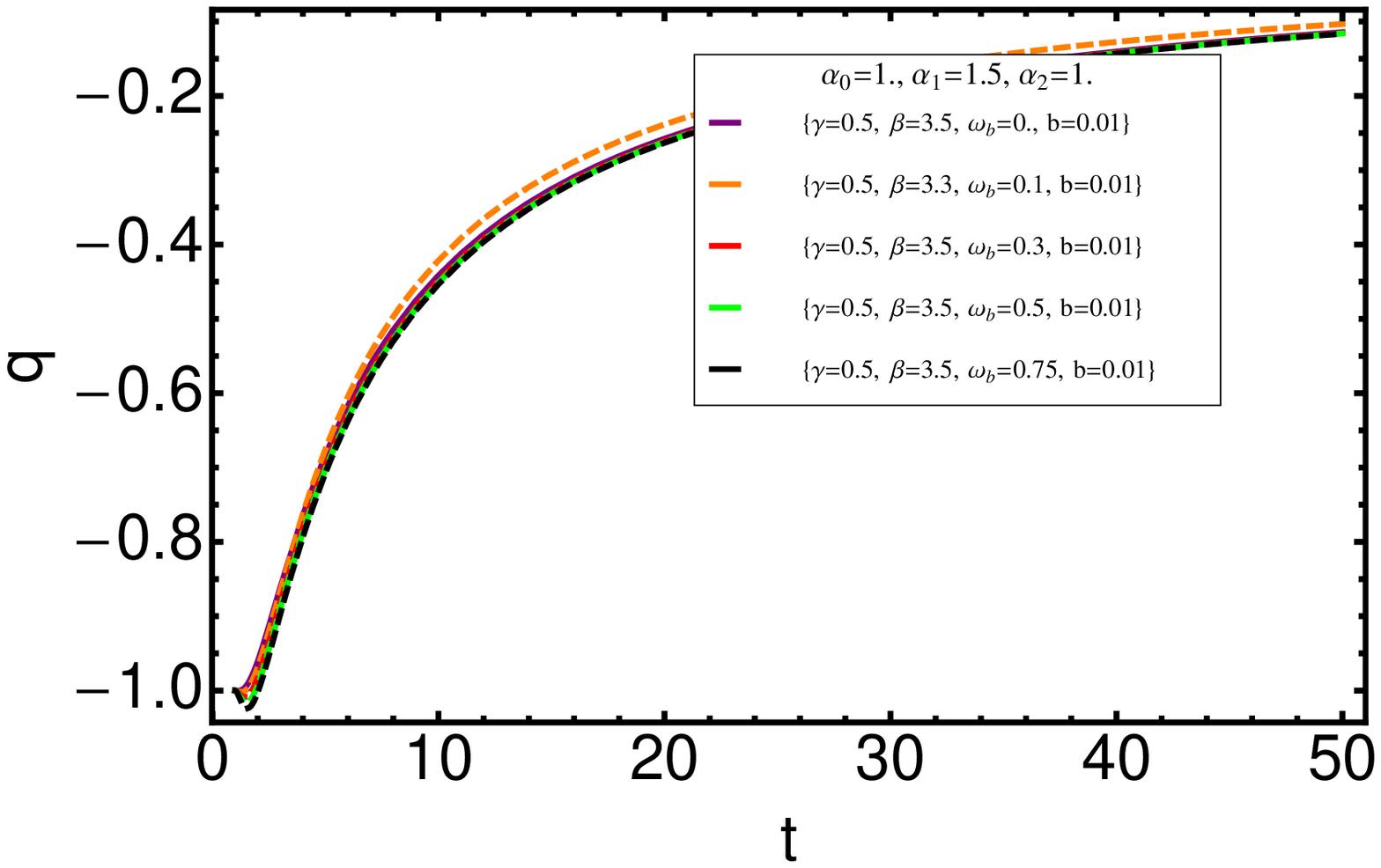}
 \end{array}$
 \end{center}
\caption{Behavior of deceleration parameter $q$ against $t$ for Model 3.}
 \label{fig:9}
\end{figure}

For this model we will consider the following phenomenological form of the $\Lambda(t)$
\begin{equation}\label{eq:Lambda2}
\Lambda(t)=\rho_{b}\ln{(tH)}+\rho_{D}\sin{\left (t\frac{\dot{G}(t)}{G(t)}\right) }.
\end{equation}
Taking into account (\ref{eq: Fridmman vlambda}) we can write $\Lambda(t)$ in a different form
\begin{equation}\label{eq:Lambda2new}
\Lambda(t)=\left [ 1+\frac{\ln(tH)}{8 \pi G(t)} \right ]^{-1} \left( \frac{3H^{2}}{8 \pi G(t)}\ln{(tH)}-\rho_{D}(\ln{(tH)}-\sin{\left (t\frac{\dot{G}(t)}{G(t)}\right) })\right ).
\end{equation}
\begin{equation}\label{eq:G2}
\frac{\dot{G}(t)}{G(t)}+\frac{\dot{\Lambda}(t)}{3H^{2}-\Lambda(t)}=0.
\end{equation}
Equation (\ref{eq:G2}) with (\ref{eq:Lambda2new}) will give us the
behavior of $G(t)$. This model also includes several interesting
facts about the behavior of the cosmological parameters. After
recovering the $G(t)$ we observe that $G(t)$ is an increasing
function, and its graphical behavior for the different cases are
given in Fig.(\ref{fig:7}). For instance with increasing $\beta$ and
$\gamma$ with $\alpha_{0}=\alpha_{2}=1$, $\alpha_{1}=1.5$,
$\omega_{b}=0.3$ and $b=0.01$ we have the following picture:
$\gamma=0.1$ and $\beta=2.5$ (a blue line at left-bottom plot) we
have a decreasing behavior for $G(t)$, while for the higher values
for $\gamma$ and $\beta$ we have increasing behavior for later
stages of evolution. With increasing $\omega_{b}$ we decrease the
value of $G(t)$ (right-bottom). We also observe that there is a
period in history of the evolution where $G(t)$ can be a constant.
With $\alpha_{0}=\alpha_{2}=1$, $\alpha_{1}=1.5$, $\gamma=0.5$,
$\beta=3.5$ and $\omega_{b}=0.3$ we see that for non interacting
case, when $b=0$ (a blue line at right-top plot) at later stages of
evolution $G(t)=const \approx 1.36$, while when we include the
interaction and increase the value of $b$, increase in the value of
$G(t)$ is observed. Behavior of $G(t)$ from $\alpha_{0}$,
$\alpha_{1}$ and $\alpha_{2}$ can be found at the left-top plot of
Fig.(\ref{fig:7}). Other cosmological parameter that we have
investigated for this model is a $\omega_{tot}$ describing
interacting DE and DM two component fluid model. From
Fig.(\ref{fig:8}) we can make conclusion about the behavior of the
parameter. We observe that as a function of $\alpha_{0}$,
$\alpha_{1}$ and $\alpha_{2}$, while the other parameters are being
fixed, we have a decreasing function for the initial stages of
evolution, while for the later stages we have a constant value for
$\omega_{tot}$. With increasing $\alpha_{1}$ and $\alpha_{2}$ we
will increase $\omega_{tot}$ and we have a possibility to obtain
decreasing-increasing-constant behavior (left-top plot). On the
right-top plot we see the role of the interaction $Q$. Starting with
the non interacting case $b=0$ and increasing $b$ we observe the
increasing value of $\omega_{tot}$. Bottom panel of
Fig.{\ref{fig:8}} represents graphical behavior of $\omega_{tot}$
from $ \{ \gamma, \beta \} $ and $\omega_{b}$. The last parameter
discussed in this section will be the deceleration parameter $q$
recovered for this specific $\Lambda(t)$. Investigating the behavior
we conclude that for this model, $\gamma > 0.1$ and $\beta>2.5$
should be taken in order to get $q>-1$ ( Fig.(\ref{fig:9})
left-bottom plot). It starts its evolution from $-1$ and then it is
strictly $q>-1$ for later stages of evolution. Interaction as well
as $\omega_{b}$ has a small impact on the behavior of $q$. Left-top
plot of Fig.{\ref{fig:9}} represents the behavior of $q$ as a
function of $\alpha_{0}$, $\alpha_{1}$ and $\alpha_{2}$. As for the
other models, additional information about other cosmological
parameters of this model can be found in Appendix.

\begin{figure}[h!]
 \begin{center}$
 \begin{array}{cccc}
\includegraphics[width=50 mm]{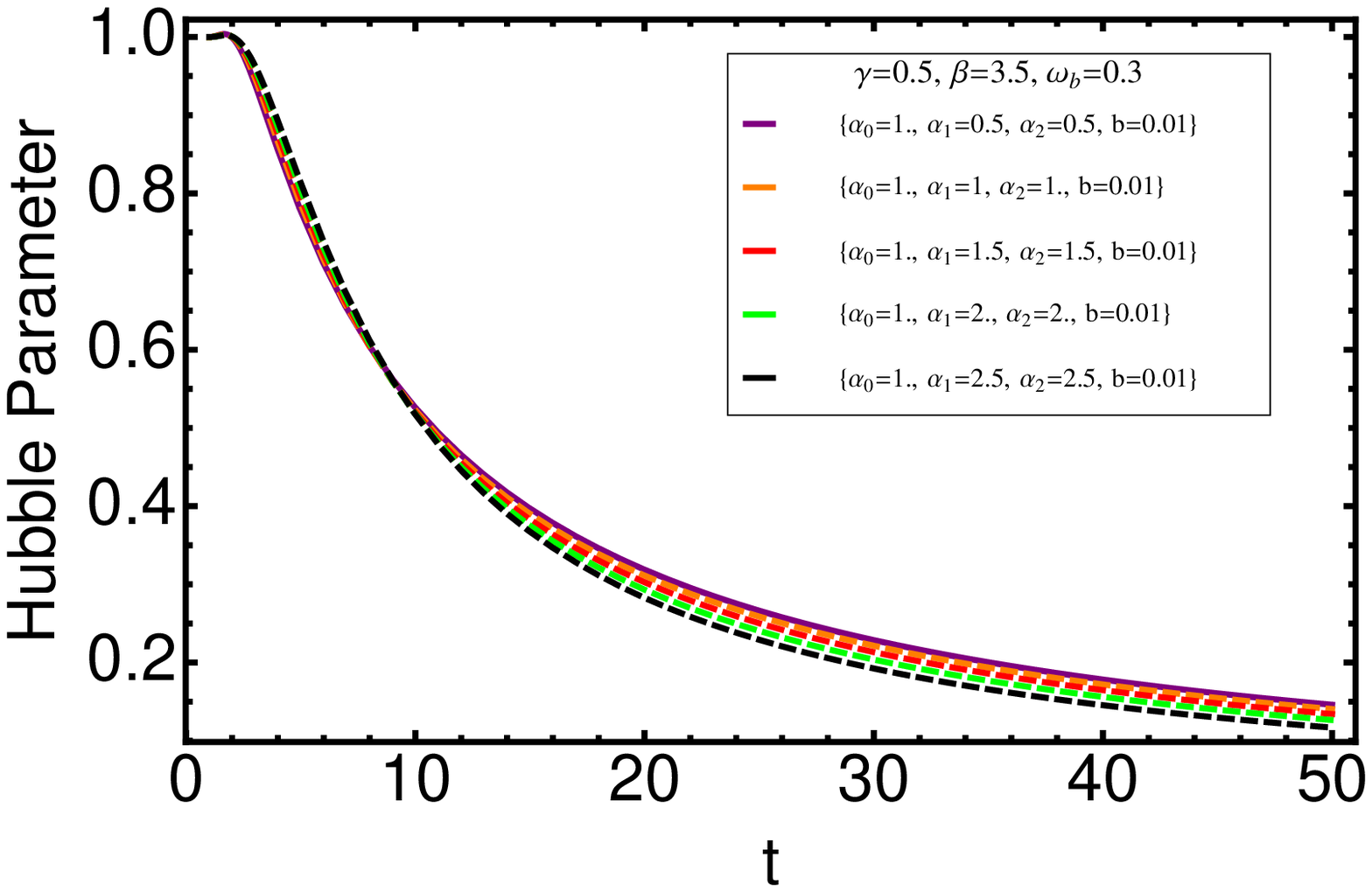} &
\includegraphics[width=50 mm]{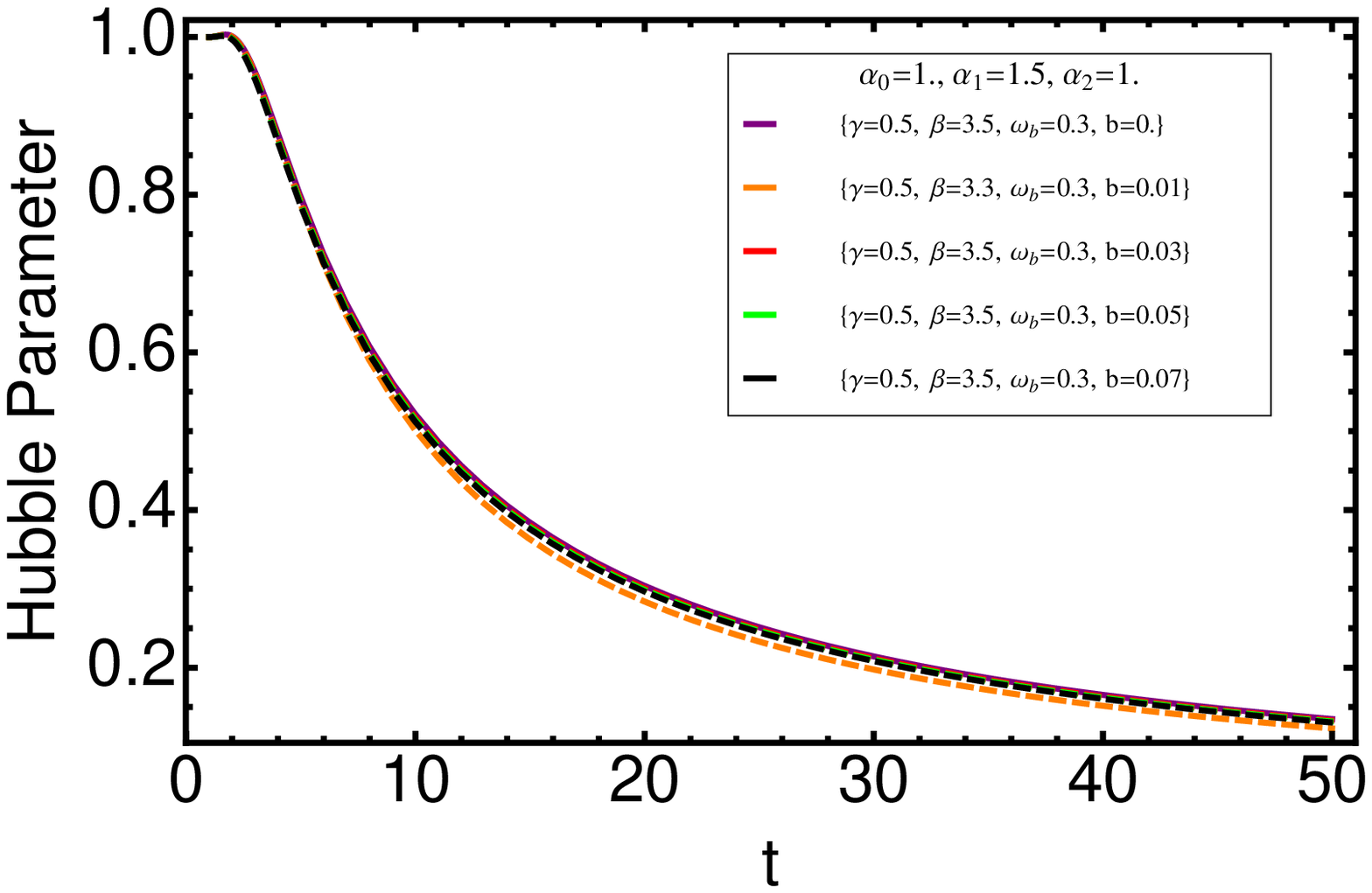}\\
\includegraphics[width=50 mm]{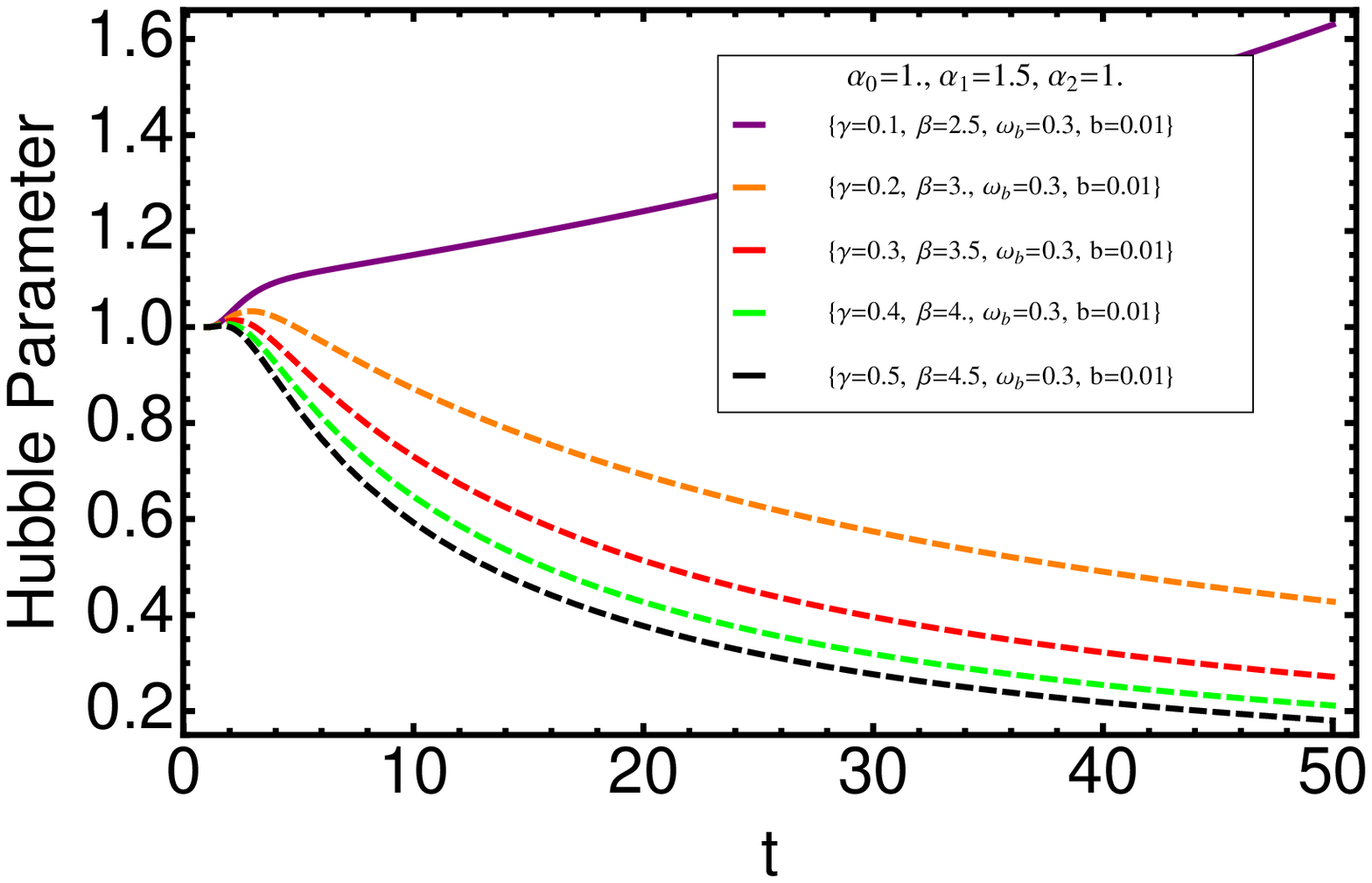} &
\includegraphics[width=50 mm]{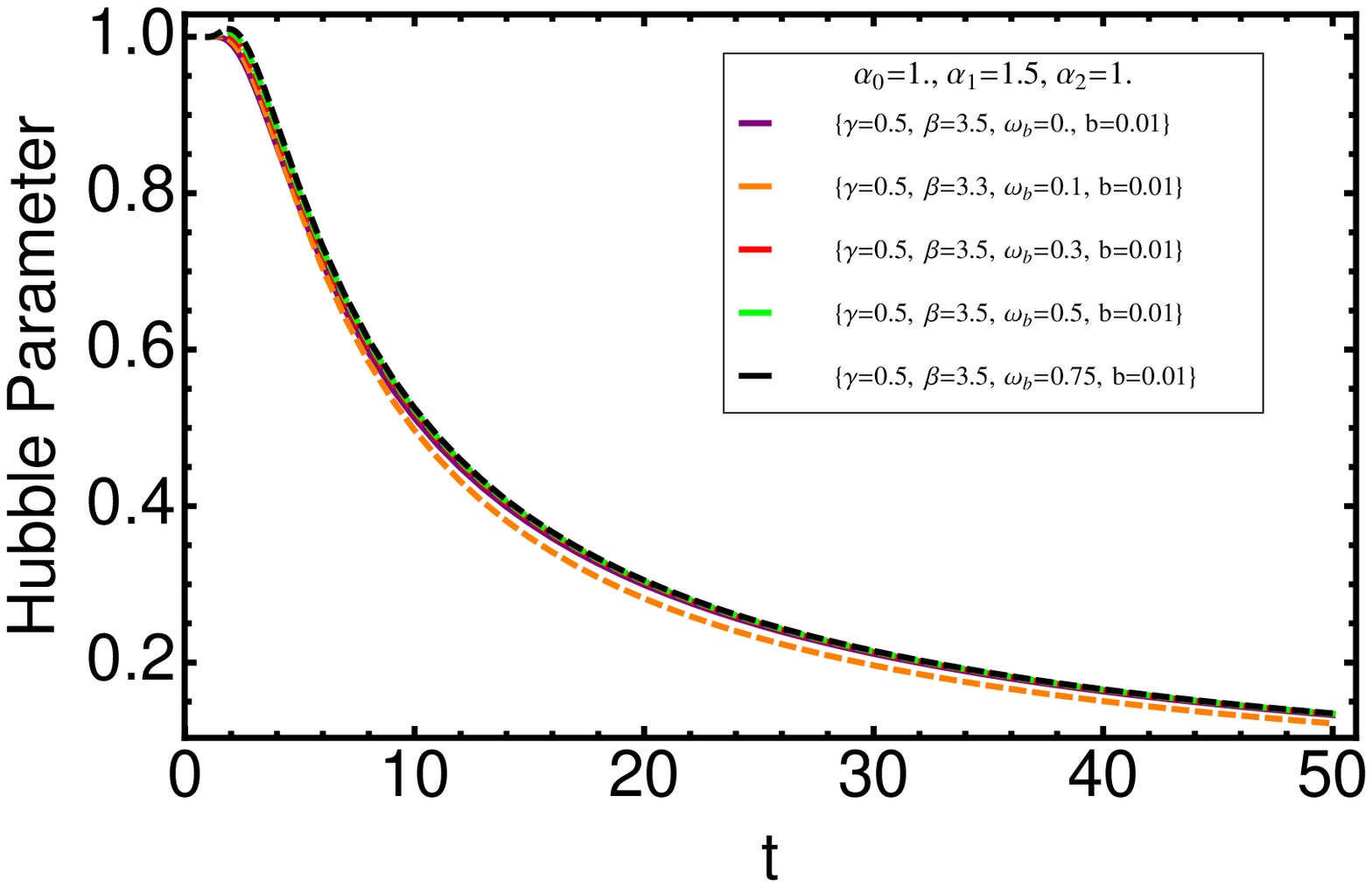}
 \end{array}$
 \end{center}
\caption{Behavior of Hubble parameter $H(t)$ against $t$ for Model 3.}
 \label{fig:18}
\end{figure}

\begin{figure}[h!]
 \begin{center}$
 \begin{array}{cccc}
\includegraphics[width=50 mm]{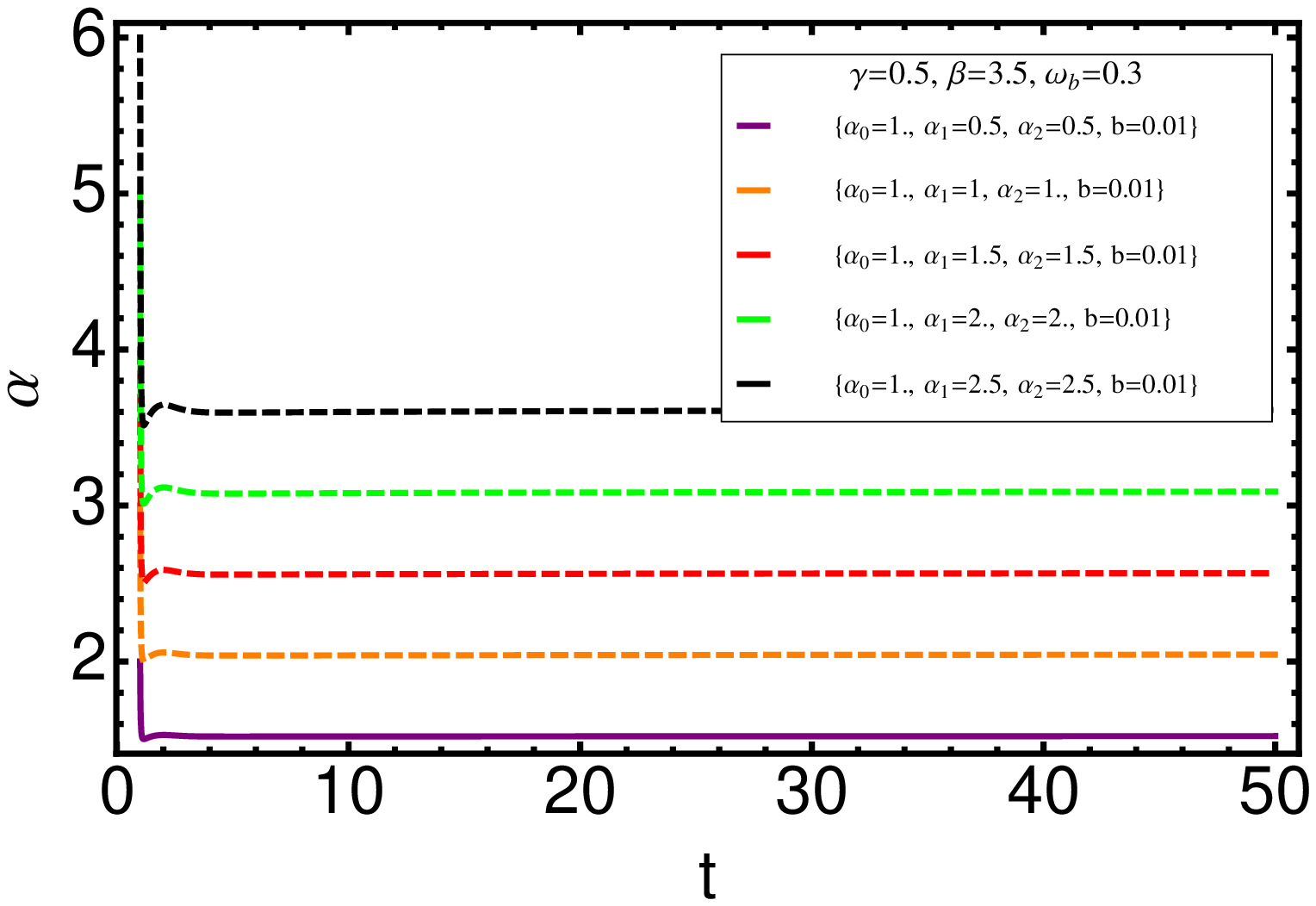} &
\includegraphics[width=50 mm]{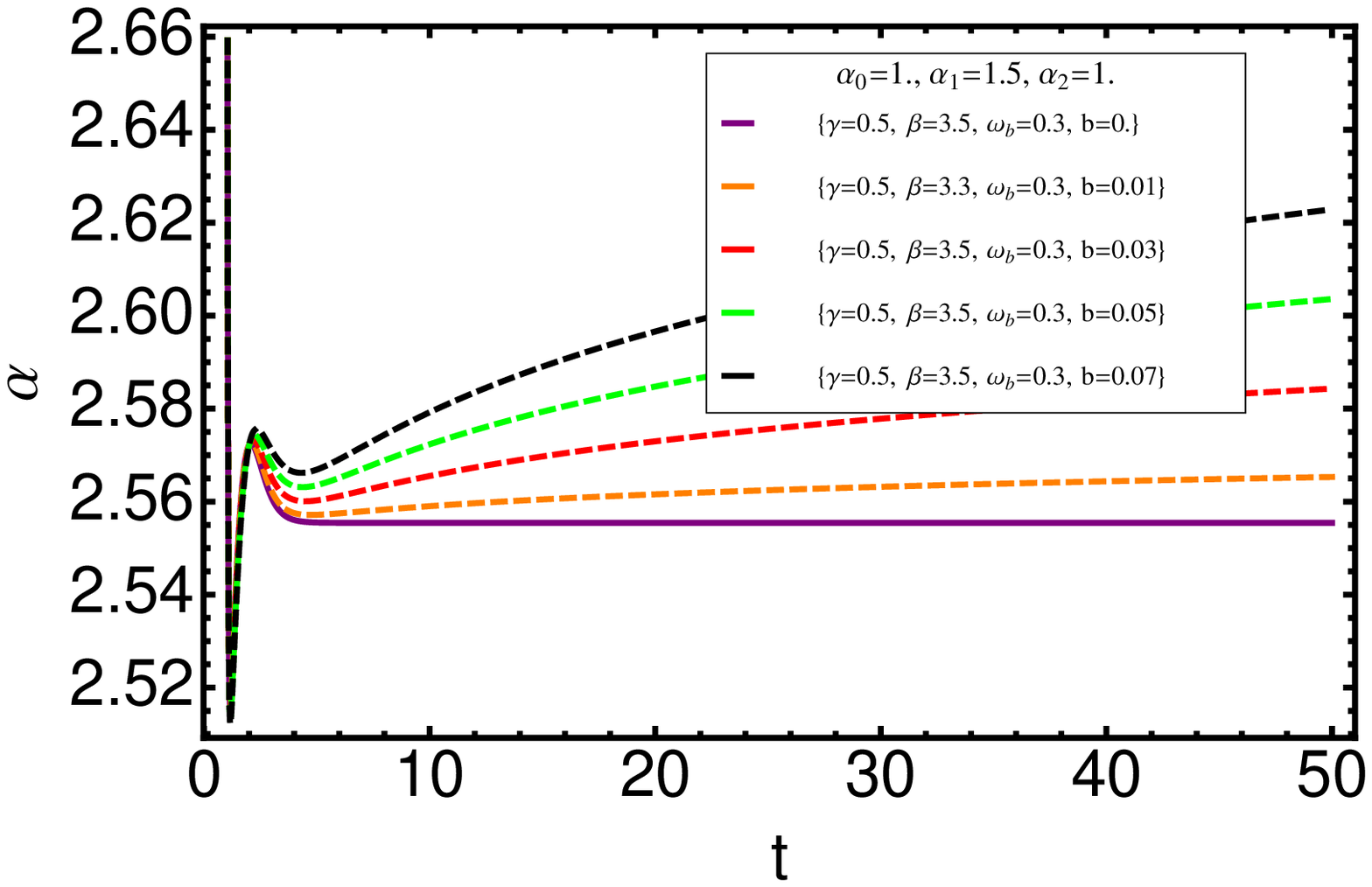}\\
\includegraphics[width=50 mm]{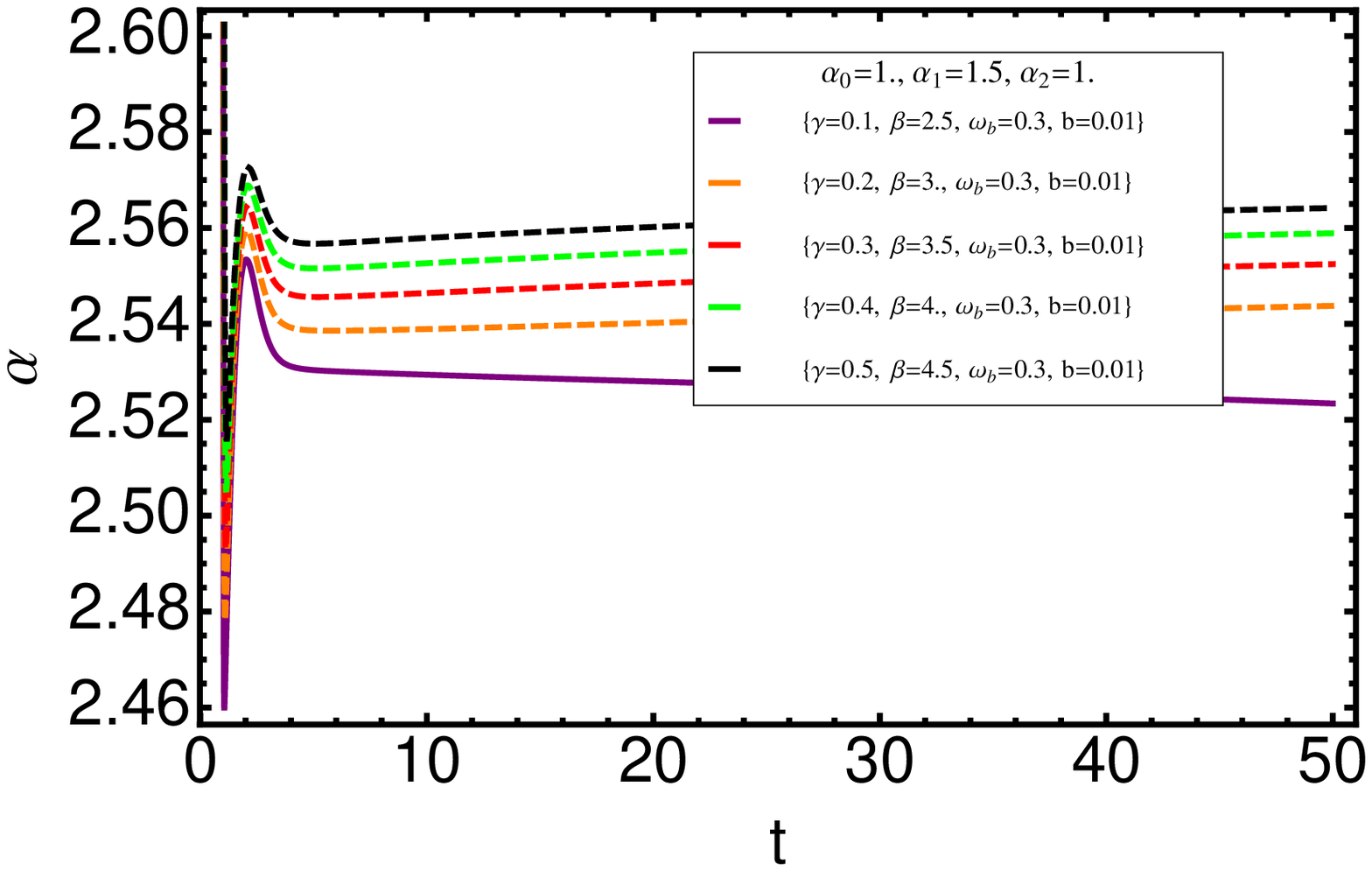} &
\includegraphics[width=50 mm]{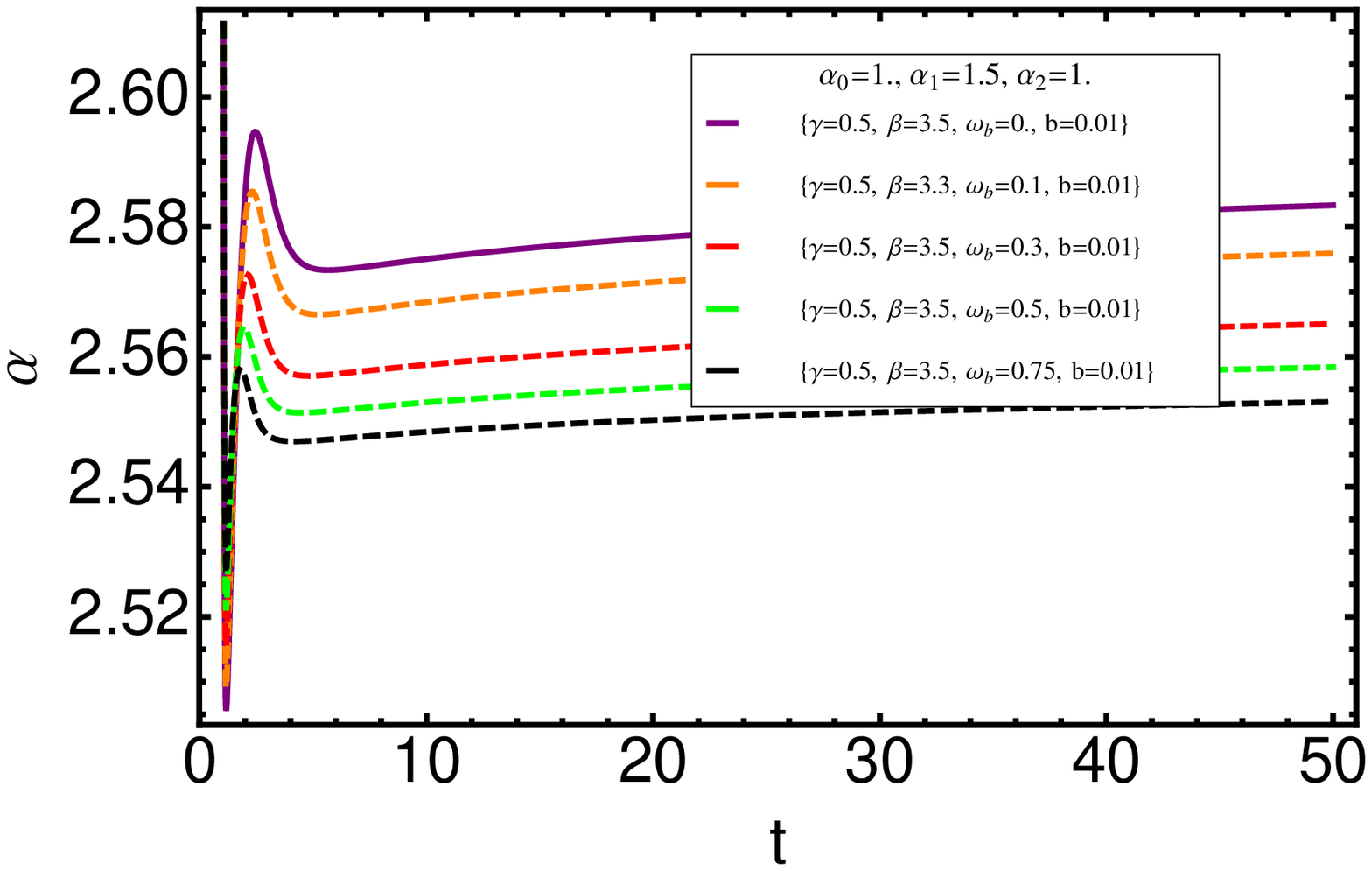}
 \end{array}$
 \end{center}
\caption{Behavior of $\alpha$ against $t$ for Model 3.}
 \label{fig:19}
\end{figure}
\begin{figure}[h!]
 \begin{center}$
 \begin{array}{cccc}
\includegraphics[width=50 mm]{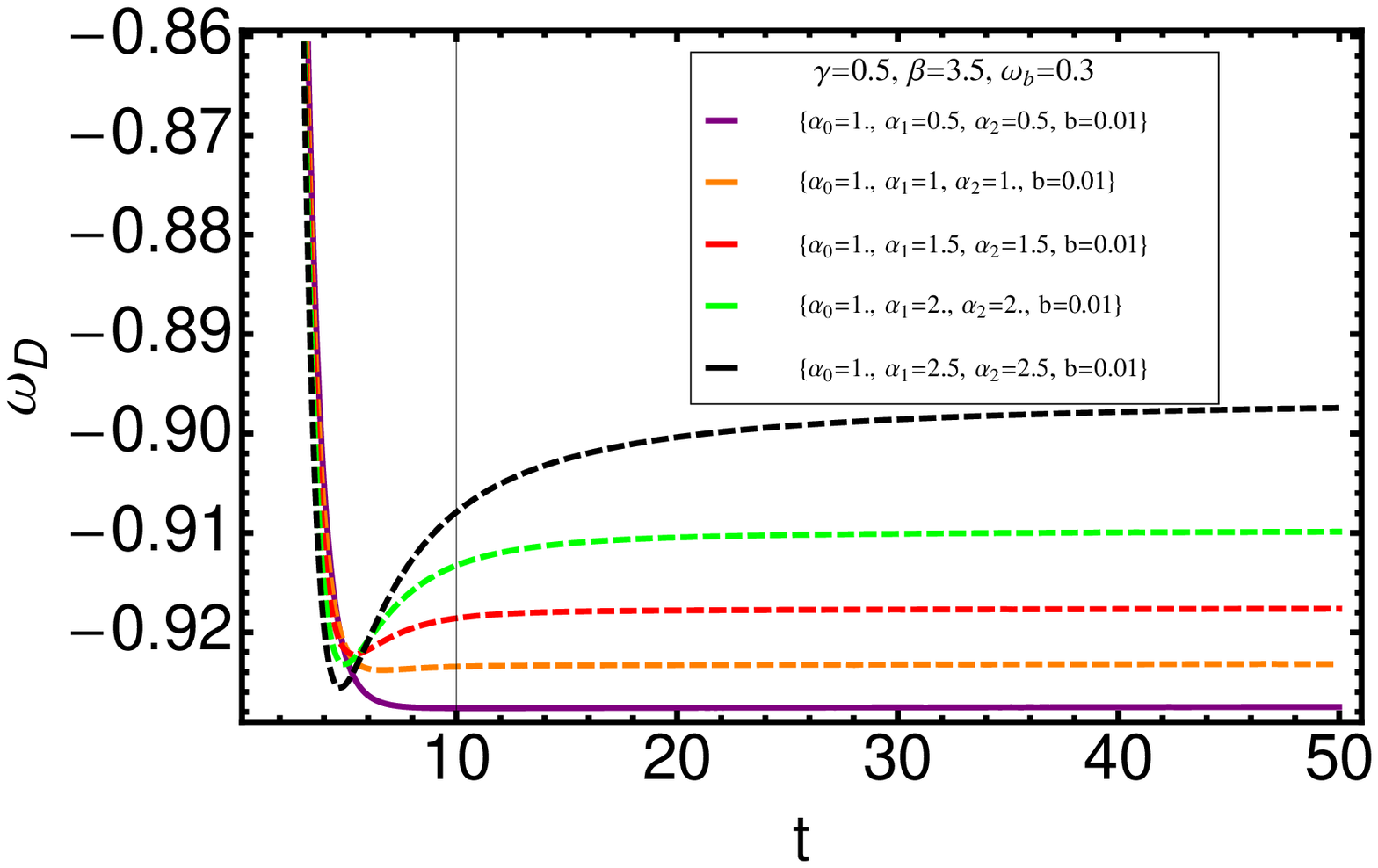} &
\includegraphics[width=50 mm]{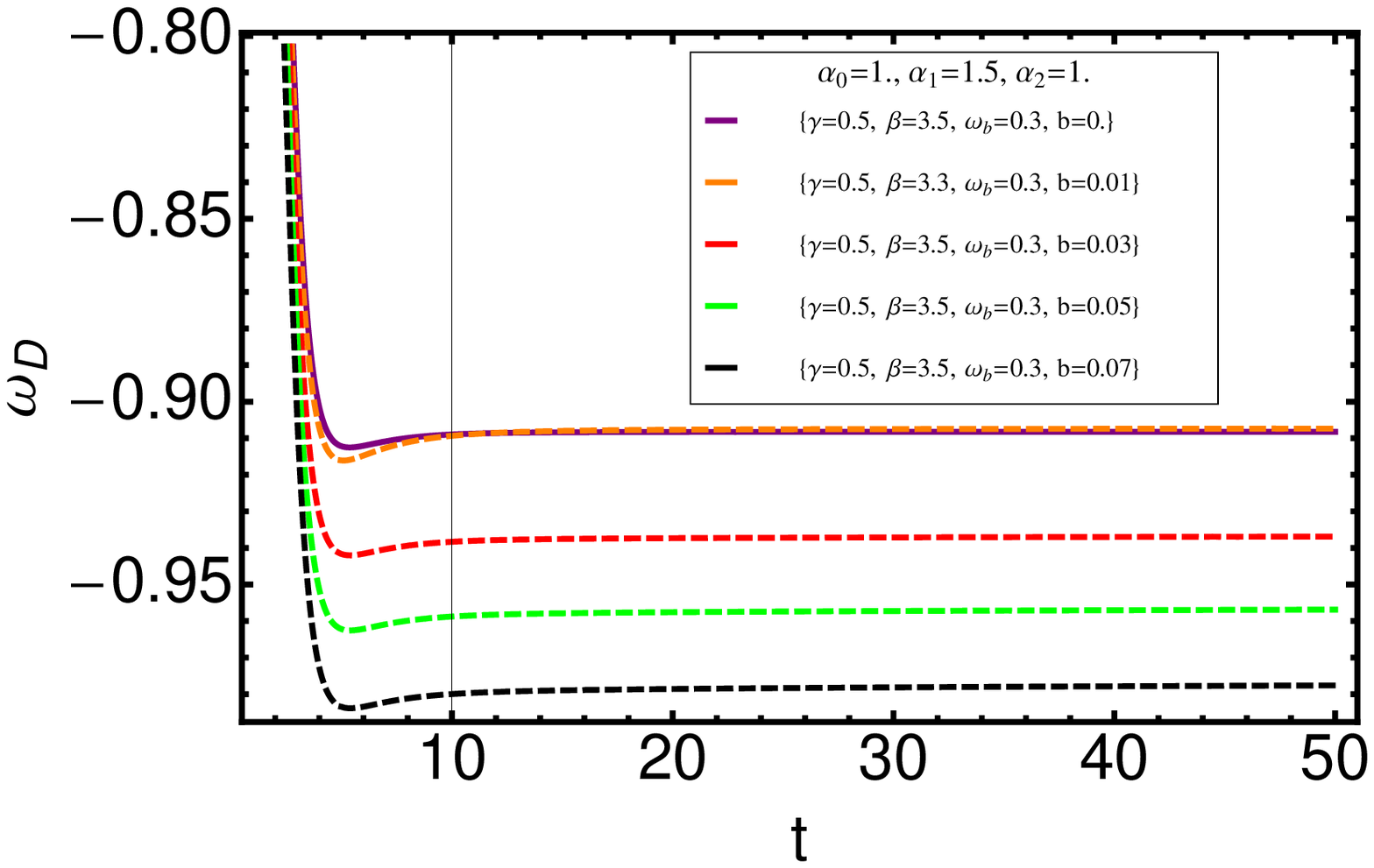}\\
\includegraphics[width=50 mm]{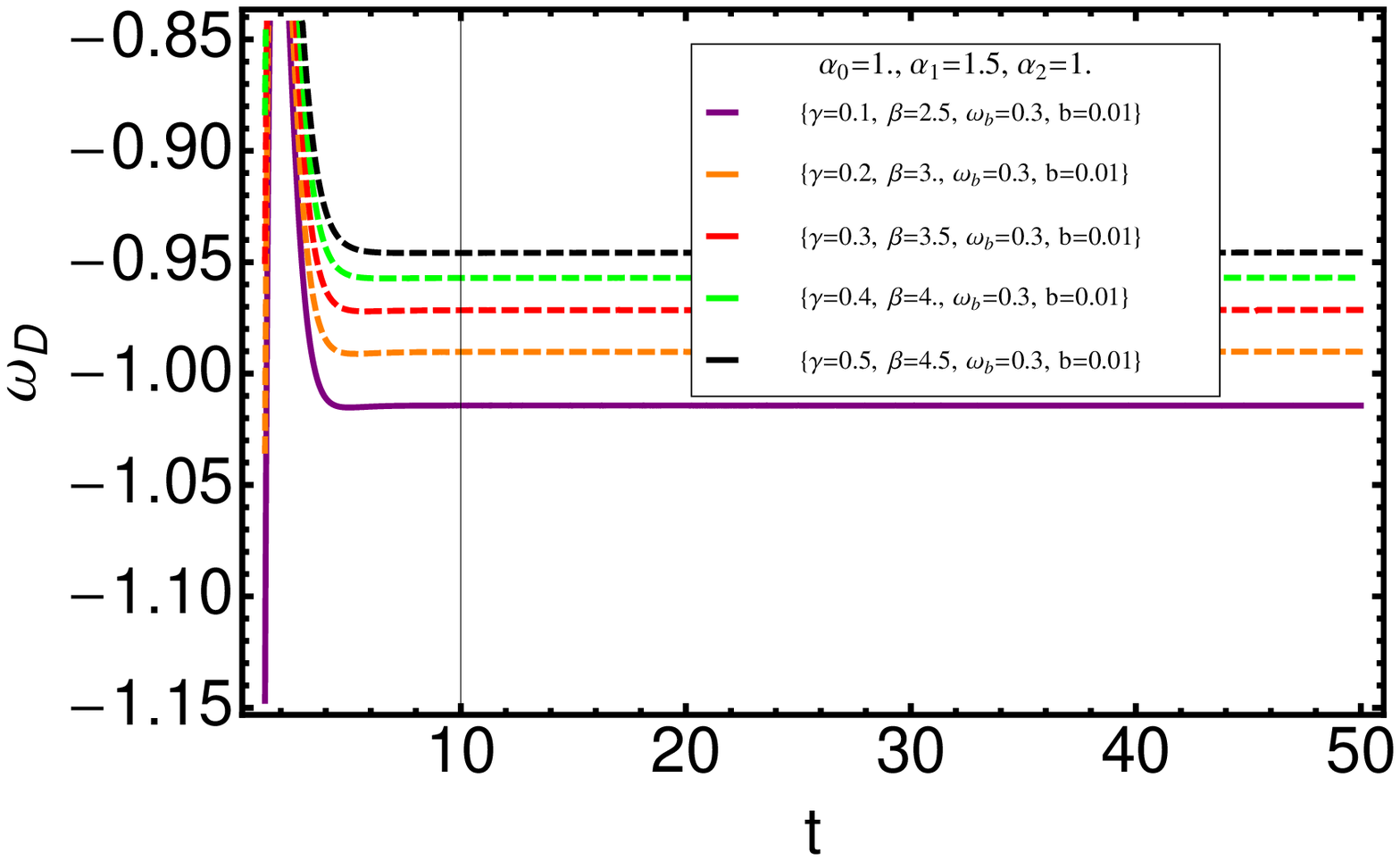} &
\includegraphics[width=50 mm]{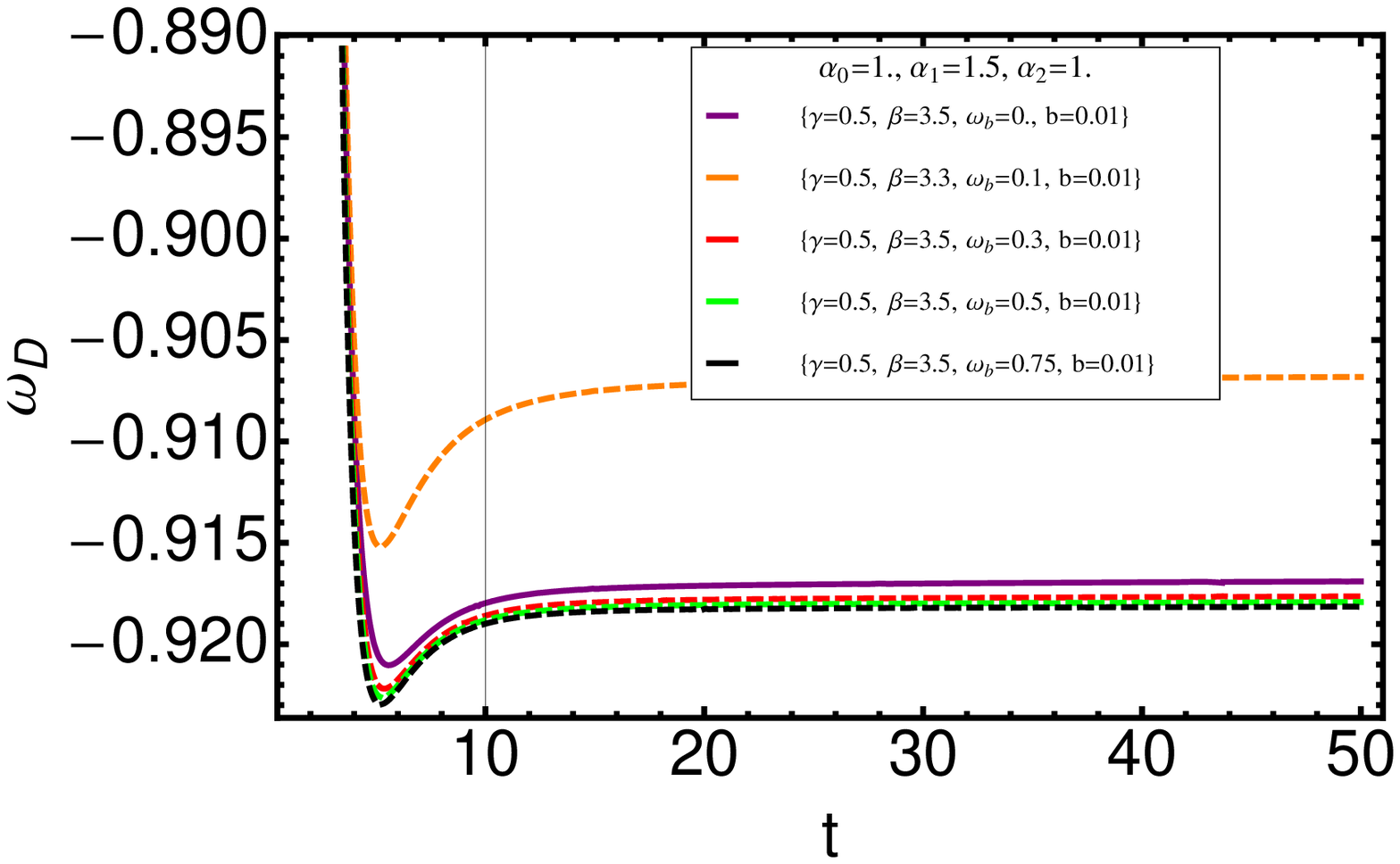}
 \end{array}$
 \end{center}
\caption{Behavior of $\omega_{D}$ against $t$ for Model 3.}
 \label{fig:20}
\end{figure}
\begin{figure}[h!]
 \begin{center}$
 \begin{array}{cccc}
\includegraphics[width=50 mm]{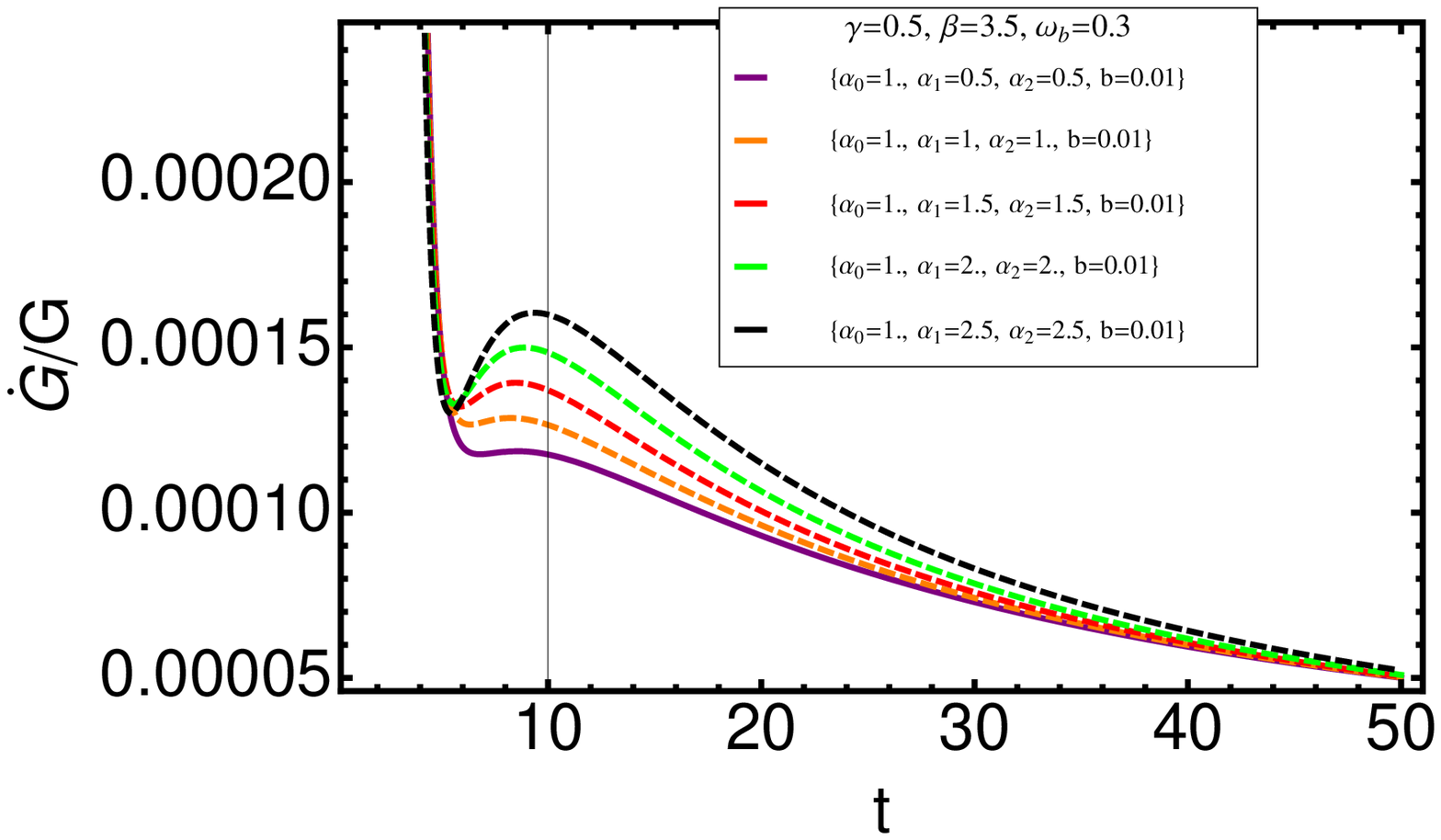} &
\includegraphics[width=50 mm]{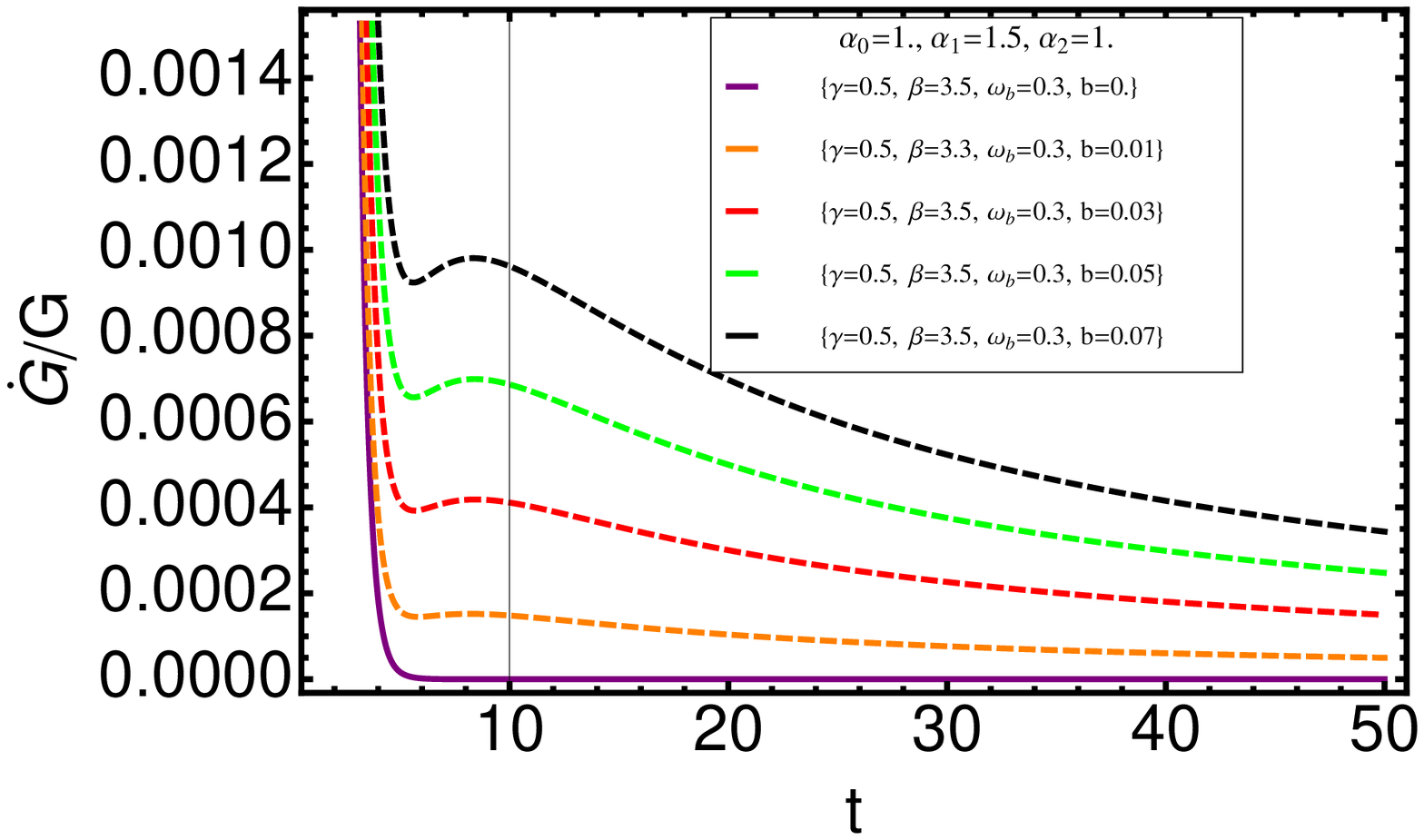}\\
\includegraphics[width=50 mm]{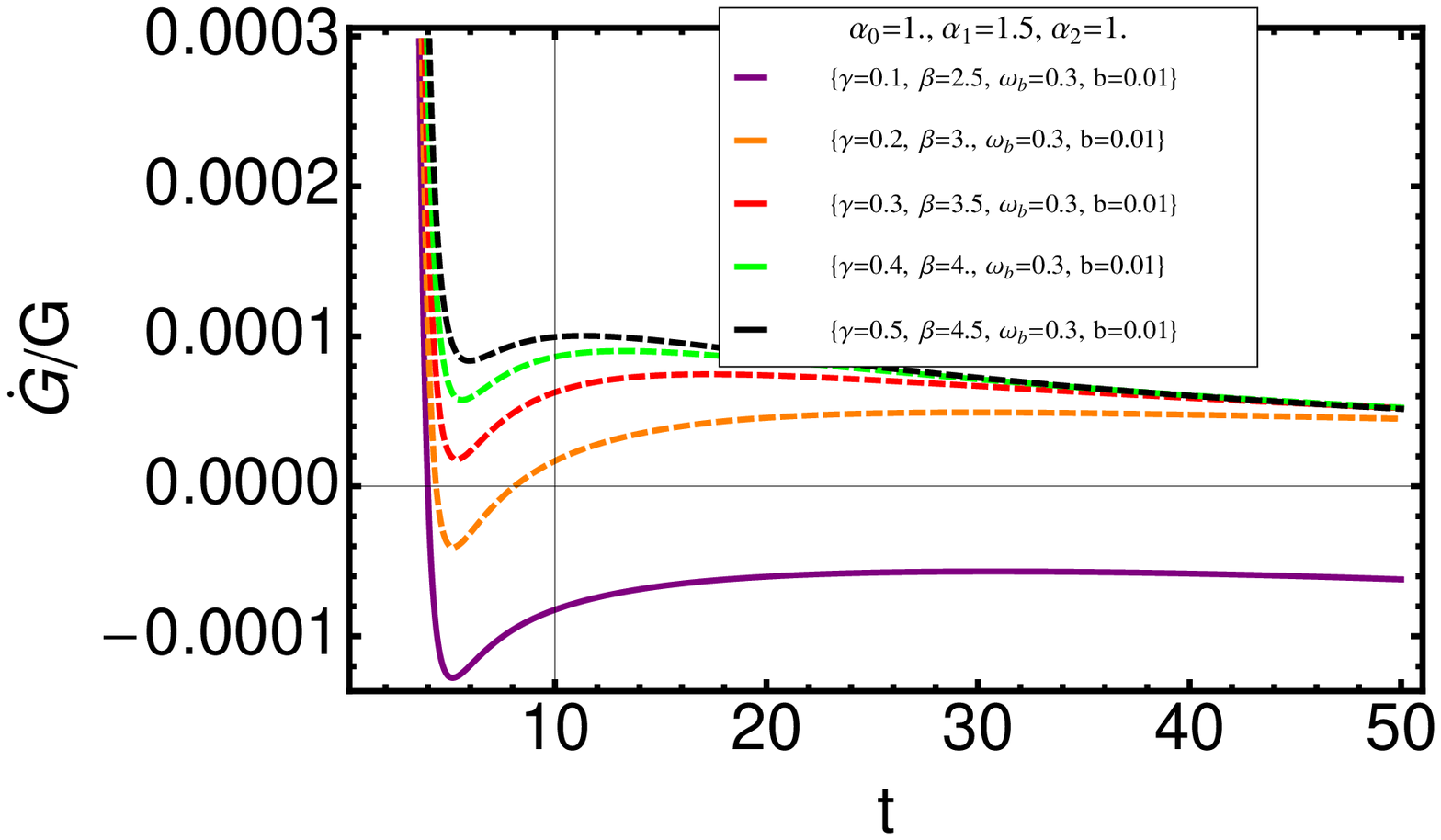} &
\includegraphics[width=50 mm]{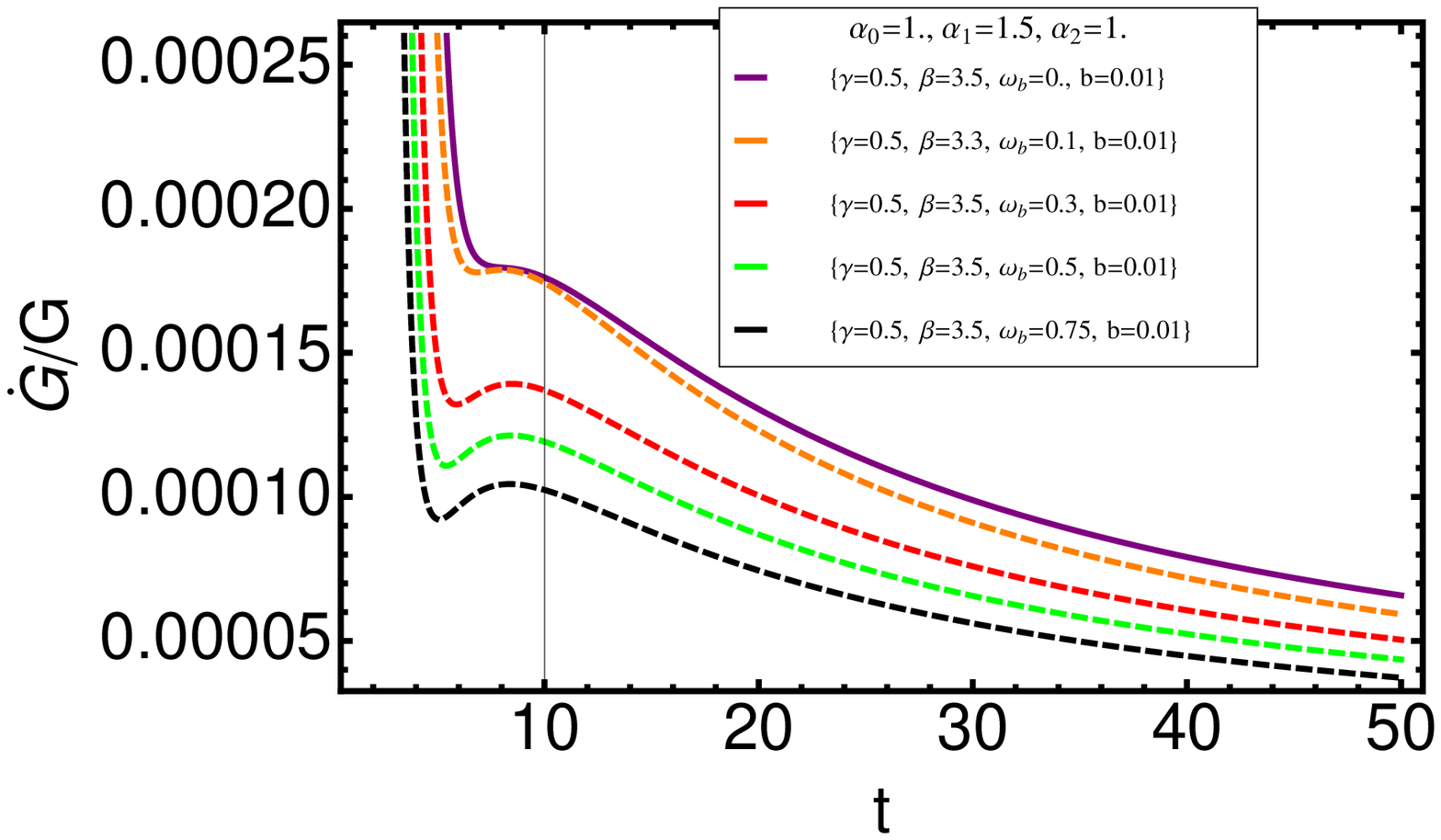}
 \end{array}$
 \end{center}
\caption{Behavior of $\dot{G}(t)/G(t)$ against $t$ for Model 3.}
 \label{fig:21}
\end{figure}

\section{State finder diagnostic}

\begin{figure}[h!]
 \begin{center}$
 \begin{array}{cccc}
\includegraphics[width=50 mm]{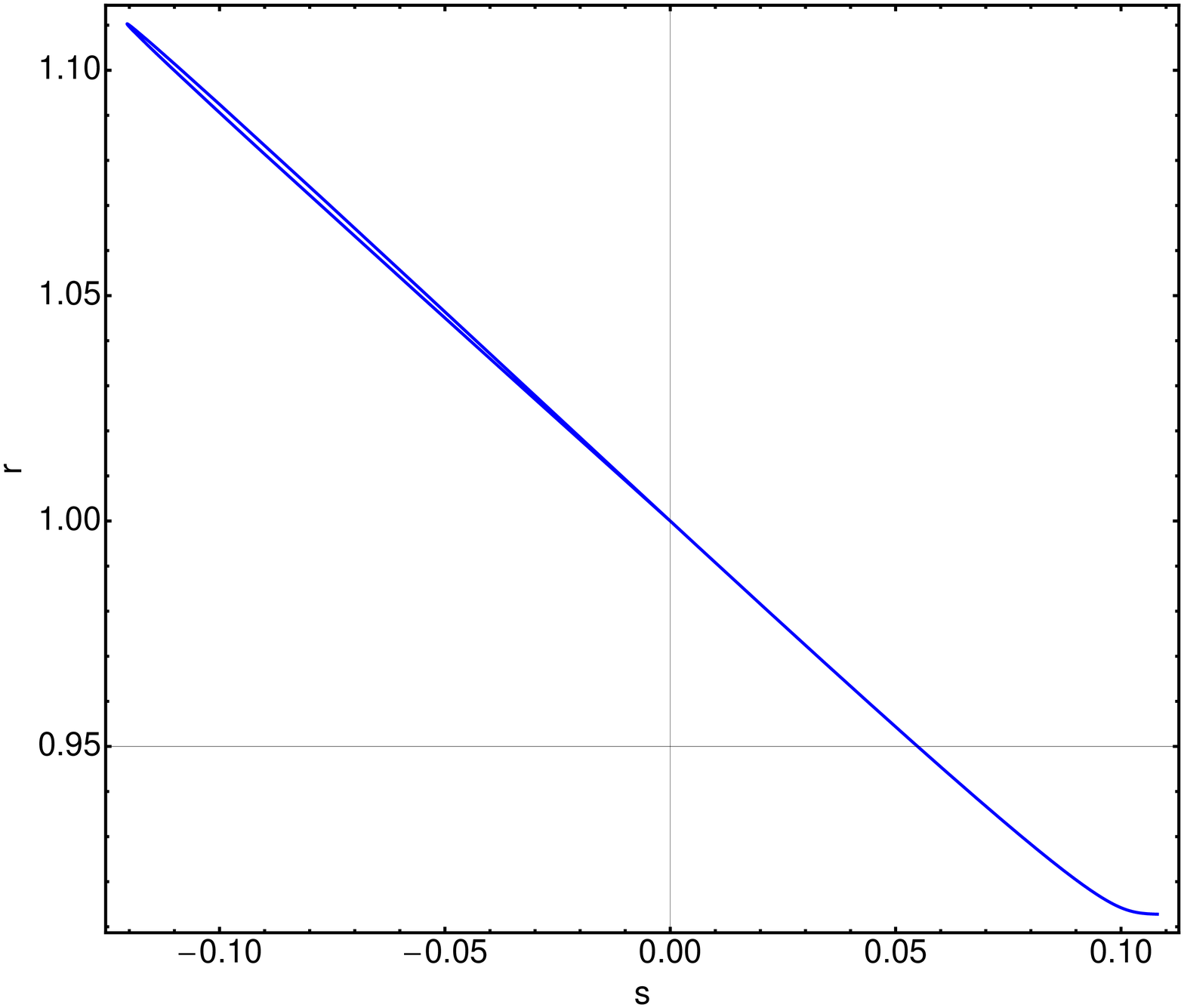} &
\includegraphics[width=50 mm]{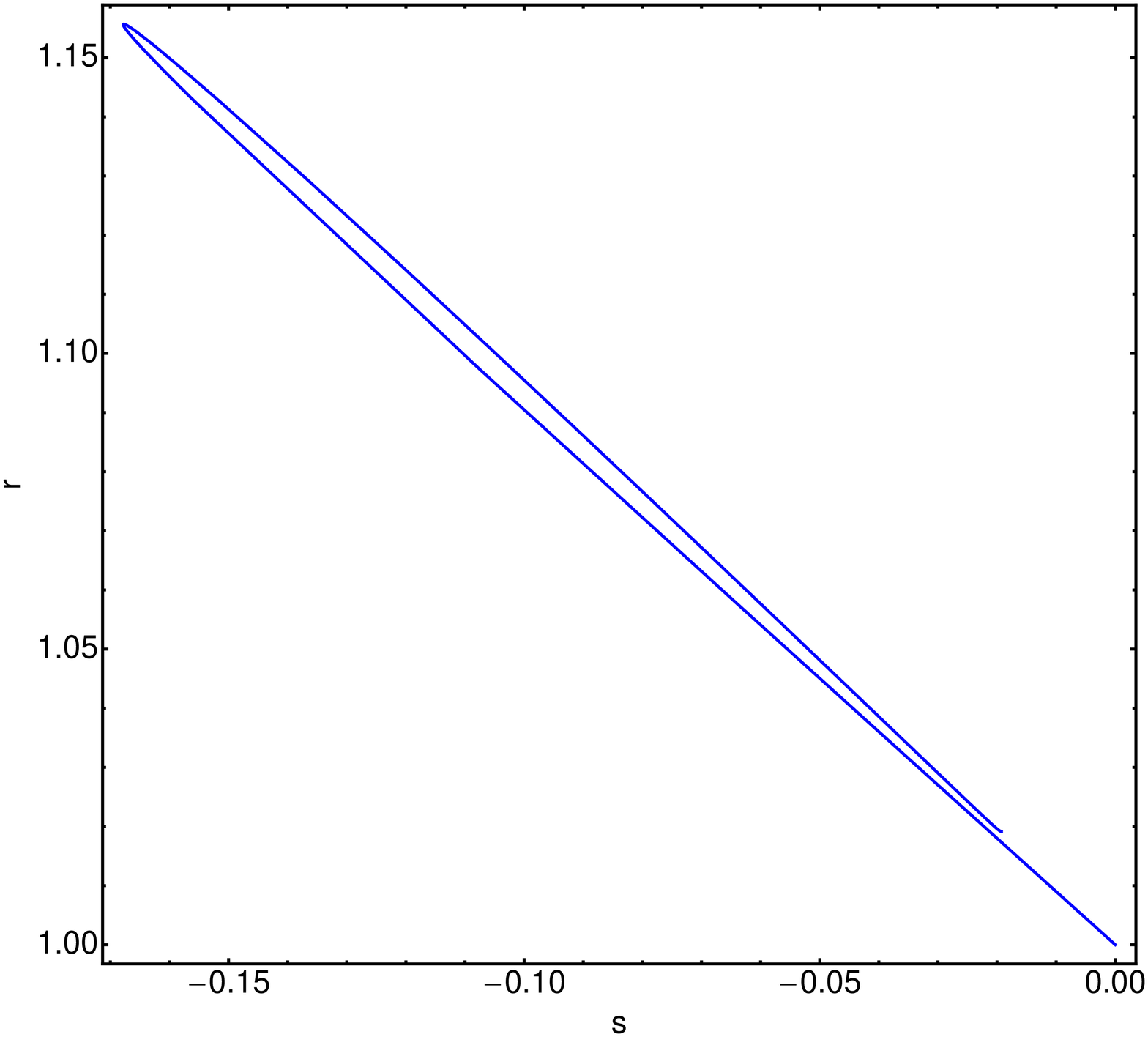}\\
 \end{array}$
 \end{center}
\caption{r-s for model 3. $\beta=2.5$ and $\gamma=0.1$ for the left plot. $\beta=3.5$ and $\gamma=0.3$ for the right plot. $\alpha_{0}=1.0$, $\alpha_{1}=1.5$, $\alpha_{2}=1.0$, $\omega_{b}=0.3$ and $b=0.01$.}
 \label{fig:rs}
\end{figure}

In the framework of GR,  Dark energy can explain the present cosmic
acceleration. Except cosmological constant many other candidates of
dark energy(quintom, quintessence, brane, modified gravity etc.) are
proposed. Dark energy is model dependent and to differentiate
different models of dark energy, a sensitive diagnostic tool is
needed.  Since $\dot{a}>0$, hence $H>0$ means the expansion of
the universe. Also, $\ddot{a}>0$ implies $q<0$. Since, the various dark
energy models give $H>0$ and $q<0$, they cannot provide enough
evidence to differentiate the more accurate cosmological
observational data and the more general models of dark energy. For
this aim we need  higher order of time derivative of scale factor
and geometrical tool. Sahni \emph{et.al.} \cite{Sahni} proposed
geometrical statefinder diagnostic tool, based on dimensionless
parameters $(r, s)$ which are function of scale factor and its time
derivative. These parameters are defined as

\begin{equation}\label{eq:statefinder}
r=\frac{1}{H^{3}}\frac{\dddot{a}}{a} ~~~~~~~~~~~~
s=\frac{r-1}{3(q-\frac{1}{2})}.
\end{equation}
For $8\pi G =1$ and $\Lambda=0$ we can obtain another form of
parameters $r$ and $s$:
\begin{equation}\label{eq:rsrhop}
r=1+\frac{9(\rho+P)}{2\rho}\frac{\dot{P}}{\dot{\rho}}, ~~~ s=\frac{(\rho+P)}{P}\frac{\dot{P}}{\dot{\rho}}.
\end{equation}
For the model 3 of our consideration, we presented the $\{r,s\}$ in
Fig.(\ref{fig:rs}) as a function of $\beta$ and $\gamma$.

As we know the pair $\{r,s\}=\{1,0\}$ corresponds to the $\Lambda$ CDM model. It is indicated on our graphs for both models. Further, $\{1,0\}$ which shows the CDM model, is present in our models. But we obsaerve the absence of Einstein static universe due to this fact that our models never mimic the pair $\{-\infty,+\infty\}$. So, our models fit the $\Lambda CDM$ and CDM perfectly.

\section{Observational constraints }
To use the $SNIa$ data, we define distance modulus $\mu$ as a
function of the luminosity distance $D_L$ as the following:
\begin{equation}
\mu=m-M=5\log_{10}{D_L},
\end{equation}
Here $D_{L}$ is in the following form:
\begin{equation}
D_{L}=(1+z)\frac{c}{H_{0}}\int_{0}^{z}{\frac{dz'}{\sqrt{H(z')}}}.
\end{equation}
Here  $m$ and $M$ denote the apparent magnitude and absolute
magnitude, respectively. Due to the photon-baryon plasma, Baryonic
acoustic oscillations exist in the decoupling redshift $z = 1.090$.
A major for scaling is the following quantity
\begin{equation}
A=\frac{\sqrt{\Omega_{m0} } }{H(z_{b})^{1/3}} \left[ \frac{1}{z_{b}} \int_{0}^{z_{b}}{\frac{dz}{H(z)}} \right ]^{2/3}.
\end{equation}
From WiggleZ-data \cite{Blake} we know that $A = 0.474 \pm 0.034$,
$0.442 \pm 0.020$ and $0.424 \pm 0.021$ at the redshifts $z_{b} =
0.44$, $0.60$ and $0.73$. The major statistical analysis parameter
is:
\begin{equation}
\chi^{2}{(x^{j})}=\sum_{i}^{n}\frac{(f(x^{j})_{i}^{t}-f(x^{j})_{i}^{0})^{2}}{\sigma_{i}},
\end{equation}
Here $f(x^{j})_{i}^{t}$ is the theoretical function of the model's
parameters. To conclude the work and model analysis we perform
comparison of our results with observational data. SNeIa data
allowed us to obtain the following observational constraints for our
models. For the Model 1, we found that the best fit can occurred
with $\Omega_{m0}=0.24$ and $H_{0}=0.3$. For $\alpha_{0}=0.3$,
$\alpha_{1}=0.5$, $\alpha_{2}=0.4$ and $\beta=4.0$, $\gamma=1.4$,
$\omega_{b}=0.5$, while for interaction parameter $b=0.02$. For the
Model 2, we found that the best fit we can obtain with $H_{0}=0.5$
and $\Omega_{m}=0.4$. Meanwhile for $\alpha_{0}=1.0$,
$\alpha_{1}=1.5$, $\alpha_{2}=1.3$ and $\beta=3.5$, $\gamma=0.5$,
$\omega_{b}=0.3$, while for interaction parameter $b=0.01$. Finally
we present the results obtained for Model 3, which say that the best
fit is possible when $H_{0}=0.35$ and $\Omega_{m0}=0.28$. For the
parameters $\alpha_{0}$, $\alpha_{1}$, $\alpha_{2}$, $\beta$,
$\gamma$, $\omega_{b}$ and $b$ we have the numbers $0.7$,$1.0$,
$1.2$, $3$, $0.8$, $0.75$ and $0.01$ respectively. Finally, we would
like to discuss the constraints resulted from $SNeIa+BAO+CMB$
\cite{obs} .

\begin{center}
    \begin{tabular}{ | l | l | l | l | l | l | l | l | l | l |}
    \hline
    M & $\alpha_{0}$ & $\alpha_{1}$ & $\alpha_{2}$ & $\beta$  & $\gamma$ & $\omega_{b}$ & $b$ & $H_{0}$ & $\Omega_{m0}$ \\ \hline
    1 & $0.3^{+0.35}_{-0.15}$ & $0.5^{+0.35}_{-0.4}$ & $0.4^{+0.35}_{-0.1}$ & $4.0^{+1.3}_{-2.7}$ & $1.4^{+0.25}_{-0.25}$ &  $0.5^{+0.4}_{-0.5}$ &  $0.01^{+0.07}_{-0.01}$ & $0.25^{+0.35}_{-0.05}$ &  $0.26^{+0.04}_{-0.03}$ \\ \hline
    2 & $1.2^{+0.2}_{-0.5}$ & $1.1^{+0.4}_{-0.3}$ & $0.7^{+0.55}_{-0.15}$ & $3.0^{+1.5}_{-0.8}$ & $0.7^{+0.12}_{-0.3}$ &  $0.4^{+0.6}_{0.2}$ &  $0.02^{+0.03}_{0.005}$ & $0.4^{+0.35}_{-0.2}$ &  $0.3^{+0.2}_{-0.05}$ \\ \hline
    3 & $0.7^{+0.5}_{-0.3}$ & $1.0^{+0.2}_{0.4}$ & $1^{+0.1}_{-0.3}$ & $3.0^{+1.0}_{0.3}$ & $0.8^{+0.1}_{-0.4}$ &  $0.7^{+0.6}_{-0.1}$ &  $0.01^{+0.09}_{-0.01}$ & $0.3^{+0.3}_{-0.1}$ & $ 0.21^{+0.15}_{-0.01} $\\ \hline
    \end{tabular}
\end{center}
\begin{figure}[h!]
 \begin{center}$
 \begin{array}{cccc}
\includegraphics[width=50 mm]{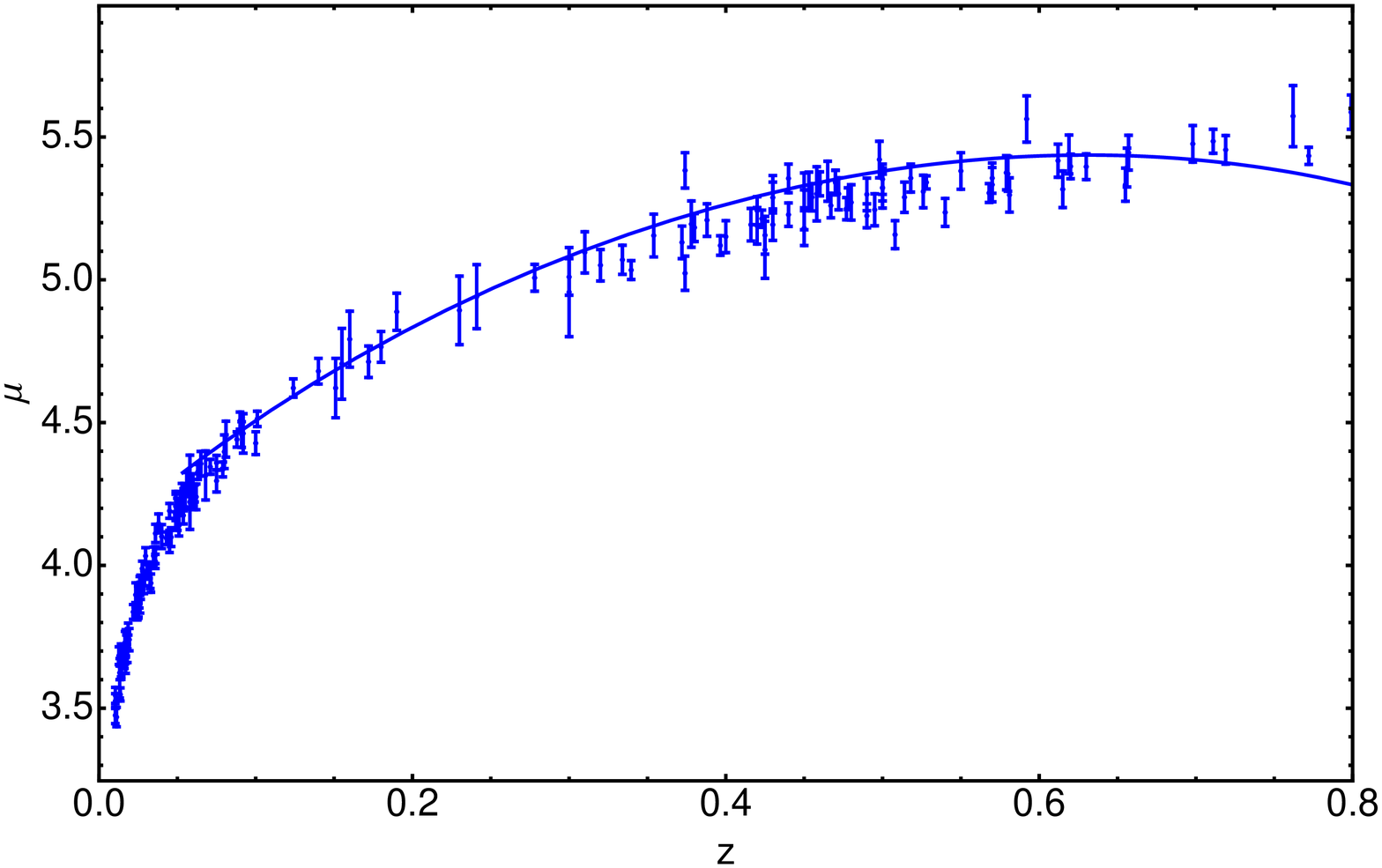} &
\includegraphics[width=50 mm]{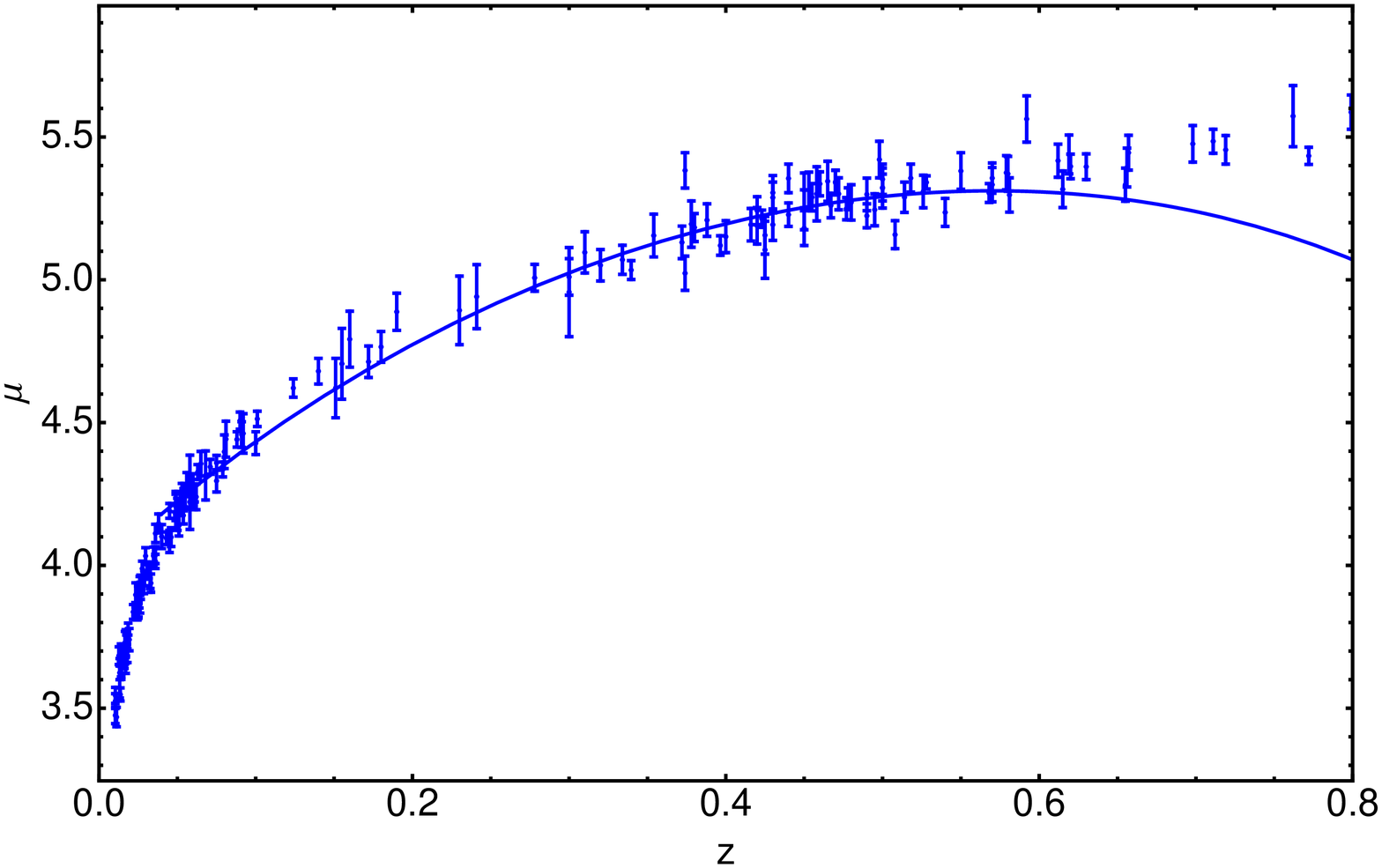}\\
 \end{array}$
 \end{center}
\caption{Observational data  $SneIa+BAO+CMB$ for distance modulus versus our theoretical results for models 1 and 2.}
 \label{fig:muz}
\end{figure}
From the graph of  luminosity distance versus zm we learn that how $\mu$ depends on the values of the parameters for different redshifts $z$. For different values of $\Omega_M,\Omega_D=0$ and i the regime of low redshifts $0.001<z<0.01$ , this graph has linearity. For $z>0.4$ the graph has typical form of models with $\Omega_M$. Hubble parameter $H$ has a centeral role in the behavior of $\mu(z)$ for different ranges of $z$.  We can use it to investigate the cosmological parameters.

\newpage
\section{Summary}
Time varying cosmological models with gravitational and cosmological
constant have been studied frequently. Nevertheless, in view of
cosmological data, rate of change of G is small. So the first order
correction terms are more important. Our approach to $\Lambda(t),
G(t)$ models is slightly different and more general than any other
previous work. As a proper generalization of general relativity,
scalar-tensor-vector gravity model has been proposed to explain the
structure of galaxies and dark matter problem. If we assume small
changes in the variation of the scalar fields, MOG model at the
level of action becomes equivalent to Einstein-Hilbert model, of
course it is necessary that we consider $G(t)$ as a slowly varying
scalar field. We proposed three models of generalized Ricci dark
energy including $\Lambda(t), G(t)$ to complete the time evolution
of dark energy. Due to the complexity of the model equations, the
numerical algorithms with cosmological parameters have been used.
Gravitational acceleration region and time evolution of state finder
parameters $\{r,s\}$ compared with $\Lambda$CDM model are
numerically studied with high accuracy.  We obtained the fit range
of data models by comparing the free parameters of dark energy
models and cosmological data $SNeIa+BAO+CMB$. Our model is a model
that is consistent with cosmological data while the other
theoretical models are not.

\begin{acknowledgments}

The authors thank J.W.Moffat for useful comments about MOG.

\end{acknowledgments}

\end{document}